\documentclass[prb,reprint,superscriptaddress,onecolumn]{revtex4-2}

\usepackage{amsmath,amssymb,bm,braket,mathrsfs,mathtools}
\usepackage{slashed}
\usepackage{hyperref}
\hypersetup{
	setpagesize=false,
	bookmarksnumbered=true,%
	bookmarksopen=true,%
	colorlinks=true,%
	linkcolor=magenta,
	citecolor=magenta,
}
\usepackage[dvipdfmx]{graphicx}
\usepackage{color}
\usepackage{threeparttable}
\usepackage{array}

\DeclareMathOperator{\tr}{Tr}

\DeclareMathOperator{\diag}{diag}

\begin{document}

\title{Electromagnetic response of spinful Majorana fermions}

\author{Shingo Kobayashi}
\affiliation{RIKEN Center for Emergent Matter Science, Wako, Saitama, 351-0198}
\author{Masatoshi Sato}
\affiliation{Yukawa Institute for Theoretical Physics, Kyoto University, Kyoto 606-8502, Japan}

\begin{abstract}
A remarkable feature of topological superconductors is the emergence of Majorana fermions in electron systems. Whereas the emergent Majorana fermions share the self-anti-particle property with Majorana fermions in particle physics, they may have essentially different electromagnetic properties.
In this paper, we argue the electromagnetic response of spinful Majorana fermions in topological superconductors. We present a general theory of the electromagnetic response of spinful Majorana fermions in topological superconductors and clarify how the pairing symmetry is encoded in the electromagnetic response.
As an application, we predict the sublattice-dependent dipole (Ising)-type magnetic response of corner Majorana fermions in iron-based superconductors.
\end{abstract}

\maketitle

\section{Introduction}
Topological superconductivity sheds new light on the investigation of superconductors \cite{Qi11, Tanaka12, Sato17}.
In particular, it has been revealed that topological superconductors (TSCs) may host self-conjugate fermionic excitations, named Majorana fermions (MFs).
Whereas such an interesting possibility was first pointed out in fully spin-polarized (spinless) $p$-wave superconductors \cite{Read00, Kitaev01}, it becomes more realistic after the discovery that $s$-wave superconductors with spin-orbit interaction also may support MFs \cite{Sato03, Fu08, SFT09, Jason10, Sau10, Lutchyn10, Oreg10}. 
Now, we have promising evidence of MFs in $s$-wave iron-based superconductors \cite{Wang15, Wu16, Xu16, Zhang18, Wang18, Machida19}:
The scanning tunneling microscope data for the vortex state clearly shows the zero-bias peak of the tunnel conductance, which is well-separated from other peaks of the mini-gap states \cite{Machida19}. 
The zero-bias peak is consistent with the existence of the Majorana zero mode localized at the vortex.

Nodal structures in unconventional superconductors are also known to be topologically protected and result in topological surface bands. For instance, line nodes in high-$T_{\rm c}$ cuprate are protected by the winding number,  giving surface flat bands \cite{Ryu02, Sato11}.
The surface flat band in high-$T_{\rm c}$ cuprates has been observed by the tunnel conductance measurement even before the recent progress of topological superconductors \cite{Kashiwaya2000}.  
A similar tunnel conductance measurement has now become the standard method to verify topological superconductivity.

As we mentioned above, the most distinct feature of TSCs is the possible realization of emergent MFs in various superconductors. 
However, we should note that there are two different types of MFs.
One is spinless MFs, and another is spinful MFs.
As we argue immediately, they have different properties, so they have different roles in studies of superconductors.

Let us first consider spinless MFs.
Remarkably, spinless MFs do not require any particular pairing symmetry except for the superconducting gap. Indeed, for weak Cooper pairing, only the Fermi surface topology in the normal state determines the presence or absence of a spinless MF. 
More precisely, we have the following criterion for spinless MFs (See Appendix \ref{sec:appendix_A} for the proof):
\begin{itemize}
    \item A spinless MF is realized if and only if the corresponding superconductor has an odd number of non-spin-degenerate Fermi surfaces.
\end{itemize}
Actually, this simple criterion explains spinless MFs in various systems.
For instance, a two-dimensional Dirac fermion with $s$-wave condensate supports a spinless Majorana zero mode in a vortex \cite{Sato03,Fu08}, which 
is consistent with the fact that the two-dimensional Dirac fermion has a single non-spin-degenerate Fermi surface. 
It has also been demonstrated that a spinless MF appears in Rashba $s$-wave superconducting state when the number of the non-spin-degenerate Fermi surfaces becomes odd by applying Zeeman magnetic fields \cite{SFT09}.
These spinless MFs may exhibit the non-Abelian anyon statistics and thus possibly apply to topological qubits.

On the other hand, the above criterion also implies that most superconductors may support only spinful MFs because they usually have spin-degenerate Fermi surfaces.
Remarkably, in contrast to spinless MFs, one needs particular pairing symmetries to realize spinful MFs.
In other words, spinful MFs may contain information on pairing symmetries, which
opens the possibility of detecting paring symmetries through spinful MFs.
This new application of spinful MFs should be explored because the determination of pairing symmetries is an essential but difficult problem in unconventional superconductors:
Pairing symmetries in most unconventional superconductors have not been established yet, except for a few, like the $d$-wave pairing symmetry in the high $T_{\rm c}$ cuprates. 

In this article, we examine electromagnetic responses of spinful MFs as a probe to identify pairing symmetries of the underlying unconventional superconductors.
We present a general theory of quantum responses for spinful MFs and point out that the information of pairing symmetries is compactly encoded in the quantum responses.
In particular, there exists a direct relation between pairing symmetries and electromagnetic responses for a particular class of spinful MFs.
As an application, we consider corner spinful MFs in iron-based superconductors and show that they show a sublattice-dependent dipole (Ising) spin behavior in accordance with the pairing symmetry.

\section{Electromagnetic response for spinful MFs of elementary particles}

First, we argue the electromagnetic response for the original spinful MFs in particle physics. 
Whereas elementary MFs in particle physics and emergent MFs in TCSs share the self-anti-particle nature, they show very different electromagnetic responses. This difference comes from the fundamental theorem in quantum field theory, that is, the $CPT$ theorem \cite{Streater00}: $CPT$ is a fundamental symmetry of relativistic quantum field theory, where $C$ is charge conjugation, $P$ is parity (inversion), and $T$ is time-reversal. The $CPT$ theorem tells us that any reasonable relativistic quantum field theory must be invariant under $CPT$. Since elementary MFs respect Lorentz invariance and the relativistic quantum field theory, they obey the $CPT$ theorem. This means that elementary MFs are self-conjugate under $CPT$, not merely under the charge conjugation $C$.\footnote{A physical Majorana neutrino can not be invariant under $C$ once $C$-violating weak interactions are turn on~\cite{Kayser82}.
Instead, it should be an eigenstate of $CPT$, which is not broken by the $CPT$-theorem.} Remarkably, this fundamental invariance of elementary MFs gives a strong constraint on electromagnetic responses: Using the Lorentz invariance and $CPT$ theorem, we can restrict the possible electromagnetic response of elementary MFs in the following manner~\cite{Kayser83}. 

We start with a state vector $|f({\bm p},J,s) \rangle$ describing elementary MFs with momentum ${\bm p}$, spin $J$, and $J_z=s$. 
From the $CPT$ theorem, 
we suppose that $|f({\bm p},J,s) \rangle$ is self-conjugate under $CPT\equiv \zeta$,
\begin{align}
 \zeta |f({\bm p},J,s) \rangle = \eta(s) |f({\bm p},J,-s) \rangle, \label{eq:cpt_eigen}
\end{align}
where we may have a phase factor $\eta(s)$ on the right-hand side because of the phase ambiguity of states.
Here, we allow the possible dependence of $s$ in $\eta$.
Then, for ${\bm p}=0$, considering a new operator $b \equiv \zeta e^{i \pi J_y}$, we also have
\begin{align}
b |f(0,J,s) \rangle = \mu(s) |f(0,J,s) \rangle,
\end{align}
with a phase factor $\mu(s)$. 
Since $\zeta$, hence, $b$ is anti-unitary\footnote{Note that the charge conjugation $C$ in the quantum field theory is unitary.}, the above equation leads to
\begin{align}
b^2 |f(0,J,s) \rangle =|f(0,J,s) \rangle.  
\end{align}
Therefore, using the relations $[\zeta, e^{i\pi J_y}]=0$ and $e^{i2\pi J_y}=(-1)^{2J}$,
we have
\begin{align}
(-1)^{2J}\zeta^2 |f(0,J,s)\rangle=|f(0,J,s)\rangle.
\label{eq:cpt_eigenv2}
\end{align}
From Eq.~(\ref{eq:cpt_eigen}), $\zeta$ also satisfies
\begin{align}
\zeta^2 |f(0,J,s) \rangle = (\eta(s))^{\ast} \eta(-s)  |f(0,J,s) \rangle.
\end{align}
Thus, Eq.~(\ref{eq:cpt_eigenv2}) implies
\begin{align}
 \eta(-s) =(-1)^{2J} \eta(s). \label{eq:eta_rel}
\end{align}
For $J=1/2$, we have $\eta(-1/2)=-\eta(1/2)$, hence $(\eta(s))^*\eta(s')=(-1)^{s-s'}$.

We now examine $CPT$ constraints on matrix elements of $\langle f | J_{\mu}| f \rangle$, where $J_{\mu}$ is the electromagnetic current. Since the electromagnetic interaction is $J_{\mu} A^{\mu}$ and $A^{\mu} (x=0)$ is $CPT$-odd, $J_{\mu} (x=0)$ must be $CPT$-odd too. Thus, one satisfies
\begin{align}
 \langle f({\bm p}_f,J,s_f) |J_{\mu}(0)| f({\bm p}_i,J,s_i) \rangle = - (\eta(s_i))^{\ast} \eta(s_f) \langle f({\bm p}_i,J,-s_i) |J_{\mu}(0)| f({\bm p}_f,J,-s_f) \rangle, \label{eq:cpt_constrain}
\end{align}
where we have used $\langle a | b \rangle = \langle \zeta b | \zeta a \rangle $.
For spin-$1/2$ relativistic fermions, the Lorentz invariance and the current conservation lead to the following general form of one-particle electromagnetic coupling~\cite{Kayser83}
 \begin{align}
 &\langle f({\bm p}_f,J,s_f) |J_{\mu}(0)| f({\bm p}_i,J,s_i) \rangle \nonumber \\
 &= i \bar{u}({\bm p}_f,s_f)[F  (q^2 \gamma_{\mu} -\slashed{q} q_{\mu}) + M \sigma_{\mu \nu} q^{\nu}+E \sigma_{\mu \nu} q^{\nu} \gamma_5 + G (q^2 \gamma_{\mu} -\slashed{q} q_{\mu})\gamma_5] u({\bm p}_i,s_i), \label{eq:dirac_em}
\end{align}
where $u$ is a Dirac spinor, and $q \equiv p_f-p_i$ is the energy-momentum four-vector.
The Dirac gamma matrices $\gamma_\mu$ ($\mu=0,1,2,3$) are given by
\begin{align}
\gamma_0 = \begin{pmatrix} \bm{1}_2 & 0 \\ 0 & -\bm{1}_2 \end{pmatrix}, \, \bm{\gamma} = \begin{pmatrix} 0 & \bm{\sigma}  \\ -\bm{\sigma} &0  \end{pmatrix}, \,  \gamma_5 \equiv  i \gamma_0 \gamma_1 \gamma_2 \gamma_3 =  \begin{pmatrix} 0 & \bm{1}_2 \\ \bm{1}_2 & 0 \end{pmatrix}, \,  \sigma_{\mu \nu} \equiv \frac{i}{2}[\gamma_{\mu},\gamma_{\nu}]. 
\end{align}
where $\bm{\sigma} = (\sigma_1, \sigma_2 ,\sigma_3)$ are the Pauli matrices and $\bm{1}_2$ is the $2 \times 2$ identity matrix, and $\bar{u}$ is defined as $\bar{u}=u^\dagger\gamma_0$.
The form factors $F$, $M$, $E$, and $G$ are functions of $q^2$, and these terms correspond to the electric charge, magnetic dipole moment, electric dipole moment, and toroidal moment, respectively. 

Using the relation
\begin{align}
 u({\bm p},-s) = (-1)^{s+\frac{1}{2}} \gamma_1 \gamma_3 \bar{u}^T ({\bm p},s), \label{eq:dirac_spinor_rel}
\end{align}
the right-hand side of Eq.~(\ref{eq:cpt_constrain}) is  rewritten as 
\begin{align}
 &-(\eta(s_i))^*\eta(s_f)
 \langle f({\bm p}_i,J,-s_i) |J_{\mu}(0)| f({\bm p}_f,J,-s_f) \rangle \nonumber \\
 &=-i(\eta(s_i))^*\eta(s_f)
\bar{u}({\bm p}_i,-s_i)[F (q^2 \gamma_{\mu} -\slashed{q} q_{\mu}) - M \sigma_{\mu \nu} q^{\nu}
 \nonumber\\
 &\hspace{30ex}
 -E \sigma_{\mu \nu} q^{\nu} \gamma_5 + G (q^2 \gamma_{\mu} -\slashed{q} q_{\mu})\gamma_5] u({\bm p}_f,-s_f) \notag \\
 &=-i(\eta(s_i))^*\eta(s_f)
 u^T({\bm p}_f,-s_f)[F (q^2 \gamma_{\mu} -\slashed{q} q_{\mu})^T - M \sigma_{\mu \nu}^T q^{\nu}
 \nonumber\\
 &\hspace{30ex}
 -E \gamma_5^T \sigma_{\mu \nu}^T q^{\nu}  + G \gamma_5^T(q^2 \gamma_{\mu} -\slashed{q} q_{\mu})^T ] \bar{u}^T({\bm p}_i,-s_i) \notag \\
 &=-i(\eta(s_i))^*\eta(s_f)
 (-1)^{s_f-s_i +1} \bar{u}({\bm p}_f,s_f) \gamma_3^T \gamma_1^T [F (q^2 \gamma_{\mu} -\slashed{q} q_{\mu})^T - M \sigma_{\mu \nu}^T q^{\nu}
 \nonumber\\
 &\hspace{30ex} 
 -E  \sigma_{\mu \nu}^T \gamma_5 q^{\nu}  
- G (q^2 \gamma_{\mu} -\slashed{q} q_{\mu})^T \gamma_5] \gamma_3\gamma_1 u({\bm p}_i,s_i) \notag  \\
 &= -i(\eta(s_i))^*\eta(s_f)
 (-1)^{s_f-s_i +1} \bar{u}({\bm p}_f,s_f)  [-F (q^2 \gamma_{\mu} -\slashed{q} q_{\mu}) - M \sigma_{\mu \nu} q^{\nu}
  \nonumber\\
 &\hspace{30ex} 
 -E  \sigma_{\mu \nu} \gamma_5 q^{\nu} + G (q^2 \gamma_{\mu} -\slashed{q} q_{\mu}) \gamma_5]  u({\bm p}_i,s_i), \label{eq:dirac_em2}
\end{align}
where we have used the properties of the Dirac gamma matrices:
\begin{align}
\gamma_3^T \gamma_1^T \gamma_\mu \gamma_3 \gamma_1=- \gamma_\mu, \ \ \gamma_3^T \gamma_1^T \sigma_{\mu\nu} \gamma_3 \gamma_1= \sigma_{\mu \nu}.
\end{align}
Since $(\eta(s_i))^{\ast} \eta(s_f) =(-1)^{s_i-s_f}$, as mentioned above, 
Eq.~(\ref{eq:cpt_constrain}) implies that $F$, $M$, and $E$ must vanish, and thus 
Eq.~(\ref{eq:dirac_em}) reduces to
 \begin{align}
 \langle f({\bm p}_f,J,s_f) |J_{\mu} (0)| f({\bm p}_i,J,s_i) \rangle = i \bar{u}({\bm p}_f,s_f)[ G (q^2 \gamma_{\mu} -\slashed{q} q_{\mu})\gamma_5] u({\bm p}_i,s_i).
\end{align}
In other words, spin-$1/2$ elementary MFs have neither electric charges, magnetic dipole moments, nor electric dipole moments. They only have toroidal moments.

Whereas the vanishing of electric charge directly follows from the self-anti-particle nature of MFs, the vanishing of magnetic and electric dipole moments is a consequence of the $CPT$ theorem. As shown immediately, this strong constraint can be derived in a more intuitive manner.

To see the strong constraint, let us suppose the elementary MF has a magnetic dipole moment $\bm{\mu}$. Then, the MF gets the energy
\begin{align}
 E = -\bm{\mu} \cdot \bm{B}  \label{eq:dipole_energy}
\end{align}
under a magnetic field $\bm{B}$.
We now take into account the $CPT$ theorem. The magnetic field is invariant under $C$ and $P$ but its sign flips under $T$ like $\bm{B} \to -\bm{B}$, so $\bm{B}$ is $CPT$-odd. 
Then, from the $CPT$ theorem requireing
the $CPT$-invariance of Eq.(\ref{eq:dipole_energy}),  $\bm{\mu}$ becomes $CPT$-odd. 
On the other hand, the elementary MF is self-conjugate under $CPT$, so ${\bm \mu}$ should coincide with its $CPT$ partner $-{\bm \mu}$, hence $\bm{\mu}$ must be zero.

The vanishing of the electric dipole moment of elementary MFs follows similarly.
For an MF with the electric dipole moment $\bm{p}$, we have the energy 
\begin{align}
E = - \bm{p} \cdot \bm{E}
\end{align}
under an electric field $\bm{E}$. Since $\bm{E}$ change to $-\bm{E}$ under $CPT$, $\bm{p}$ must be $CPT$-odd from the $CPT$ theorem. 
Then, the self-conjugate nature of an elementary MF under $CPT$ leads to $\bm{p}=0$.

Note that the above argument does not apply to emergent MFs in TSCs: Superconductors break the Lorentz invariance, and thus, they do not obey the $CPT$ theorem. 
As we discuss below, this difference enables rich structures of the electromagnetic response of emergent MFs.

\section{Electromagnetic response for MFs of TSCs}

We now turn to electromagnetic response for MFs of TSCs. In contrast to elementary MFs, spinful MFs of TSCs are not subject to the constraint of the $CPT$ theorem. 
Instead of the Lorentz invariance,
spinful MFs respect the crystalline symmetry of the underlying materials. 
In addition, they are self-conjugated just under $C$, not $CPT$, where $C$ is particle-hole symmetry (PHS) in the context of superconductivity. (See below.)  From these differences, spinful MFs may have a variety of electromagnetic responses, such as a magnetic dipole. In the following, we will show that their electromagnetic responses are closely related to the pairing symmetry of Cooper pairs, which offer valuable information about pairing mechanisms of superconductivity.    

\subsection{pairing symmetry}
First, let us see how the pairing symmetry is encoded in the topological classification.
In the topological classification~\cite{Schnyder08,Kitaev09,Schnyder09,Ryu10}, all the information of symmetry is compactly encoded in the following three relations~\cite{Shiozaki22}:
\begin{align}
 &U_g (\bm{k}) H(\bm{k}) U_g^{-1}(\bm{k}) = c(g) H(g \bm{k}), \ \ c(g) = \pm 1, \label{eq:rel1} \\
 &U_g(\bm{k}) i = \phi(g) i U_g(\bm{k}), \ \ \phi(g) =\pm1, \label{eq:rel2} \\
 &U_g(g'\bm{k})U_{g'}(\bm{k}) = e^{i \tau_{g,g'}(gg'\bm{k})} U_{gg'} (\bm{k}), \label{eq:rel3}
\end{align}
where $U_g(\bm{k})$ is the unitary operator of symmetry operation $g$ and $H(\bm{k})$ is the Hamiltonian in the momentum space: (i) The first equation specifies the type of symmetry. For $c(g) = +1$, $U_g$ is ordinary symmetry, and for $c(g) = -1$, $U_g$ is anti-symmetry like chiral symmetry. (ii) The second equation specifies unitarity or anti-unitarity of symmetry $g$. For $\phi(g)=+1$, the symmetry operator $U_g$ commutes with the imaginary unit $i$, so $U$ is unitary, and for $\phi(g)=-1$, $U_g$ anti-commute with $i$, so $U$ is anti-unitary. (iii) The last relation determines the commutation relation between symmetry operators, and the phase factor $e^{i \tau_{g,g'}(gg'\bm{k})} $ is called {\it twist} between $g$ and $g'$. 
Most cases show the trivial twist determined by the commutation relations between point groups, but some may have a non-trivial twist. Representative examples are unconventional superconductors and nonsymmorphic crystals. In particular, as we show immediately, the twist between PHS and crystalline symmetry encodes the information on the pairing symmetry of superconductors.

Consider the Hamiltonian of a superconductor
\begin{align}
H= \frac{1}{2} \sum_{\bm{k}, \alpha, \beta} \hat{\Psi}_{\alpha}^{\dagger} (\bm{k}) \tilde{H}_{\alpha \beta}(\bm{k}) \hat{\Psi}_{\beta} (\bm{k})    
\end{align}
with the Nambu spinor
\begin{align}
 \hat{\Psi}^T_{\alpha} (\bm{k}) = (\hat{c}_{\bm{k},\alpha},\hat{c}_{-\bm{k},\alpha}^{\dagger}).\label{eq:nambu}
\end{align}
The Bogoliubov-de Gennes (BdG) Hamiltonian $\tilde{H} (\bm{k})$ consists of the normal-state Hamiltonian ${\cal E}(\bm{k})$ and the gap function $\Delta (\bm{k})$, represented as  
\begin{align}
 \tilde{H}_{\alpha \beta}(\bm{k}) = \begin{pmatrix} {\cal E}_{\alpha \beta}(\bm{k})  & \Delta_{\alpha \beta}(\bm{k}) \\ \Delta_{\alpha \beta}^{\dagger}(\bm{k}) & -{\cal E}_{\alpha \beta}^{T} (-\bm{k}) \end{pmatrix}. \label{eq:bdg}
\end{align}
Here, $\hat{c}_{\bm{k},\alpha}$  ($\hat{c}_{\bm{k},\alpha}^{\dagger}$) is an annihilation (a creation) operator of electron with momentum $\bm{k}$ and internal degrees of freedom of electron $\alpha$ such as spin, orbital, and sublattice. $\delta_{\alpha \beta}$ is the Kronecker delta. From the Fermi statistics, the gap function satisfies $\Delta^T(\bm{k}) = -\Delta(-\bm{k}) $. 
Because the Nambu spinor $\hat{\Psi}_\alpha({\bm k})$ satisfies the self-conjugate relation,
\begin{align}
\hat{\Psi}^\dagger_\alpha({\bm k})\tau_x=\hat{\Psi}_\alpha^T(-{\bm k}),    
\quad
\tau_x=
\begin{pmatrix}
0&1\\
1&0
\label{eq:C_Nambu}
\end{pmatrix},
\end{align}
$\tilde{H}^{\rm BdG} (\bm{k})$ automatically has the following symmetry
\begin{align}
 C \tilde{H}(\bm{k}) C^{-1} = -\tilde{H}(-\bm{k}), \ \ C_{\alpha \beta}= \begin{pmatrix} 0 & \delta_{\alpha \beta}\\ \delta_{\alpha \beta} & 0\end{pmatrix} K,
\end{align}
where $K$ is the complex conjugate operator. This particular symmetry in superconductors is called PHS and enables TSCs to support MFs.

Conventionally, the transformation law of $\Delta (\bm{k})$ under crystalline symmetry specifies the pairing symmetry.
However, the information on pairing symmetry can also be encoded in twists between PHS and crystalline symmetry operators. For instance, let us consider odd-parity SCs (e.g., $p$-wave pairing). In this case, the gap function satisfies~\cite{Sato09,Fu10,Sato10}
\begin{align}
P \Delta(\bm{k}) P^{T} = - \Delta(-\bm{k}), \label{eq:gap-inversion}
\end{align}
under inversion $P$. Because of the additional minus sign on the right-hand side of Eq.~(\ref{eq:gap-inversion}), the inversion operation in the Nambu space should be
\begin{align}
 \tilde{P} = \begin{pmatrix} P & 0 \\ 0 & -P^{\ast}\end{pmatrix},
\end{align}
 which gives inversion symmetry of the BdG Hamiltonian, $ \tilde{P} \tilde{H}(\bm{k}) \tilde{P}^{\dagger} = \tilde{H}(-\bm{k})$. 
This equation implies that the electron and hole states in odd-parity superconductors behave in a different manner under inversion, and thus, inversion and PHS do not commute with each other. Thus, we have the non-trivial twist,
\begin{align}
 C\tilde{P} = -\tilde{P}C.
\end{align}

In a similar manner, we can encode the information of pairing symmetry in the twists
when the gap function does not spontaneously break the corresponding crystalline symmetry.
Namely, if the gap function obeys 
\begin{align} 
U_g(\bm{k}) \Delta(\bm{k}) U_g^T(\bm{k}) = \eta_g \Delta(g \bm{k}),
\label{eq:gap_g}
\end{align}
with a phase factor $\eta_g$ under crystalline symmetry operation $U_g({\bm k})$, 
the BdG Hamiltonian retains the crystalline symmetry,
\begin{align}
\tilde{U}_g({\bm k})\tilde{H}({\bm k})\tilde{U}_g^\dagger({\bm k})=\tilde{H}(g{\bm k}).    
\end{align}
where the crystalline symmetry operator in the Nambu basis is given by
\begin{align}
 \tilde{U}_g(\bm{k}) = \begin{pmatrix} U_g(\bm{k}) & 0 \\ 0 & \eta_g U_g^{\ast}(\bm{k})\end{pmatrix}. \label{eq:op_nambu}
\end{align}
Then, we have an additional factor in the commutation relation between PHS and crystalline symmetry operators:
 \begin{align}
 C\tilde{U}_g(\bm{k}) = \eta_g^{\ast} \tilde{U}_g(-\bm{k})C.
\label{eq:twist_UC}
\end{align}

\subsection{electromaginetic responses}
We now consider the electromagnetic responses of MFs. To evaluate the possible response of MFs, we consider a general local operator of the electrons. In the Nambu basis defined by Eq.~(\ref{eq:nambu}), the local operator in the real space reads
\begin{align}
 \hat{O} &= \sum_{\alpha \beta} \hat{c}_{\alpha}^{\dagger}(\bm{x}) O_{\alpha \beta} \hat{c}_{\beta} (\bm{x}) \nonumber \\
  &= \frac{1}{2} \hat{\Psi}^{\dagger}(\bm{x}) \mathcal{O} \hat{\Psi}(\bm{x}),
\end{align}
where $O$ is an arbitrary Hermitian operator $O$ and ${\cal O}$ is its representation in the Nambu basis, 
\begin{equation}
 \mathcal{O} = \begin{pmatrix} O & 0 \\ 0 & -O^T \end{pmatrix}.
\end{equation}
Here, we have neglected an irrelevant constant term. We also note that the Hermiticity of $O$ implies $\{\mathcal{O}, C\}=0$.  
To extract the contribution of MFs, we first use the self-conjugate nature in Eq.(\ref{eq:C_Nambu}) of the Nambu spinor, which is given by 
\begin{align}
\hat{\Psi}^{\dagger}({\bm x}) \tau_x = \hat{\Psi}^T({\bm x})    
\end{align}
in the real space, so the local operator is recast into
 \begin{align}
 \hat{O} = \frac{1}{2} \hat{\Psi}^{T}(\bm{x}) [\tau_x \mathcal{O}] \hat{\Psi}(\bm{x}).
\end{align}
Then, we perform the mode expansion of the Nambu spinor as
\begin{align}
 \hat{\Psi}(x) = \sum_a \hat{\gamma}^{(a)} |u_0^{(a)} \rangle + \cdots,
\end{align}
where $\hat{\gamma}^{(a)}$ is a Majorana operator, $|u_0^{(a)} \rangle$ represents a wave function of $a$th zero energy state, and $\cdots$ implies states with finite energy. In the low-energy limit, the finite energy states can be neglected. Thus, we obtain the local operator in terms of MFs~\cite{Kobayashi2019}
\begin{align}
 \hat{O}_{\rm MF} =-\frac{1}{8} \sum_{ab}[\hat{\gamma}^{(a)},\hat{\gamma}^{(b)}] \tr[\mathcal{O} \rho^{(ab)}(\bm{x})], \label{eq:em_mf}
\end{align}
with
\begin{align}
 \rho^{(ab)} =  |u_0^{(a)} \rangle \langle C u_0^{(b)}| - |u_0^{(b)} \rangle \langle C u_0^{(a)}| ,
\end{align}
where we have used $|Cu_0^{(a)} \rangle = \tau_x |u_0^{\ast \, (a)} \rangle$. 
This equation implies that  
the local operator has a nonzero coupling to  MFs only when the trace part in Eq.~(\ref{eq:em_mf}) is nonzero.
Then, from the group theoretical argument \cite{Kobayashi2019}, we can conclude that $\hat{O}_{\rm MF}$ is nonzero only when ${\cal O}$ shares the same irreducible representation of symmetry as $\rho^{(ab)}$.

Interestingly, $\rho^{(ab)}$ transforms as a bi-product of electrons like Cooper pairs under the crystalline symmetry: When applying a crystalline symmetry operation to the MF state $|u_0^{(a)}\rangle$, we have an MF state again, and thus, the following relation obeys
\begin{align}
 \tilde{U}_g |u_0^{(a)} (g \bm{x}) \rangle = \sum_b |u_0^{(b)} (\bm{x}) \rangle \mathcal{U}_{ba}(g). \label{eq:mf_trans}
\end{align} 
Therefore, MFs are a (projective) representation of crystalline symmetry in a manner similar to electrons. 
Similarly, for the PHS partner state $|Cu_0^{(a)} \rangle$, we have
\begin{align}
 \tilde{U}_g |Cu_0^{(a)} (g \bm{x}) \rangle &=\eta_g C \tilde{U}_g |u_0^{(a)} (g \bm{x}) \rangle \nonumber \\
 &= \eta_g \sum_b |C u_0^{(b)} (\bm{x}) \rangle \mathcal{U}_{ba}^{\ast}(g). \label{eq:mfc_trans}
\end{align}
where we have used Eq.(\ref{eq:twist_UC}) and the anti-unitarity of $C$. 
Therefore, the PHS partner behaves like a hole state under the crystalline symmetry. 
Finally, from Eqs.~(\ref{eq:mf_trans}) and (\ref{eq:mfc_trans}), $\rho^{(ab)}$ transforms, under the crystalline symmetry operation,  as
\begin{align}
\tilde{U}_g \rho^{(ab)}(g \bm{x}) \tilde{U}_g^{\dagger} = \sum_{cd} \rho^{(cd)} (\bm{x}) \eta_g^{\ast} \mathcal{U}_{ca}(g) \mathcal{U}_{db}(g), \label{eq:rho_trans}
\end{align}
which means that $\rho^{(ab)}$ transforms like a bi-product of electrons. 
Note that non-trivial factor $\eta_g$ in pairing symmetry appears in Eq.~(\ref{eq:rho_trans}), so MFs know the pairing symmetry of Cooper pairs. 

Below, we focus on spinful MFs in time-reversal invariant TSCs and assume they form a single Majorana Kramers pair (MKP). In this case, we have a direct relation between the MKP and pairing symmetry under $g$. Since MKP states $|u_0^{(1)}\rangle$ and $|u_0^{(2)}\rangle$ form a Kramers pair, they satisfy
\begin{align}
 T|u_0^{(1)} (\bm{x}) \rangle = e^{i\alpha}|u_0^{(2)} (\bm{x}) \rangle,
\end{align}
where $T$ is a time-reversal operator with $T^2=-1$, and $\alpha$ is a real constant.
Thus, $\rho^{(12)}$ for the MKP is recast into
\begin{align}
 \rho^{(12)}_{\rm MKP} = e^{i\alpha}(|u_0^{(1)} \rangle \langle CT u_0^{(1)}| - |Tu_0^{(1)} \rangle \langle C u_0^{(1)}|).
\end{align}
This equation implies that $\rho^{(12)}$ behave like a ``spin-singlet'' state under $g$
since $|u_0^{(1)} \rangle$ and $T|u_0^{(1)} \rangle$ corresponds to spin-up  and spin-down states and $\rho^{(12)}$ is their antisymmetric product. (Note that $C$ does not affect the spin.)
In particular, for $g$ preserving the position of MFs, which is a necessary condition for symmetry protection by $g$,  $\mathcal{U}_{ca}(g) \mathcal{U}_{db}(g)$ represents a rotation in the spin space. Thus, this factor becomes trivial for the MKP since the spin-singlet state is invariant under any spin rotation. Consequently, we have 
\begin{align}
\tilde{U}_g\rho^{(12)}_{\rm MKP}\tilde{U}_g^\dagger=\eta_g^*\rho^{(12)}_{\rm MKP}    
\label{eq:rho_U}
\end{align}
which is essentially the same as the transformation law of the gap function in Eq.(\ref{eq:gap_g}).
Moreover, we can show that $\rho^{(12)}_{\rm MKP}$ is odd under $T$ operation ~\cite{Kobayashi2021},
 \begin{align}
  T [\rho_{\rm MKP}^{(12)}]^{\dagger} T^{-1} = -\rho_{\rm MKP}^{(12)},
\label{eq:rho_T} 
 \end{align}
which means that $\rho_{\rm MKP}^{(12)}$ has a non-zero trace only with a magnetic operator (=a time-reversal odd operator) \cite{Kobayashi2021}. 
Therefore, a single MKP protected by $g$ can couple only to a magnetic operator that has the same representation of $g$ as the gap function.

When we apply a magnetic field ${\bm B}$ to the MKP, the above coupling generally induces the magnetic multipole term,
\begin{align}
H_{\rm m}=\int d{\bm x}g_{\rm m}({\bm B})\hat{O}_{\rm MF}({\bm x}),   
\end{align}
where $g_{\rm m}({\bm B})$ is a function of ${\bm B}$.
From Eqs.(\ref{eq:rho_U}) and (\ref{eq:rho_T}), $\hat{O}_{\rm MF}$ transforms as
\begin{align}
\hat{O}_{\rm MF}({\bm x}) \xrightarrow{g}
\eta^*_g\hat{O}_{\rm MF}({\bm x}),
\quad
\hat{O}_{\rm MF}({\bm x}) \xrightarrow{T}-\hat{O}_{\rm MF}({\bm x}),
\end{align}
under $g$ and $T$. 
Then, because the system is invariant under these symmetries once we transform the applied magnetic field at the same time, $g_{\rm m}({\bm B})$ must transform as
\begin{align}
g_{\rm m}({\bm B}) \xrightarrow{g}g_{\rm m}(g{\bm B})=\eta_g g_{\rm m}({\bm B}),
\quad
g_{\rm m}({\bm B}) \xrightarrow{T}g_{\rm B}(-{\bm B})=-g_{\rm m}({\bm B}),
\end{align}
which determines possible forms of $g_{\rm m}({\bm B})$ in the induced multipole term. The magnetic multipole term is bi-linear in the Majorana operator $\hat{\gamma}^{(a)}$, and thus, it gives finite energy to the Majorana zero modes.

\begin{table}[ht]
 \caption{Possible magnetic multipoles of a single MKP protected by two-dimensional point groups (PGs) and material candidates of TSCs. ``N/A'' means no MF exists; instead, a symmetry-protected nodal point appears. The rotation axis is parallel to the $z$ axis. The irreducible representations (IRs) are labeled in the Mulliken notation~\cite{Bradley72}.  }
 \label{table:mulipole}
 \centering
\begin{threeparttable}
  \begin{tabular}{cccccc}
   \hline
PG &  Spin of MFs & IR of $\Delta$ & Magnetic multipole $g_{\rm m}(\bm{B})$ & Type & Material Candidates \\
\hline
$C_1$ & $1/2$  & A& $B_x,B_y,B_z$ & Dipole &\\
$C_2$ & $1/2$  & A& $B_z$ & Dipole & $^3$He-B~\cite{Chung09,Mizushima12}, UBe$_{13}$~\tnote{a} \\
         & $1/2$  & B& $B_x,B_y$ & Dipole &\\
$C_3$ & $1/2$ or $3/2$  & A& $B_z$ & Dipole &\\
$C_4$ & $1/2$ or $3/2$  & A& $B_z$ & Dipole &\\
         & $1/2$ or $3/2$  & B& N/A & &\\
$C_6$ & $1/2$ or $5/2$  & A& $B_z$ & Dipole&\\ 
         & $1/2$ or $5/2$  & B& N/A & &\\
         & $3/2$  & A& $B_z$ & Dipole &\\
         & $3/2$  & B& $B_x^3-3B_xB_y^2,B_y^3-3B_yB_x^2$ & Octapole &\\
$C_{s}$ & $1/2$  & A& $B_z$ & Dipole & Monolayer FeSe~\cite{Qin2022} \\
           & $1/2$  & B& $B_x,B_y$ & Dipole &\\
$C_{2v}$ & $1/2$  & A$_1$& N/A & &\\
           & $1/2$  & A$_2$& $B_z$ & Dipole & UTe$_2$~\cite{Tei2023}~\tnote{b} \\
           & $1/2$  & B$_1$& $B_y$ & Dipole & UTe$_2$~\cite{Tei2023}~\tnote{b} \\
           & $1/2$  & B$_2$& $B_x$ & Dipole & UTe$_2$~\cite{Tei2023}~\tnote{b} \\
$C_{3v}$ & $1/2$  & A$_1$& N/A & &\\
           & $1/2$  & A$_2$& $B_z$ & Dipole & Cu$_x$Bi$_2$Se$_3$~\cite{Fu14}\\
           & $3/2$  & A$_1$& $B_x^3-3B_xB_y^2$ & Octapole & YPtBi~\cite{Brydon16}\\
           & $3/2$  & A$_2$& $B_z$ & Dipole &\\
$C_{4v}$ & $1/2$ or $3/2$ & A$_1$& N/A & & \\
           & $1/2$ or $3/2$  & A$_2$& $B_z$ & Dipole &\\
           & $1/2$ or $3/2$  & B$_1$& N/A & &\\
           & $1/2$ or $3/2$  & B$_2$& N/A & &\\
$C_{6v}$ & $1/2$ or $5/2$ & A$_1$& N/A & &\\
           & $1/2$ or $5/2$  & A$_2$& $B_z$ & Dipole &\\
           & $1/2$ or $5/2$  & B$_1$& N/A & &\\
           & $1/2$ or $5/2$  & B$_2$& N/A & &\\
           & $3/2$ & A$_1$& N/A & &\\
           & $3/2$ & A$_2$& $B_z$ & Dipole &\\
           & $3/2$ & B$_1$& $B_y^3-3B_yB_x^2$ & Octapole &\\
           & $3/2$ & B$_2$& $B_x^3-3B_xB_y^2$ & Octapole &\\
   \hline
  \end{tabular}
   \footnotesize            
   \setlength{\extrarowheight}{2pt}
   \begin{tablenotes}[normal]
    \item[a] For pairing symmeteries of UBe$_{13}$, there are two scenarios: the degenerate $E_{u}$ scenario \cite{Shimizu2017,Machida2018,Mizushima2018} and accidental scenario \cite{Sigrist1989}. Both scenarios predict that UBe$_{13}$ is a candidate of fully-gapped TSCs with a single MKP.
    \item[b]  Some pairing symmetries lead to double or more MKPs at a high-symmetry point on a surface Brillouin zone. In these cases, the magnetic dipole appears as the leading order contribution to the magnetic response~\cite{Yamazaki2021prb}. 
    \end{tablenotes}
\end{threeparttable}
\end{table}

In Table~\ref{table:mulipole}, we summarize the possible induced magnetic multipole terms of a single MKP protected by two-dimensional point groups. 
Note that a different representation of the gap function $\Delta$, namely a different $\eta_g$, gives a different magnetic multipole term.
Thus, by detecting the  magnetic response of a MKP, 
we can identify the pairing symmetry $\eta_g$. 
We also have a similar result in three dimensions.

For systems with double or more MKPs, Eq.~(\ref{eq:rho_trans}) constrains the electromagnetic structures. Since $\rho^{(ab)}$ transforms as an antisymmetric product representation under the action of $g$, Eq.~(\ref{eq:rho_trans}) is recast into
\begin{align}
 \tilde{U}_g \rho^{(ab)}(g\bm{x}) \tilde{U}_g^{\dagger} = \sum_{cd} \rho^{(cd)}(\bm{x}) [\Omega_g]_{(cd)(ab)},
\end{align}
with 
\begin{align}
    [\Omega_g]_{(cd)(ab)} = \frac{\eta_g^\ast}{2} \left[ \mathcal{U}_{ca}(g)\mathcal{U}_{db}(g)- \mathcal{U}_{cd}(g)\mathcal{U}_{ab}(g) \right].
\end{align}
Then, the character of $\Omega_g$ determines the representation of $\rho^{(ab)}$. Taking the trace of $\Omega_g$, the character reads
\begin{align}
    \chi_g^{\Omega} = \frac{\eta_g^\ast}{2} \left\{ \left( \tr\left[ \mathcal{U}(g)\right] \right)^2 - \tr\left[ \mathcal{U}^2(g)\right] \right\}. \label{eq:chig}
\end{align}
By decomposing $\mathcal{U}(g)$ into $\oplus_{\alpha} \mathcal{U}_{\alpha}(g)$ with $\alpha$ being the label of irreducible representations of crystalline symmetry, we eventually have
\begin{align}
     \chi_g^{\Omega} = \frac{\eta_g^\ast}{2} \left\{ \left( \sum_\alpha \tr\left[ \mathcal{U}_{\alpha}(g)\right] \right)^2 - \sum_{\alpha} \tr\left[ \mathcal{U}_{\alpha}^2(g)\right] \right\}. \label{eq:chig2}
\end{align}
Then, one can calculate the right-hand side of Eq.~(\ref{eq:chig2}) from the character table for $g$ without referring to the explicit form of $\mathcal{U}(g)$. 
The resultant character determines the representation of $\rho^{(ab)}$, and thus, it also determines possible electromagnetic coupling $\hat{O}_{\rm MF}$, as described above. 

Generally, $\rho^{(ab)}$ for multiple MKPs contains both time-reversal odd and even representation, giving electronic and magnetic responses, respectively.

\section{Applications to various TSCs}
\label{sec:extended_s}

Finally, we argue the application of our theory to candidate materials of TSCs. In the previous studies~\cite{Shiozaki14,Xiong2017,Kobayashi2019,Kobayashi2021}, we and the collaborators have revealed that the magnetic response of an MKP can identify various pairing symmetry. For instance, our theory predicts magnetic dipole response for an MKP in superconducting topological insulators such as {\it A}$_{x}$Bi$_2$Se$_3$ ($A$=Cu, Sr, Nb)~\cite{Hor10,Sasaki11,Matano16,Yonezawa17}. We have shown that the direction of the magnetic dipole is fully determined by the pairing symmetry of unconventional Cooper pairs. Another nontrivial example is high-spin TSCs in YPtBi~\cite{Goll08,Butch11,Tafti13,GXu16,Bay12,Kim18}. In this material, a proposed gap function has multipole structure due to higher spin~\cite{Brydon16}, which gives rise to magnetic octupole response on $(111)$ surface on which an MKP protected by $C_{3v}$ symmetry resides~\cite{Kobayashi2019} (See also Table~\ref{table:mulipole}). Moreover, the theory has been extended to the cases with nonsymmorphic crystalline symmetry and double MKPs in Refs.~\cite{Kobayashi2021,Yamazaki2019,Yamazaki2021prb,Yamazaki2021jpsj} and a large variety of responses of MFs beyond magnetic dipole and octapole responses have been proposed. Examples include magnetic-quadrupole-like response~\cite{Yamazaki2021prb} in nonsymmorphic topological crystalline superconductor UCoGe~\cite{Daido2019} and strain-induced-electric response~\cite{Yamazaki2021jpsj} in high-spin topological crystalline superconductor Sr$_3$SnO~\cite{Kawakam2019}.

\subsection{Corner MKP in higher-order TSC}
Our theory is also applicable to MKPs in higher-order TSCs~\cite{Langbehn2017,Geier2018,Volpez19}. 
Here, we apply our theory to corner MKPs in iron-based superconductors.

Recently, an extended $s$-wave superconducting state in iron-based materials has been proposed as a member of higher-order TSCs~\cite{Qin2022}. This finding is surprising because, in the presence of time-reversal and inversion symmetries, most $s$-wave superconductors are known to be topologically trivial. Nevertheless, the authors of Ref.~\cite{Qin2022} have discovered that an extended $s$-wave SC is not always the case, where the gap function changes its sign depending on the Fermi surfaces. They have found that an extended $s$-wave superconductor with a nonsymmorphic crystal of iron-based superconductors
may realize a second-order TSC, which hosts a corner MKP protected by mirror-reflection symmetry. 
Thus, iron-based superconductors such as monolayer FeSe~\cite{Wang2012interface,Wang2017high} are expected to be a good platform for studying the magnetic response of spinful MFs.

\begin{figure}[tb]
\centering\includegraphics[width=6in]{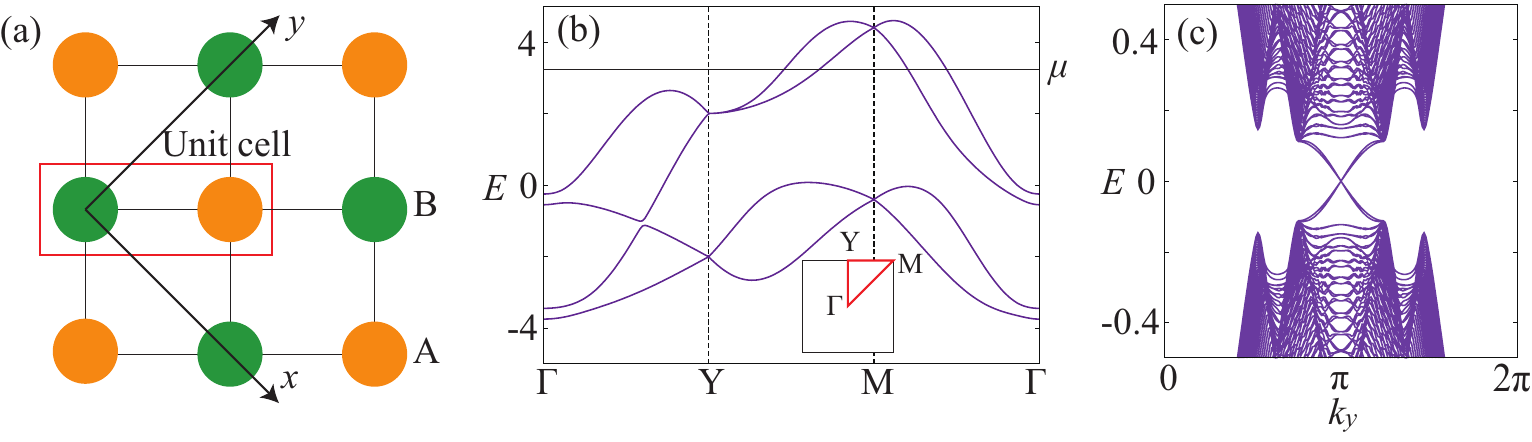}
\caption{(a) Square lattice with two sites A and B in the unit cell. (b) The band structure of the tight-binding model~(\ref{eq:normal}) along the red line in the inset, where we choose the parameters as $(t,t_1,t_2,\lambda,\lambda_{\rm R}) = (-1.0, 0.4, 0.6, 0.3, 0.75)$. (c) Surface energy spectrum of the BdG Hamiltonian (\ref{eq:bdg}) in the (10) surface as a function of $k_y$, where the order parameters and chemical potential are chosen as $(\Delta_0,\Delta_1,\mu)=(-0.58, -0.2, 3.6)$. }
\label{fig:band}
\end{figure}

To examine the magnetic response in detail, we consider a 2D tight-binding model proposed in Ref.~\cite{Qin2022}, with space group symmetry $P4/nmm$ (SG \#. 129). 
The unit cell of the model consists of $A$ and $B$ sites, and electrons in $p_x$ and $p_y$ orbitals reside in each site. The lattice structure is shown in Fig.~\ref{fig:band} (a). The normal state Hamiltonian is given by, in the momentum space,
\begin{align}
    {\cal E}(\bm{k}) = \begin{pmatrix} h^A(\bm{k}) & h^{AB} (\bm{k}) \\ [h^{AB} (\bm{k})]^{\dagger} & h^B(\bm{k})\end{pmatrix}_\eta-\mu\eta_0s_0\sigma_0, 
\label{eq:normal}
\end{align}
with 
\begin{align}
    h^A(\bm{k}) = &t \cos (k_x) s_0 (\sigma_0+\sigma_3) + t \cos (k_y) s_0 (\sigma_0-\sigma_3) \nonumber \\
    &- \lambda_{\rm R} \sin(k_x) s_2 (\sigma_0+\sigma_3) + \lambda_{\rm R} \sin (k_y)s_1(\sigma_0-\sigma_3)+\lambda/2 s_3 \sigma_2, \\
    h^B(\bm{k}) = &t \cos (k_x) s_0 (\sigma_0+\sigma_3) + t \cos (k_y) s_0 (\sigma_0-\sigma_3) \nonumber \\
    &+ \lambda_{\rm R} \sin(k_x) s_2 (\sigma_0+\sigma_3) - \lambda_{\rm R} \sin (k_y)s_1(\sigma_0-\sigma_3) +\lambda/2 s_3 \sigma_2, \\
    h^{AB}(\bm{k}) = &t_1[1+e^{ik_x}+e^{ik_y}+e^{i(k_x+k_y)}]s_0 \sigma_0 -t_2[1-e^{ik_x}-e^{ik_y}+e^{i(k_x+k_y)}]s_0 \sigma_1, 
\end{align}
where $s_i$ and $\sigma_i$ ($i=1,2,3$) are the Pauli matrices in the spin and orbital spaces, respectively, and $\eta_0$, $s_0$, $\sigma_0$ are the $2 \times 2$ identity matrices in the sublattice, spin, and orbital spaces. In Eq.~(\ref{eq:normal}), $h^A$ ($h^B$) describes the intra-sublattice hopping terms for the A (B) site and $h^{AB}$ describes the inter-sublattice hopping terms. The subscript $\eta$ implies the matrix in the sublattice space. Here, $t$ is the intra-sublattice hopping amplitude, $t_1,t_2$ are the inter-sublattice hopping amplitudes, $\mu$ is the chemical potential, $\lambda$ is the atomic spin-orbit coupling, and $\lambda_{\rm R}$ is the Rashba-type spin-orbit coupling allowed by the symmetry group $P4/nmm$. 

The normal-state Hamiltonian (\ref{eq:normal}) is invariant under the time-reversal symmetry $T$:
\begin{align}
 &T {\cal E}(\bm{k}) T^{-1} = {\cal E}(-\bm{k}), \ \ T = i\eta_0 s_2 \sigma_0 K, \label{eq:trs_model}
 \end{align}
 where $K$ is the complex conjugation operation and $\eta_i$ (i=0,1,2,3) are the identity matrix and Pauli matrices in the sublattice space.
 The Hamiltonian is also invariant under the space group symmetry $P4/nmm$. The generators of $P4/nmm$ are given by the inversion $\{P|\hat{\bm x}/2+\hat{\bm y}/2\}$, mirror-reflection of the $xz$ plane $\{M_{xz}|\bm{0}\}$, mirror-reflection of the $yz$ plane $\{M_{yz}|\bm{0}\}$, two-fold rotation about  the $z$ axis $\{C_{2z}|\bm{0}\}$, and four-fold-rotation about the $z$ axis $\{C_{4z}|\bm{0}\}$~\footnote{Here the Seitz symbol $\{p|\bm{a}_p\}$ represents a space group operation consisting of a point-group operation $p$ and a translation $\bm{a}_p$.}, where $\hat{\bm{x}}$ ($\hat{\bm{y}}$) is the unit vector in the $x$ ($y$) direction. For the normal state Hamiltonian, they are represented as 
\begin{align}
 &P {\cal E}(\bm{k}) P^{\dagger}= {\cal E}(-\bm{k}), \ \ P = \eta_1 s_0 \sigma_0, \label{eq:is_model} \\
 &M_{xz}(\bm{k}) {\cal E}(k_x,k_y) M_{xz}^{\dagger}(\bm{k}) = {\cal E}(k_x,-k_y), \ \ M_{xz}(\bm{k}) = i \begin{pmatrix}
     1 & 0 \\ 0 & e^{i k_y}
 \end{pmatrix}_\eta s_2 \sigma_3 , \label{mrs_model} \\
 &M_{yz}(\bm{k}) {\cal E}(k_x,k_y) M_{xz}^{\dagger}(\bm{k}) = {\cal E}(-k_x,k_y), \ \ M_{yz}(\bm{k}) = i \begin{pmatrix}
     1 & 0 \\ 0 & e^{i k_x}
 \end{pmatrix}_\eta s_1 \sigma_3 , \label{mrs_yz_model} \\
 &C_{2z}(\bm{k}) {\cal E}(\bm{k}) C_{2z}^{\dagger}(\bm{k}) = {\cal E}(-\bm{k}), \ \ C_{2z}(\bm{k}) = i \begin{pmatrix}
     1 & 0 \\ 0 & e^{i (k_x+k_y)}
 \end{pmatrix}_\eta s_3 \sigma_0 , \label{c2z_model} \\
 &C_{4z}(\bm{k}) {\cal E}(k_x,k_y) C_{4z}^{\dagger}(\bm{k}) = {\cal E}(k_y,-k_x), \ \ C_{4z}(\bm{k}) = -i \begin{pmatrix}
     1 & 0 \\ 0 & e^{i k_x}
 \end{pmatrix}_\eta e^{i s_3 \pi/4} \sigma_2 . \label{screw_model} 
\end{align}
Here, the mirror and rotation operators have non-trivial ${\bm k}$-dependence since $P4/nmm$ is nonsymmorphic.
The band structure of ${\cal E}$ is shown in Fig.~\ref{fig:band} (b). All the bands are doubly degenerate due to the Kramers theorem. The band crossings at the $Y$ and $M$ points come from the nature of nonsymmorphic symmetry~\cite{Qin2022}.

We consider an extended $s$-wave superconducting state of the system. The superconducting state is described by the BdG Hamiltonian~(\ref{eq:bdg})
with the gap function given by
\begin{equation}
 \Delta(\bm{k}) = [\Delta_0 + 2 \Delta_1 (\cos (k_x) + \cos (k_y))] i\eta_0 s_2 \sigma_0, \label{eq:pair_po}
\end{equation}
where $\Delta_0$ is the on-site intraorbital pairing and $\Delta_1$ is the nearest neighbor intrasublattice intraorbital pairing. The second term in Eq.~(\ref{eq:pair_po}) depends on momentum, which gives rise to the sign change of the gap function. The BdG Hamiltonian~(\ref{eq:bdg}) satisfies PHS,
\begin{align}
 C \tilde{H}(\bm{k}) C^{-1} = -\tilde{H}(-\bm{k}), \ \ C = \tau_1 \eta_0 s_0 \sigma_0 K,
\end{align} 
where $\tau_i$ represents the Pauli matrix in the Nambu space.
The combination of the PHS and $T$ operators defines the chiral operator $\Gamma$, which satisfies  
\begin{align}
 \Gamma \tilde{H}(\bm{k}) \Gamma^{-1} = -\tilde{H}(\bm{k}), \ \ \Gamma \equiv -i \tilde{T}C = \tau_1 \eta_0  s_2 \sigma_0,
\end{align} 
where $\tilde{T} = \diag[T,T^{\ast}]=i\tau_0\eta_0s_2\sigma_0K$.
In addition, the symmetry properties of the normal-state Hamiltonian in Eqs.~(\ref{eq:trs_model}), (\ref{eq:is_model}), (\ref{mrs_model}), and (\ref{screw_model}) are also satisfied in the superconducting state.  The symmetry operations in the Nambu space are described as in Eq.~(\ref{eq:op_nambu}), labeled by $\tilde{P}$, $\tilde{M}_{xz}$, $\tilde{M}_{yz}$, $\tilde{C}_{2z}$, and $\tilde{C}_{4z}$, respectively. In $s$-wave and extended $s$-wave superconductors, we have a trivial twist; that is, the PHS operator commutes with every crystalline symmetry operator.   

Following Ref.\cite{Qin2022}, we now discuss topological superconductivity in the extended $s$-wave superconductor. 
Whereas the time-reversal invariant even-parity superconductor does not have 2D topological numbers~\cite{SatoFujimoto16}, it may have a non-trivial 1D topological number due to crystalline symmetry~\cite{Shiozaki14, Qin2022}: Let us focus on 1D mirror-invariant subspaces of Brillouin zone satisfying $M_{xz} \bm{k} =\bm{k}+{\rm G}$ with a reciprocal lattice vector ${\bm G}$, i.e., the lines of $k_y=0,\pi$. In the 1D subspace, the BdG Hamiltonian commutes with the mirror-reflection operator $\tilde{M}_{xz}$, and thus it has a block-diagonal form with mirror eigenvalues $m_y = \pm i $: $H_{\rm sc} \to H_{+i} \oplus H_{-i}$. Then, from $[\Gamma, \tilde{M}_{xz}]=0$, each mirror subspace $H_{\pm i}$ has chiral symmetry, so we can define the 1D winding number in each mirror subspace:
\begin{align}
w_{m_y}(k_y) = \int_{-\pi}^{\pi} \frac{k_x}{2\pi} \tr [\Gamma_{m_y} H_{m_y}^{-1}(\bm{k}) \partial_{k_x} H_{m_y}(\bm{k})] \in \mathbb{Z}, 
\label{eq:1D_winding}
\end{align}
 where $k_y=0,\pi$ and $\Gamma_{m_y}$ is the chiral operator in the mirror eigenspace of $m_y$. 
Using the formula in Ref.\cite{Sato11}, the 1D winding number is recast into 
 \begin{align}
  w_{m_y}(k_y) = \frac{1}{2} \sum_{\bm{k}_{\rm F}} {\rm sgn}[v_{m_y}(\bm{k}_{\rm F}) \Delta_{m_y}(\bm{k}_{\rm F})],
\label{eq:1D_winding_Fermi}
 \end{align}
where $v_{m_y}(\bm{k}_{\rm F})$ and $\Delta_{m_y}(\bm{k}_{\rm F})$ are the Fermi velocity and the gap function at the Fermi point $\bm{k}_{\rm F}$. Here the Fermi point ${\bm k}_{\rm F}$ is given by $\det[{\cal E}(\bm{k}_{\rm F}) -\mu] =0$ in the mirror eigenspace of $m_y$ with the fixed momenta $k_y=0,\pi$.

Let us evaluate $w_{m_y}(k_y)$ for the extended $s$-wave superconducting state.
For $k_y=0$, we have $[\tilde{P},\tilde{M}_{xz}]=0$, so the mirror eigenspace keeps inversion symmetry. 
Thus, the Fermi points on the right-hand side of Eq.(\ref{eq:1D_winding_Fermi}) appear in a pair $(-{\bm k}_{\rm F}, {\bm k}_{\rm F})$, and
 $v_{m_y}(-\bm{k}_{\rm F}) = -v_{m_y}(\bm{k}_{\rm F})$ and $\Delta_{m_y} (-\bm{k}_{\rm F}) = \Delta_{m_y}(\bm{k}_{\rm F})$ lead to $w_{m_y}(0)=0$. On the other hand, for $k_y=\pi$, from $\{\tilde{P},\tilde{M}_{xz}\}=0$, 
the mirror subsector does not have inversion symmetry, and thus, we have no such constraint. 
Then, we find that $|w_{m_y}(\pi)|=2$ when the Fermi pockets around $M$ in Fig.~\ref{fig:band} (b) have opposite signs of $\Delta$.
Note that the sign change is possible for the extended $s$-wave superconducting state, and the Kramers degeneracy due to $PT$ symmetry leads to the even parity of $w_{m_y}(\pi)$.  (The mirror subspace keeps $PT$ symmetry.) 
The non-trivial winding number results in two helical Majorana edge states in each mirror subspace when one considers the boundary keeping the mirror symmetry, {\it i.e.}, the (10) edge. In Fig.~\ref{fig:band} (c), we numerically demonstrate the boundary states on the (10) edge.
Because of $|w_{m_y}(\pi)|=2$,  
the helical Majorana edge states support two zero energy MKPs at $k_y=\pi$.
A similar analysis is also applicable to the 1D mirror-invariant momenta for $M_{yz}$, i.e., $k_x = 0, \pi$, leading to two Majorana edge modes on the (01) edge. 

\begin{figure}[tb]
\centering\includegraphics[width=6in]{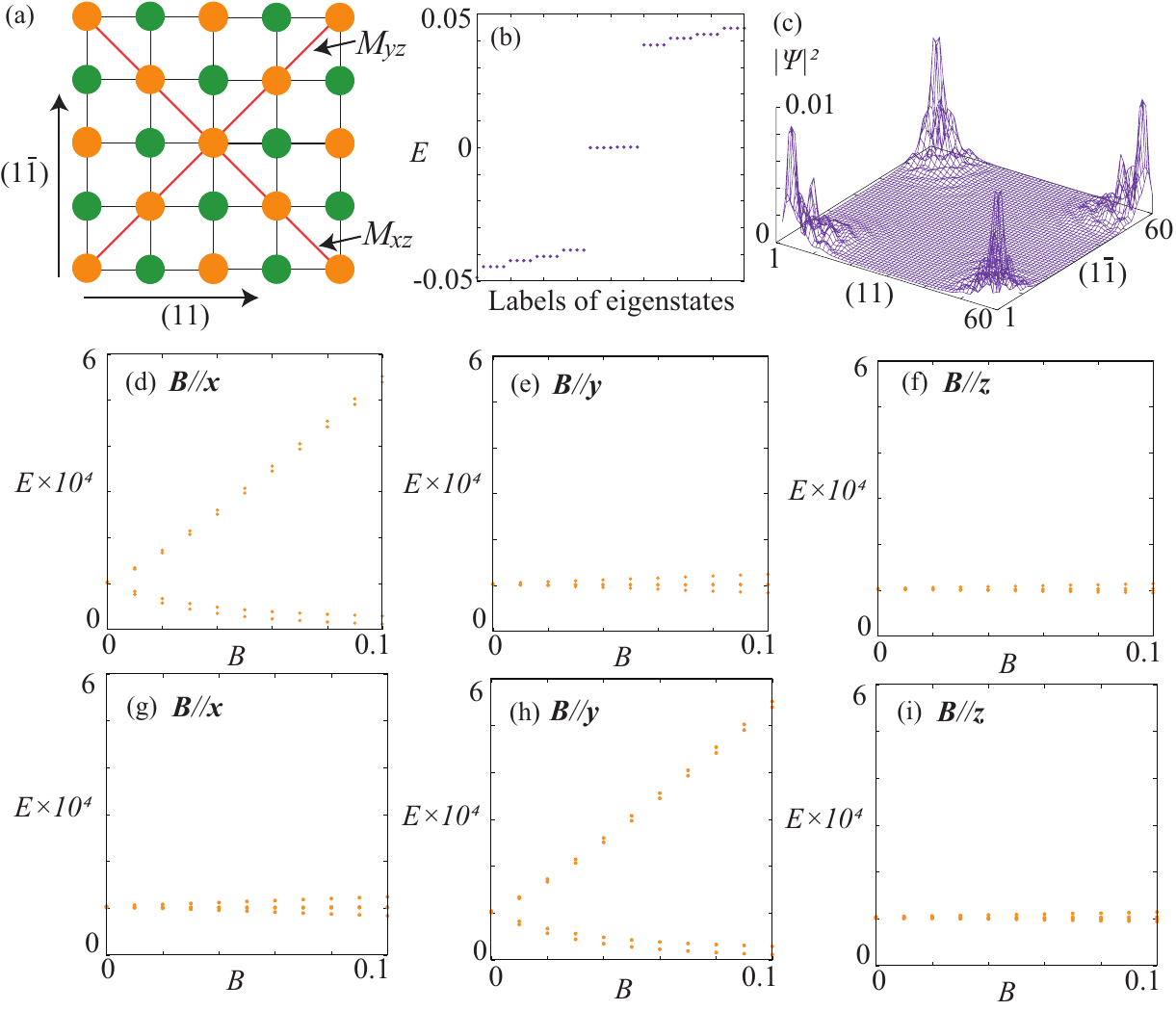}
\caption{(a) Schematic illustration of the system with the $(11)$ and $(1\bar{1})$ edges. The red lines describe the mirror symmetric lines with respect to $M_{yz}$ or $M_{xz}$. (b) Eigenvalues of the BdG Hamiltonian in Eq. (\ref{eq:bdg}) with the full open boundary conditions as shown in (a) with the lattice size being $N_{(11)}=N_{(1\bar{1})}=60$. We used the same parameters as in Fig.~\ref{fig:band} (c). There are eight zero energy states corresponding to an MKP at the four corners. Note that a very tiny gap opens at the zero energy states due to finite-size effects. (c) Density of zero energy states in the real space, where $|\Psi|^2$ is the normalized density of states.  (d), (e), and (f) show the energy spectra ($E>0$) of Majorana corner modes as a function of the Zeeman magnetic field $B=|\bm{B}|$ for $\bm{B} \parallel \hat{x}$, $\bm{B} \parallel \hat{y}$, and $\bm{B} \parallel \hat{z}$, respectively. The Zeeman term is added only at the corners in the $M_{yz}$ symmetric line. Similarly, (g), (h), and (i) show the energy spectra of corner modes when the Zeeman magnetic field is applied only to the corner in the $M_{xz}$ symmetric line.
}
\label{fig:corner}
\end{figure}

These helical Majorana edge states give a corner MKP \cite{Qin2022} of mirror symmetry-protected second-order topological superconductivity~\cite{Langbehn2017,Geier2018}.
To see this, let us consider a corner, say the top-left corner, in Fig.~\ref{fig:corner} (a), which is formed by the $(1 1)$ and $(1 \bar{1})$ edges.
For a while, we neglect the orbital degrees of freedom for simplicity. (Later, we come back to the model with orbital degrees of freedom.)  The simplified extended $s$-wave superconductor also hosts a helical Majorana edge state, which is solved analytically as shown in Appendix~\ref{app:effectivemodel}. 
Since one can obtain this corner configuration by adiabatically bending the $(10)$ edge, the helical Majorana edge states on the $(10)$ edge give the following low-energy effective Hamiltonian near the corner (see Appendix~\ref{app:effectivemodel} for the derivation) 
\begin{align}
 H_{\rm corner}(y) = -i v \partial_y \mu_1 \kappa_0 + m(y) \mu_2 \kappa_1,
\end{align}
where $v$ is the Fermi velocity of the edge mode, $\mu_\mu$ the 
the Pauli matrix in the mirror subspaces, i.e. $\mu_3=\pm1$ corresponds to the mirror subspaces $\tilde{M}_{xz}=\pm i$, and $\kappa_\mu$ the Pauli matrix labeling the two helical Majorana states.
Here, $y$ is a real space parameter along the boundary near the corner, where $y<0$, $y=0$, and $y>0$ correspond to
the $(1\bar{1})$ edge, the top-left corner, and the $(11)$ edge respectively.
The mass term $m(y)$ appears because $\tilde{M}_{xz}$ exchanges the $(11)$ and $(1\bar{1})$ edges, and thus the symmetry-protection of helical edge states does not work on each of these edges; the two helical edge modes can mix except at the corner, {\it i.e.} at $y=0$. 
Still, the whole system can retain the mirror reflection symmetry because the corner configuration is invariant under mirror reflection.
To satisfy this symmetry constraint, $m(y)$ must be an odd function of $y$, $m(y)=-m(-y)$.
Then, the effective Hamiltonian has time-reversal symmetry, PHS, and mirror-reflection symmetry:
\begin{align}
 T_{\rm eff} H_{\rm corner} (y) T_{\rm eff}^{-1} = H_{\rm eff} (y), \ \ T_{\rm eff} =  \mu_2 \kappa_3 K, \\
 C_{\rm eff} H_{\rm corner} (y) C_{\rm eff}^{-1} = -H_{\rm eff} (y), \ \ C_{\rm eff} =  \mu_1 \kappa_3 K, \\
 M_{\rm eff} H_{\rm corner} (y) M_{\rm eff}^{-1} = H_{\rm eff} (-y). \ \ M_{\rm eff} = i \mu_3 \kappa_0. 
 \label{eq:mirror_eff}
\end{align}
Note that $M_{\rm eff}$ commutes with $C_{\rm eff}$, which is consistent with the fact that the extended $s$-wave pairing is mirror-even. 
For simplicity, we assume $v>0$ and $m(y>0)>0$ below.

We obtain a corner MKP by solving the BdG equation for zero modes:
\begin{equation}
 [ -i v \partial_y \mu_1 \kappa_0 + m(y) \mu_2 \kappa_1] |u(y)\rangle =0.
\end{equation}
Multiplying $i \mu_1 \kappa_0$ to the above from the left, we have
 \begin{equation}
 [ v \partial_y + m(y) \mu_3 \kappa_1] |u(y)\rangle =0,
\end{equation}
which leads to two zero-mode solutions,
 \begin{align}
| u_0^{(1)} (y)\rangle = f(y)\left(\begin{array}{@{\,}c @{\,}} 1 \\ -1 \\ 0 \\ 0\end{array}\right)_{\mu \otimes \kappa},  \quad
|u_0^{(2)} (y)\rangle = i f(y)\left(\begin{array}{@{\,}c @{\,}} 0 \\ 0 \\ 1 \\ 1 \end{array}\right)_{\mu \otimes \kappa}, 
 \end{align}
 with 
  \begin{equation}
 f(y) = \frac{C}{\sqrt{2}} \exp \left[ - \int_0^y dy' m(y')/v \right].
 \end{equation}
Here, $C$ is a real constant, and we have used the standard definition 
of the Kronecker product $\mu_\mu \kappa_\nu$.
Since $|u_0^{(2)}\rangle = T_{\rm eff} |u_0^{(1)}\rangle$, these zero modes form a Kramers pair, and thus they are a corner MKP.

Using the above corner MKP, we can confirm the validity of our theory for 
the electromagnetic response:
First, the zero modes satisfy the following relation
\begin{align}
M_{\rm eff}|u_0^{(1)}(-y)\rangle=i|u^{(1)}_0(y)\rangle,
\quad
M_{\rm eff}|u_0^{(2)}(-y)\rangle=-i|u^{(2)}_0(y)\rangle,
\end{align}
in the form of Eq.~(\ref{eq:mf_trans}).
Then, the corner MKP gives
 \begin{equation}
 \rho_{\rm MKP}^{(12)}(y) = i f^2(y) \begin{pmatrix} 1 & -1 &0 & 0 \\ -1 & 1 & 0 & 0 \\ 0 & 0 & -1 & -1 \\ 0 & 0 & -1 & -1 \end{pmatrix}_{\mu \otimes \kappa},
 \end{equation}
 which obeys the same transformation law as an $s$-wave pairing under the mirror reflection,
 \begin{align}
 M_{\rm eff}\rho_{\rm MKP}^{(12)}(-y)M_{\rm eff}^\dagger=\rho_{\rm MKP}^{(12)}(y).
 \end{align}
 One can also check the relation in Eq.(\ref{eq:rho_T}),
 \begin{align}
 T_{\rm eff}[\rho^{(12)}_{\rm MKP}(y)]^\dagger
 T_{\rm eff}^{-1}=-\rho_{\rm MKP}^{(12)}(y).    
 \end{align}
 
 According to our theory, an operator $\mathcal{O}$ satisfying $\tr[\rho_{\rm MKP}^{(12)}(x) \mathcal{O}]\neq 0$ must be odd under time-reversal and share the same irreducible representation of $\rho_{\rm MKP}^{(12)}$ under mirror reflection. 
 We can also confirm these properties directly: 
 As the hermiticity of $O$ requires $\{\mathcal{O},C_{\rm eff}\}=0$, possible $\mathcal{O}$ are restricted to
 \begin{align}
  \mu_1 \kappa_1, \ \ \mu_2 \kappa_1, \ \ \mu_0 \kappa_1, \ \ \mu_3 \kappa_0, \ \ \mu_3 \kappa_3, \ \ \mu_3 \kappa_2.
 \end{align} 
 Then, we find that only $\mu_0 \kappa_1$ and $\mu_3 \kappa_0$ satisfy $\tr[\rho^{(12)}(x) \mathcal{O}]\neq 0$. Thus, the MKP can couple only to $\mathcal{O}_{\rm eff} = A \mu_0 \kappa_1 + B \mu_3 \kappa_0$, where $A$ and $B$ are real coefficients. 
 Since we have $[\mathcal{O}_{\rm eff}, M_{\rm eff}]=\{\mathcal{O}_{\rm eff}, T_{\rm eff}\}=0$,  $\mathcal{O}_{\rm eff}$ is a mirror-even magnetic operator like $\rho^{(12)}_{\rm MKP}$.

As shown in Appendix \ref{app:effectivemodel}, the operators $\eta_1s_3$ and $\eta_3s_2$ in the normal state Hamiltonian of the simplified extended $s$-wave superconductor give these
$\mu_0\kappa_1$ and $\mu_3\kappa_0$ terms in the effective model,
\begin{align}
\eta_1 s_3\to \mu_0\kappa_1, 
\quad 
\eta_3 s_2\to \mu_3\kappa_0.
\end{align}
Therefore, the Majorana corner modes can couple only to the following sub-lattice-dependent Zeeman magnetic fields,
\begin{align}
B_z\eta_1 s_3, \quad B_y\eta_3s_2.    
\end{align}
Among them, one can easily obtain the latter type of the Zeeman field in experiments:
A local magnetic field $B_y$ on, say, the A-site induces the latter type of the sub-lattice-dependent Zeeman magnetic term. Note that it also induces the term $B_y\eta_0s_2$, but the sub-lattice-independent Zeemam term does not couple to the Majorana corner modes.
As shown in Appendix \ref{app:effectivemodel}, the operator $\eta_0s_2$ gives $\mu_3\kappa_3$ in the effective Hamiltonian, which obeys $\tr[\rho_{\rm MKP}^{(12)}(y)\mu_3\kappa_3]=0$.  
 
Finally, we come back to the original model with orbital degrees of freedom and 
discuss the magnetic response of the corner MKP in the BdG Hamiltonian with Eqs.~(\ref{eq:normal}) and (\ref{eq:pair_po}). 
Here, we focus on the magnetic response for the following Zeeman magnetic fields, 
\begin{align}
{\bm B}\cdot \eta_0 {\bm s} \sigma_0,
\quad
{\bm B}\cdot \eta_3 {\bm s} \sigma_0,
\end{align}
which are feasible in experiments. As mentioned above, the corner MKP protected by the mirror reflection symmetry $M_{xz}$ can couple only to mirror-even operators, and thus, it can respond only to
$
B_y\eta_0 s_2\sigma_0 
$
and 
$
B_y\eta_3 s_2\sigma_0. 
$
Furthermore, the analysis of the simplified model implies that the sublattice-independent Zeeman term $B_y\eta_0 s_2\sigma_0$ rarely couples to the corner MKP: As the original model consists of a pair of simplified models with couplings between them, we can construct the corner MKP in the original model from the corner MKPs of the paired simplified models.  
Since the latter MKPs do not couple to $B_y\eta_0s_2$, we do not have a direct coupling between the former MKP and $B_y\eta_0s_2\sigma_0$.
We also have a similar argument for a corner MKP protected by $M_{yz}$.
In summary, we can expect that the corner MKP protected by $M_{xz}$ ($M_{yz}$) responds only to the sublattice-dependent Zeeman magnetic field $B_y\eta_3s_2\sigma_0$ ($B_x\eta_3s_1\sigma_0$).

We confirm this property numerically.
We diagonalize the BdG Hamiltonian under the open boundary conditions with $(11)$ and $(1\bar{1})$ edges. See Fig. ~\ref{fig:corner} (a).
In the absence of the magnetic field, each corner hosts a single MKP, as shown in Figs~\ref{fig:corner} (b) and (c).
Then, we add the sublattice-dependent Zeeman term 
\begin{align}
 \bm{B} \eta_3\cdot \bm{s} \sigma_0
\label{eq:bdg_mag}
\end{align}
to the normal state Hamiltonian in Eq.(\ref{eq:normal}), and examine the spectrum of the corner MKP. 
We apply the Zeeman magnetic fields (i) at corners preserving $M_{yz}$, or (ii) corners at the corners preserving $M_{xz}$. 
As shown in Fig.~\ref{fig:corner}, the energy spectra of the corner modes confirm the aforementioned anisotropic magnetic responses.   
    
\section{Conclusion}
We have discussed the electromagnetic response of spinful MFs. Whereas the $CPT$ theorem gives a strong constraint on the electromagnetic coupling and forbids electric and magnetic dipole momenta for elementary MFs in high-energy physics, MFs in TSCs do not suffer such a strong constraint. The associated electromagnetic responses are determined by underlying TSCs, which provide information on the pairing mechanism. In particular, a magnetic operator of a MKP (spinful MF of TSCs) and the Cooper pairs share the same property for crystalline symmetry.
Therefore, we can identify the pairing symmetry of TSCs by measuring a magnetic response of a single MKP through spin-sensitive measurements such as spin-resolved tunneling spectroscopy~\cite{Jeon17,Cornils17}, spin relaxation rate~\cite{Chung09}, spin susceptibility~\cite{Nagato09}, and dynamical spin response~\cite{Ominato2024}. 

In addition to surface or edge MFs, our theory is applicable to spatially localized MFs, such as corner MFs or hinge MFs. 
As a concrete example, we have studied the magnetic response of corner MFs in iron-based SCs~\cite{Qin2022} proposed as second-order TSCs. Using the effective low-energy Hamiltonian and the numerical calculation, we have found that the corner MKPs exhibit magnetic dipole response to applied magnetic fields. Measuring the magnetic response of the corner MKPs offers a clue for the pairing mechanism of iron-based SCs.

Another possible direction is to study the magnetic response of Majorana zero modes localized at the vortex in the interface of $s$-wave superconductor/topological crystalline insulator hybrid systems.  Recent theoretical works~\cite{CFang2014,XJLiu2014,Kobayashi2020,Hu2022Competing,Hu2023topological} have revealed that the interplay between crystalline symmetry and vortex topology leads to the topological superconducting states having multiple Majorana fermions at the vortex. In this phase, although time-reversal symmetry is broken in the presence of the vortex, the stability of Majorana fermions is ensured by a rotation symmetry and/or magnetic-mirror-reflection symmetry, the combination of mirror-reflection and time-reversal operations. Thus, their magnetic response still obey Eq.~(\ref{eq:rho_trans}), and the irreducible representation of the magnetic multipole may encode information about an effective pairing symmetry realized in the interface. 

\section*{Acknowledgment}

 This work was supported by JSPS KAKENHI (Grants No.\ JP19K14612, No.\ JP22K03478, and No.\ JP24K00569) and JST CREST (Grants No.\ JPMJCR19T2).

\appendix

\section{Fermi surface criterion for spinless MFs}
\label{sec:appendix_A}

Here, we show the Fermi surface criterion for spinless MFs.
Before going to the argument, we would like to make clear the assumptions of the criterion.
First, we consider superconductors in class D in the Altland-Zirnbauer classification \cite{Altland97} since spinless MFs do not form Kramers pairs. 
Other classes with PHS cannot be realized without time-reversal symmetry or special crystalline symmetry. 
Second, we assume that Cooper pairs are formed
by electrons near the Fermi surface, and we can neglect the gap function except near the Fermi surface without gap-closing.
In particular, the gap function is assumed to be neglected at time-reversal invariant momenta since the Fermi surfaces are usually away from time-reversal invariant momenta. 
Finally, we suppose that if the momentum ${\bm k}$ is on the Fermi surface, its time-reversal partner $-{\bm k}$ is also on the Fermi surface. The final assumption ensures the formation of Cooper pairs with electrons at ${\bm k}$ and $-{\bm k}$.

The topological numbers in class D are summarized in Table \ref{table:class_D}\cite{Schnyder08}.
\begin{table}[t]
 \caption{Topological numbers of superconductors in class D.}
 \label{table:class_D}
 \centering
  \begin{tabular}{c|ccc}
   \hline
Dimension &  1d & 2d & 3d \\
\hline
Topological number & $\mathbb{Z}_2$  & $\mathbb{Z}$& 0 \\
   \hline
  \end{tabular}
\end{table}
Here the one-dimensional $\mathbb{Z}_2$ topological number is given by
\begin{eqnarray}
(-1)^{\nu_{\rm 1d}}={\rm sgn}
\left[{\rm Pf}H_{\rm BdG}(0)\tau_x\right]
\left[{\rm Pf}H_{\rm BdG}(\pi)\tau_x\right],
\label{eq:z2}
\end{eqnarray}
and the two-dimensional $\mathbb{Z}$ topological number is the Chern number,
\begin{eqnarray}
N_{\rm Ch}=\frac{1}{2\pi}\int_{\rm BZ} d^2k {\rm tr}[\vec{\nabla} \times \vec{a}({\bm k})]_z,
\label{eq:chern}
\end{eqnarray}
where $\vec{a}({\bm k})$ is defined as
\begin{eqnarray}
\vec{a}({\bm k})=i\sum_{n<0} \langle \phi_n({\bm k})|\vec{\nabla}\phi_n({\bm k})\rangle 
\end{eqnarray}
and the integral in Eq.(\ref{eq:chern}) is performed over the entire two-dimensional Brillouin zone. 
For the Chern number in class D, we also have \cite{Sato10}
\begin{eqnarray}
(-1)^{N_{\rm Ch}}={\rm sgn}\prod_{{\bm k}_0: {\rm TRIM}} 
{\rm Pf}(H_{\rm BdG}({\bm k}_0)\tau_x),
\label{eq:chernZ2}
\end{eqnarray}
where the product is taken over all time-reversal invariant momenta in the Brillouin zone.
Note that Eqs. (\ref{eq:z2}) and (\ref{eq:chernZ2}) share the same Pfaffian ${\rm Pf}(H_{\rm BdG}({\bm k}_0)\tau_x)$.

To evaluate the Pfaffian ${\rm Pf}(H_{\rm BdG}({\bm k}_0)\tau_x)$, we use the fact that Cooper pairs are formed by electrons near the Fermi surface and time-reversal invariant momenta are usually away from the Fermi surface. Therefore, $H_{\rm BdG}({\bm k}_0)$ can be approximated as
\begin{eqnarray}
H_{\rm BdG}({\bm k}_0)=
\left(
\begin{array}{cc}
{\cal E}({\bm k}_0) & \Delta({\bm k}_0)\\
\Delta^\dagger({\bm k}_0) & -{\cal E}^T({\bm k}_0)
\end{array}
\right)
\rightarrow
\left(
\begin{array}{cc}
{\cal E}({\bm k}_0) & 0\\
0 & -{\cal E}^T({\bm k}_0)
\end{array}
\right)
\label{eq:weak_pairing}
\end{eqnarray}
without gap closing, from which we can obtain
\begin{eqnarray}
{\rm sgn}{\rm Pf}[H_{\rm BdG}({\bm k}_0)\tau_x]
=(-1)^{N(N-1)/2}{\rm sgn}[{\rm det}{\cal E}({\bm k}_0)],
\end{eqnarray}
where $N$ is the matrix size of  $\epsilon({\bm k}_0)$.
Since the sign of ${\rm det}{\cal E}({\bm k}_0)$ is determined by the number $n_{\rm occ}({\bm k}_0)$ of occupied states of electrons at the momentum ${\bm k}_0$, 
\begin{eqnarray}
{\rm sgn}[{\rm det}\epsilon({\bm k}_0)]=(-1)^{n_{\rm occ}({\bm k}_0)},
\end{eqnarray}
we have
\begin{eqnarray}
{\rm sgn}{\rm Pf}[H_{\rm BdG}({\bm k}_0)\tau_x]
=(-1)^{N(N-1)/2}(-1)^{n_{\rm occ}({\bm k}_0)},
\end{eqnarray}
so, the topological numbers are evaluated as
\begin{eqnarray}
&&(-1)^{\nu_{\rm 1d}}=(-1)^{n_{\rm occ}(0)+n_{\rm occ}(\pi)}, 
\nonumber\\
&&(-1)^{N_{\rm Ch}}=(-1)^{\sum_{{\bm k}_0:{\rm TRIM}}n_{\rm occ}({\bm k}_0)}.
\end{eqnarray}
Now we relate the number of occupied states of electrons at time-reversal momenta to the number of non-spin-degenerate Fermi surfaces:
From our assumptions mentioned above, if a Fermi surface does not enclose any time-reversal invariant momentum, then it has a partner Fermi surface with opposite momentum. Thus, the number of such Fermi surfaces must be even. 
Then, consider Fermi surfaces enclosing a time-reversal invariant momentum ${\bm k}_0$.
Since the number of occupied states of electrons changes by $\pm 1$ when we cross a Fermi surface, the number of Fermi surfaces enclosing ${\bm k}_0$ coincides with the number of occupied states ${\bm k}_0$ in their parity. 
Combining these two results, we eventually have
\begin{eqnarray}
(-1)^{\nu_{\rm 1d}}=(-1)^{N_{\rm Fermi}}, 
\quad
(-1)^{N_{\rm Ch}}=(-1)^{N_{\rm Fermi}}.
\end{eqnarray}
where $N_{\rm Fermi}$ is the number of non-spin-degenerate Fermi surfaces.
From this equation, we conclude that a sinless MF is realized if and only if the corresponding superconductor has an odd number of non-spin-degenerate Fermi surfaces.

\section{Corner MF in simplified extended $s$-wave superconductor}
\label{app:effectivemodel}

To confirm the validity of our theory, we analytically examine corner MFs in a simplified version of the extended $s$-wave superconductor in Sec.\ref{sec:extended_s}.
For this purpose, we neglect the orbital degree of freedom $\sigma$ and terms propositional to $\lambda$ and $t_2$ in Eq.(\ref{eq:normal}).
Then, expanding the momentum around the M point with ${\bm k}=(\pi,\pi)$, we have a simplified normal state Hamiltonian,
\begin{align}
{\cal E}({\bm k})=({\bm k}^2-\mu)\eta_0 s_0+\eta_3(k_xs_2-k_y s_1)+k_xk_y\eta_1s_0
, \label{eq:simple_model}
\end{align}
where we have replaced $k_i-\pi$ by $k_i$, put $t=\lambda_{\rm R}=-t_1=1$, and redefined the chemical potential $\mu$. 
Corresponding to Eqs.(\ref{eq:trs_model})-(\ref{screw_model}), 
the normal state Hamiltonian ${\cal E}({\bm k})$ has the following symmetries:
\begin{align}
&T{\cal E}({\bm k})T^{-1}={\cal E}(-{\bm k}), \quad T=i\eta_0 s_2 K,
\nonumber\\
&P{\cal E}({\bm k})P^{-1}={\cal E}(-{\bm k}), \quad P=\eta_1s_0,   
\nonumber\\
&M_{xz}{\cal E}({\bm k})M^{-1}_{xz}={\cal E}(k_x,-k_y), \quad M_{xz}=i\eta_3s_2,
\nonumber\\
&M_{yz}{\cal E}({\bm k})M^{-1}_{yz}={\cal E}(k_x,-k_y), \quad M_{yz}=i\eta_3s_1,
\nonumber\\
&C_{2z}{\cal E}({\bm k})C^{-1}_{2z}={\cal E}(-k_x,-k_y), \quad C_{2z}=i\eta_0s_3.
\label{eq:sym0}
\end{align}
We note that the $k_xk_y\eta_1s_0$ term breaks $C_{4z}$ symmetry, but this simplification does not affect the stability of corner MFs because the corner configuration is not invariant under $C_{4z}$.
As a gap function, we consider 
\begin{align}
\Delta({\bm k})=i(\Delta_0-{\bm k}^2\Delta_1)\eta_0s_2,
\quad (\Delta_0, \Delta_1>0)
\end{align}
which represents an $s_{+-}$-wave pairing symmetry in a manner similar to Eq.(\ref{eq:pair_po}).
The gap function is time-reversal symmetric, 
\begin{align}
T\Delta({\bm k})T^{-1}=\Delta(-{\bm k}),    
\end{align}
and even under the above crystalline symmetries,
\begin{align}
&P\Delta({\bm k})P^{T}=\Delta(-{\bm k}),
\quad
M_{xz}\Delta({\bm k})M_{xz}^{T}=\Delta(k_x,-k_y),
\nonumber\\
&M_{yz}\Delta({\bm k})M_{yz}^{T}=\Delta(-k_x,k_y),
\quad
C_{2z}\Delta({\bm k})C_{2z}^{T}=\Delta(-{\bm k}).
\end{align}
Then, the BdG Hamiltonian
\begin{align}
\tilde{H}({\bm k})&=
\begin{pmatrix}
{\cal E}({\bm k}) & \Delta({\bm k})\\    
\Delta^{\dagger}({\bm k}) & -{\cal E}^{T}(-{\bm k})
\end{pmatrix}
\nonumber\\
&=({\bm k}^2-\mu)\tau_3\eta_0s_0+k_x\tau_3\eta_3s_2-k_y\tau_0\eta_3s_1-(\Delta_0-{\bm k}^2\Delta_1)\tau_2\eta_0s_2
\end{align}
keeps these symmetries,
\begin{align}
&T\tilde{H}({\bm k})T^{-1}=\tilde{H}(-{\bm k}),
\quad
T=i\tau_0\eta_0s_2 K,
\nonumber\\
&\tilde{P}\tilde{H}({\bm k})\tilde{P}^{-1}=\tilde{H}(-{\bm k}),
\quad
\tilde{P}=\tau_0\eta_1s_0,
\nonumber\\
&\tilde{M}_{xz}\tilde{H}({\bm k})\tilde{M}_{xz}^{-1}=\tilde{H}(k_x,-k_y),
\quad
\tilde{M}_{xz}=i\tau_0\eta_3s_2,
\nonumber\\
&\tilde{M}_{yz}\tilde{H}({\bm k})\tilde{M}_{yz}^{-1}=\tilde{H}(-k_x,k_y),
\quad
\tilde{M}_{yz}=i\tau_3\eta_3s_1,
\nonumber\\
&\tilde{C}_{2z}\tilde{H}({\bm k})\tilde{C}_{2z}^{-1}=\tilde{H}(-{\bm k}),
\quad
\tilde{C}_{2z}=i\tau_3\eta_0s_3,
\label{eq:sym1}
\end{align}
together with particle-hole symmetry,
\begin{align}
C\tilde{H}({\bm k})C^{-1}=-\tilde{H}(-{\bm k}),
\quad
C=\tau_1\eta_0s_0 K.
\label{eq:sym2}
\end{align}

\begin{figure}[tb]
\centering\includegraphics[width=15.5cm]{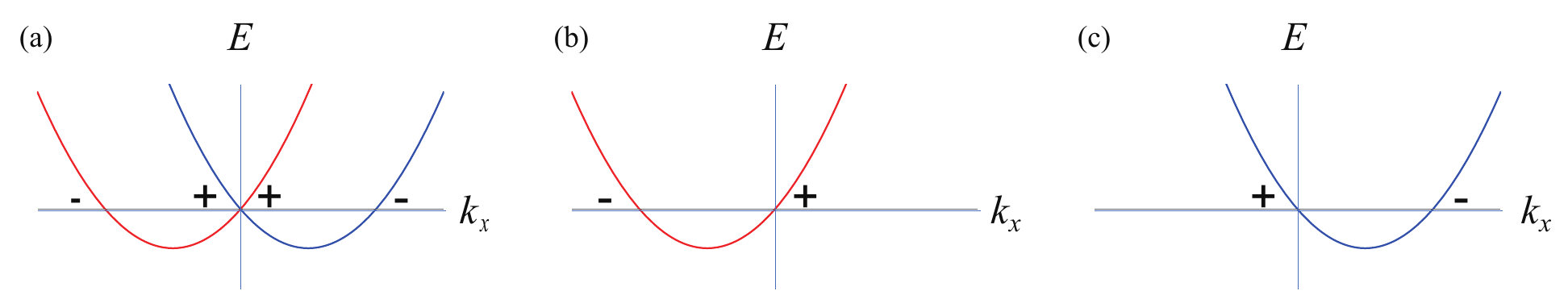}
\caption{(a) The superconducting state in Eq.~(\ref{eq:model_mirror}). The red (blue) curve is the normal state dispersion in the $\tilde{M}_{xz}=i$ ($\tilde{M}_{xz}=-i$) sector. The symbols $\pm$ indicate the sign of the gap function. (b,c) The superconducting states in the mirror subsectors effectively realize $p$-wave pairings.}
\label{fig:mirror}
\end{figure}

Remarkably, this Hamiltonian may effectively realize a $p$-wave superconducting state: 
On the mirror invariant line at $k_y=0$, the BdG Hamiltonian becomes
\begin{align}
\tilde{H}(k_x,0)=(k_x^2-\mu)\tau_3\eta_0s_0+k_x\tau_3\eta_3s_2-(\Delta_0-\Delta_1k_x^2)\tau_2\eta_0s_2.
\end{align}
Then, in the $\tilde{M}_{xz}=i$ ($\tilde{M}_{xz}=-i$) sector, we have 
$\eta_3=s_2=\pm 1$ ($\eta_3=-s_2=\pm 1$), and thus, the BdG Hamiltonian $\tilde{H}_{\pm}(k_x,0)$ in the mirror subsector $\tilde{M}_{xz}=\pm i$
reads
\begin{align}
\tilde{H}_{\pm}(k_x,0)&=(k_x^2-\mu)\tau_3 s_0\pm k_x\tau_3 s_0-(\Delta_0-   
\Delta_1k_x^2)\tau_2 s_2
\nonumber\\
&=
\begin{pmatrix}
(k_x^2\pm k_x-\mu)s_0
& i(\Delta_0-\Delta_1 k_x^2)s_2\\    
-i(\Delta_0-\Delta_1 k_x^2)s_2 & 
-(k_x^2\pm k_x-\mu)s_0
\end{pmatrix}.
\label{eq:model_mirror}
\end{align}
As shown in Fig. \ref{fig:mirror}, for an $s_{+-}$-wave gap function, each mirror subsector shows a sign change of the gap function between opposite Fermi points in a manner similar to a $p$-wave superconductor.
The sign change of the gap function results in a non-trivial one-dimensional winding number in Eq.~(\ref{eq:1D_winding_Fermi}),
and thus, we have two helical Majorana modes on the (10) edge under the semi-infinite boundary condition ($x>0$) in the $x$-direction.

Now, we construct the helical Majorana edge states for $k_y=0$ explicitly.
For simplicity, we choose the model parameters as $\mu=0$, $\Delta_0=1/4$ and $\Delta_1=1$, then, the BdG Hamiltonian is
\begin{align}
\tilde{H}({\bm k})={\bm k}^2\tau_3\eta_0s_0+k_x\tau_3\eta_3s_2-k_y\tau_0\eta_3s_1
-(1/4-{\bm k}^2)\tau_2\eta_0s_2+k_xk_y\eta_1s_0.
\end{align}
For $k_y=0$, the Majorana edge states become zero energy states $|\psi_0\rangle$, so they obey the BdG equation
\begin{align}
\left[k_x\tau_2\eta_3 s_2+k_x^2\tau_3\eta_0 s_0-(1/4-k_x^2)\tau_2\eta_0s_2\right]
|\psi_0\rangle=0.    
\end{align}
Then, by multiplying $\tau_2\eta_0s_2$ from the left, we have
\begin{align}
\left[(k_x^2-1/4)+ik_x^2\tau_1\eta_0 s_2+ik_x\tau_1\eta_3s_0\right]
|\psi_0\rangle=0.    
\end{align}
In the $\tilde{M}_{xz}=i$ sector, it holds that $\eta_3s_2=1$ hence $\eta_0s_2=\eta_3s_0$, so the above equation is simplified as
\begin{align}
\left[(k_x^2-1/4)+i(k_x^2+k_x)\tau_1\eta_3 s_0\right]|\psi_0\rangle=0.    
\end{align}
Then, we have a nontrivial zero mode $|\psi_0\rangle$ when $k_x^2-1/4+i(k_x^2+k_x)\delta=0$ with $\delta=\tau_1\eta_3s_0=\pm 1$. So $k_x$ is given by
\begin{align}
k_x=\frac{\pm \sqrt{2}-1}{4}-i\frac{\delta}{4}\equiv k_\pm -i\frac{\delta}{4}
\end{align}
Therefore, for $\delta=-1$, ${\rm Im}(k_x)$ becomes positive, so we have a zero mode
\begin{align}
|\psi_0\rangle=e^{-x/4}(e^{ik_+x}-e^{ik_-x})|\tilde{\psi}_0\rangle,    
\label{eq:zero}
\end{align}
which satisfies the semi-infinite boundary condition $x>0$.
Here $|\tilde{\psi}_0\rangle$ in Eq.~(\ref{eq:zero}) is an eigenstate with the eigenvalues $\eta_3s_2=1$ and $\tau_1\eta_3s_0=-1$, and thus
it has the two possible form
\begin{align}
\ket{\tau_1=-1}\otimes\ket{\eta_3=1}\otimes\ket{s_2=1},
\quad
\ket{\tau_1=1}\otimes\ket{\eta_3=-1}\otimes\ket{s_2=-1}.
\end{align}
In a similar manner, we also have zero modes in the $\tilde{M}_{xz}=-i$ sector,
\begin{align}
\ket{\psi_0}=e^{-x/4}(e^{-ik_+x}-e^{-ik_-x})\ket{\tilde{\psi}_0},    
\end{align}
where $\ket{\tilde{\psi}_0}$ are given by 
\begin{align}
\ket{\tau_1=-1}\otimes\ket{\eta_3=1}\otimes\ket{s_2=-1},
\quad
\ket{\tau_1=1}\otimes\ket{\eta_3=-1}\otimes\ket{s_2=1}.
\end{align}
Below, we represent these zero modes as $\ket{\psi_0^{(+1)}}$, $\ket{\psi_0^{(+2)}}$, $\ket{\psi_0^{(-1)}}$, and $\ket{\psi_0^{(-2)}}$:
\begin{align}
&|\psi^{(+1)}_0\rangle=N e^{-x/4}(e^{ik_+x}-e^{ik_-x})
\ket{\tau_1=-1}\otimes\ket{\eta_3=1}\otimes\ket{s_2=1},       
\nonumber\\
&|\psi^{(+2)}_0\rangle=N e^{-x/4}(e^{ik_+x}-e^{ik_-x})
\ket{\tau_1=1}\otimes\ket{\eta_3=-1}\otimes\ket{s_2=-1}, 
\nonumber\\
&|\psi^{(-1)}_0\rangle=N e^{-x/4}(e^{-ik_+x}-e^{-ik_-x})
\ket{\tau_1=-1}\otimes\ket{\eta_3=1}\otimes\ket{s_2=-1},       
\nonumber\\
&|\psi^{(-2)}_0\rangle=N e^{-x/4}(e^{-ik_+x}-e^{-ik_-x})
\ket{\tau_1=1}\otimes\ket{\eta_3=-1}\otimes\ket{s_2=1},
\end{align}
where the $\pm$ in the surfix specifies the mirror eigenvalue $\tilde{M}_{xz}=\pm i$, and $N=\sqrt{3/8}$ is the normalization constant.

From the zero modes, we can derive the effective Hamiltonian for the helical Majorana edge states. 
First, we approximate the helical Majorana edge states as
\begin{align}
\ket{\psi(k_y)}=u^{(+1)}\ket{\psi_0^{(+1)}}
+u^{(+2)}\ket{\psi_0^{(+2)}}
+u^{(-1)}\ket{\psi_0^{(-1)}}
+u^{(-2)}\ket{\psi_0^{(-2)}},
\label{eq:psi}
\end{align}
where $u^{(\mu \kappa)}$ ($\mu=\pm$, $\kappa=1,2$) is a function of $k_y$.
Then, the BdG equation $H({\bm k})\ket{\psi(k_y)}=E\ket{\psi(k_y)}$ leads to
\begin{align}
\sum_{\mu'\kappa'}\langle \psi_0^{(\mu\kappa)}|-k_y\tau_0\mu_3s_1|\psi_0^{(\mu'\kappa')}\rangle
u^{(\mu'\kappa')}=Eu^{(\mu\kappa)},        
\label{eq: BdG_edge}
\end{align}
where we have neglected $O(k_y^2)$ terms in $H({\bm k})$.
Using the relation
\begin{align}
s_1|s_2=\pm 1\rangle= \pm i\ket{s_2=\mp 1},   
\end{align}
we obtain
\begin{align}
-k_y\tau_0\eta_3s_1|\psi_0^{(+\kappa)}\rangle
&=-ik_y\frac{e^{ik_+ x}-e^{ik_- x}}{e^{-ik_+ x}-e^{-ik_- x}}
\ket{\psi_0^{(-\kappa)}},
\nonumber\\
-k_y\tau_0\eta_3s_1|\psi_0^{(-\kappa)}\rangle&
=ik_y\frac{e^{-ik_+ x}-e^{-ik_- x}}{e^{ik_+ x}-e^{ik_- x}}
\ket{\psi_0^{(+\kappa)}},
\end{align}
which leads to
\begin{align}
\langle \psi_0^{(\pm\kappa)}|-k_y\tau_0\eta_3 s_1|\psi_0^{(\mp\kappa)}\rangle=\frac{3}{4}k_y.
\end{align}
Thus, the matrix elements in the left-hand side of Eq.(\ref{eq: BdG_edge}) reads
\begin{align}
\langle \psi_0^{(\mu\kappa)}|-k_y\tau_0\mu_3s_1|\psi_0^{(\mu'\kappa')}\rangle=\frac{3}{4}k_y (\mu_1)_{\mu\mu'}(\kappa_0)_{\kappa\kappa'},
\end{align}
where $\mu_{\mu}$ and $\kappa_{\mu}$ are the Pauli matrices.
Therefore, the effective Hamiltonian for the helical edge states on the (10) surface is
\begin{align}
H_{\rm edge}=-iv\partial_y\mu_1\kappa_0,
\label{eq:effective_edge}
\end{align}
where $v=3/4$ and we have used the real space representation of $k_y$.
The obtained edge Hamiltonian retains $\tilde{M}_{xz}$, $\tilde{C}_{2z}\tilde{P}$, $\tilde{M}_{yz}P$, $T$ and $C$ in Eqs.(\ref{eq:sym1}) and (\ref{eq:sym2}).
Acting these symmetries on $|\psi(k_y)\rangle$ in the above, we specify the representation of these symmetries on the effective Hamiltonian as
\begin{align}
\tilde{M}_{xz}=i\mu_3\kappa_0,
\quad
\tilde{P}\tilde{M}_{yz}
=\mu_3\kappa_1,
\quad
\tilde{P}\tilde{C}_{2z}
=i\mu_0\kappa_1,
\quad
T=\mu_2\kappa_3 K,
\quad
C=\mu_1\kappa_3 K.
\end{align}

To obtain corner MFs, 
we make a corner between the $(1\bar{1})$ and $(11)$ edges by adiabatically bending the (10) edge at the right angle. 
The coordinate $y$ now parameterizes the corner configuration:
$y<0$ ($y>0$) corresponds to the $(1\bar{1})$ ((11))
edge, and $y=0$ is the corner.
On the corner configuration, the effective Hamiltonian in Eq.(\ref{eq:effective_edge}) can have a mass term because the $(1\bar{1})$ and $(11)$ edges do not preserve $M_{xz}$ and $M_{yz}P$.
The possible mass term is $\eta_2\kappa_1$, which anti-commutes with the kinetic term and preserves time-reversal and particle-hole symmetries.
$M_{xz}$ and $M_{yz}P$ exchange the $(1\bar{1})$ 
and $(11)$ edges, so they exchange the mass terms on these edges.
Therefore, the effective Hamiltonian on the corner is
\begin{align}
H_{\rm corner}= -iv\partial_y\mu_1\kappa_0
+m(y)\eta_2\kappa_1,
\end{align}
where $m(y)$ is a real odd function $m(-y)=-m(y)$.
For simplicity, we assume $m(y>0)>0$ below.

From the above corner Hamiltonian, we can immediately construct Majorana corner zero modes,  which obey the BdG equation,
\begin{align}
(-iv\partial_y \mu_1\kappa_0+m(y)\mu_2\kappa_1)\ket{u}=0.    
\end{align}
By multiplying $i\mu_1\kappa_0$ from the left, the above equation reads
\begin{align}
(v\partial_y -m(y)\mu_3\kappa_1)\ket{u}=0,   
\end{align}
which has a pair of normalized solutions when $\sigma_3\kappa_1=-1$,
\begin{align}
&\ket{u_0^{(1)}}=C\exp\left(-\int^{y} dy'm(y')/v\right)\ket{\sigma_3=1}\otimes\ket{\kappa_1=-1},    
\nonumber\\
&\ket{u_0^{(2)}}=iC\exp\left(-\int^{y} dy'm(y')/v\right)\ket{\sigma_3=-1}\otimes\ket{\kappa_1=1},    
\end{align}
with a normalization real constant $C$.
These corner zero modes form a Kramers pair, $\ket{u_0^{(2)}}=T\ket{u_0^{(1)}}$.

Now, we argue how to gap out the Majorana corner modes by perturbing the normal state Hamiltonian.
First, the perturbation must break time-reversal symmetry; otherwise, the mixing of the Majorana Kramers pair is prohibited.
Second, the perturbation should make the topological number for Majorana modes ill-defined.
In the present case, we use the one-dimensional winding number $w_{m_y}$ in Eq.(\ref{eq:1D_winding}) to obtain helical Majorana edge modes. 
From the equivalent form of $w_{m_y}$,
\begin{align}
w_{+i}(k_y)=-w_{-i}(k_y)=\int_{-\pi}^{\pi}\frac{k_x}{4\pi}\tr[\Gamma\tilde{M}_{xz}H^{-1}({\bm k})\partial_{k_x}H({\bm k})],    
\end{align}
it is evident that the mirror chiral symmetry
\begin{align}
\Gamma \tilde{M}_{xz}H({\bm k})(\Gamma \tilde{M}_{xz})^\dagger=-H(k_x,-k_y),  
\quad \Gamma \tilde{M}_{xz}=-iTC\tilde{M}_{xz},
\end{align}
is necessary to define the winding number. Thus, the time-reversal breaking perturbation should break the chiral symmetry at the same time. Since one cannot break particle-hole symmetry intrinsic to superconductors, we find that the perturbation must be even under $M_{xz}$.
From these two conditions, we have four possible perturbation terms in the normal state Hamiltonian
\begin{align}
O_{02}=\eta_0s_2,
\quad
O_{11}=\eta_1s_1,
\quad 
O_{13}=\eta_1 s_3,
\quad
O_{32}=\eta_3 s_2,
\end{align}
which give the following terms in the BdG Hamiltonian,
\begin{align}
{\cal O}_{02}=\tau_0\eta_0s_2,
\quad
{\cal O}_{11}=\tau_3\eta_1s_1,
\quad 
{\cal O}_{13}=\tau_3\eta_1 s_3,
\quad
{\cal O}_{32}=\tau_0\eta_3 s_2,
\label{eq:pert1}
\end{align}

Note that the above argument is consistent with our response theory. 
According to our theory, only a magnetic operator breaking time-reversal symmetry can coupled to a Majorana Kramers pair.
Moreover, it must share the same representation of crystalline symmetry as the gap function. 
In the present simplified model,
the relevant crystalline symmetry consists of 
\begin{align}
M_{xz}= i\eta_3s_2, \quad PM_{yz}=\eta_2s_1,\quad PC_{2z}=i\eta_1s_3,   
\end{align}
which preserve a corner between the $(11)$ and $(1\bar{1})$ edges, and the gap function is an $s_{+-}$-wave pairing.
Thus, the magnetic operator should be even under these symmetry operations. 
In particular, it should be even under $M_{xz}$.

Remarkably, our theory gives a stronger constraint of the magnetic perturbation:
In addition to $M_{xz}$, it must be even also under $PM_{xz}$ and $PC_{2z}$.
As a result, the magnetic terms coupled to the Majorana corner mode are found to be
\begin{align}
{\cal O}_{13}=\tau_3\eta_1 s_3,
\quad
{\cal O}_{32}=\tau_0\eta_3 s_2.
\label{eq:pert2}
\end{align}

We confirm this result by checking the coupling of these terms to the $\rho_{\rm MKP}^{(12)}$ operator.
The $\rho_{\rm MKP}^{(12)}$ operator of these Majorana corner modes is given by
\begin{align}
\rho_{\rm MKP}^{(12)}(y)=if^2(y)
\begin{pmatrix}
1&-1&0&0\\
-1&1&0&0\\
0&0&-1&-1\\
0&0&-1&-1
\end{pmatrix}_{\mu\otimes\kappa},
\end{align}
with
\begin{align}
f(y)=\frac{C}{\sqrt{2}} \exp\left(- \int^{y} dy'm(y')/v\right).
\end{align}
Acting the terms in Eq.(\ref{eq:pair_po}) on $\ket{\psi(k_y)}$ in Eq.(\ref{eq:psi}), we obtain the representation of these terms in the space of the effective Hamiltonian as
\begin{align}
{\cal O}_{02}=\mu_3\kappa_3
\quad
{\cal O}_{11}=\mu_3\kappa_2,
\quad 
{\cal O}_{13}=\mu_0\kappa_1,
\quad
{\cal O}_{32}=\mu_3\kappa_0.
\end{align}
Then, the trace of these terms with $\rho_{\rm MKP}^{(12)}$ reads
\begin{align}
&\tr [\mathcal{O}_{02}\rho^{(12)}_{\rm MKP}]=0, 
\quad
\tr [\mathcal{O}_{11}\rho^{(12)}_{\rm MKP}]=0, 
\nonumber\\
&\tr [\mathcal{O}_{13}\rho^{(12)}_{\rm MKP}]=-4if(y),
\quad
\tr [\mathcal{O}_{32}\rho^{(12)}_{\rm MKP}]=4if(y). 
\end{align}
Therefore, as expected from our theory, only ${\cal O}_{13}$ and ${\cal O}_{32}$ in Eq.(\ref{eq:pert2}) are coupled to the Majorana corner modes.

We can numerically obtain the same magnetic response of Majorana corner modes for the simplified s$_{+-}$-wave superconductor in Eq. (\ref{eq:simple_model}). We redefine the model on the square lattice by replacing $k_i^2 \to \cos(k_i)$ and $k_i \to \sin(k_i)$.  The tight-binding Hamiltonian is given by
\begin{align}
\mathcal{E} (\bm{k}) =& [t(\cos(k_x) + \cos(k_y)) - \mu ] \eta_0 s_0 + \lambda_{\rm R} \eta_3 [\sin(k_x)s_2 - \sin(k_y) s_1 ] \notag  \\
 &+ v \sin(k_x) \sin(k_y) \eta_1 s_0,
\end{align} 
which preserves the symmetries in Eq.(\ref{eq:sym0}). 
For the superconducting state, we consider the BdG Hamiltonian with an $s_{+-}$-wave pairing symmetry given by
\begin{align}
\tilde{H} (\bm{k}) &=  [t(\cos(k_x) + \cos(k_y)) - \mu ] \tau_3 \eta_0 s_0 + \lambda_{\rm R} [\sin(k_x) \tau_3 \eta_3 s_2 - \sin(k_y) \tau_0 \eta_3 s_1 ] \notag  \\
 &+ v \sin(k_x) \sin(k_y) \tau_3 \eta_1 s_0 - [\Delta_0+\Delta_1 (\cos(k_x)+\cos(k_y))] \tau_2 \eta_0 s_2,
\end{align}
which is invariant under the symmetries in Eq.(\ref{eq:sym1}) and particle-hole symmetry. It is verified that the corner MKP appears when imposing the open boundary conditions with $(11)$ and $(1\bar{1})$ edges and choosing the parameters as $(t,\lambda_{\rm R},v,\mu,\Delta_0,\Delta_1)=(2,0.5,0.5,1.5,0.7,-1)$. In the following, we use this parameter.

\begin{figure}[tb]
\centering\includegraphics[width=6in]{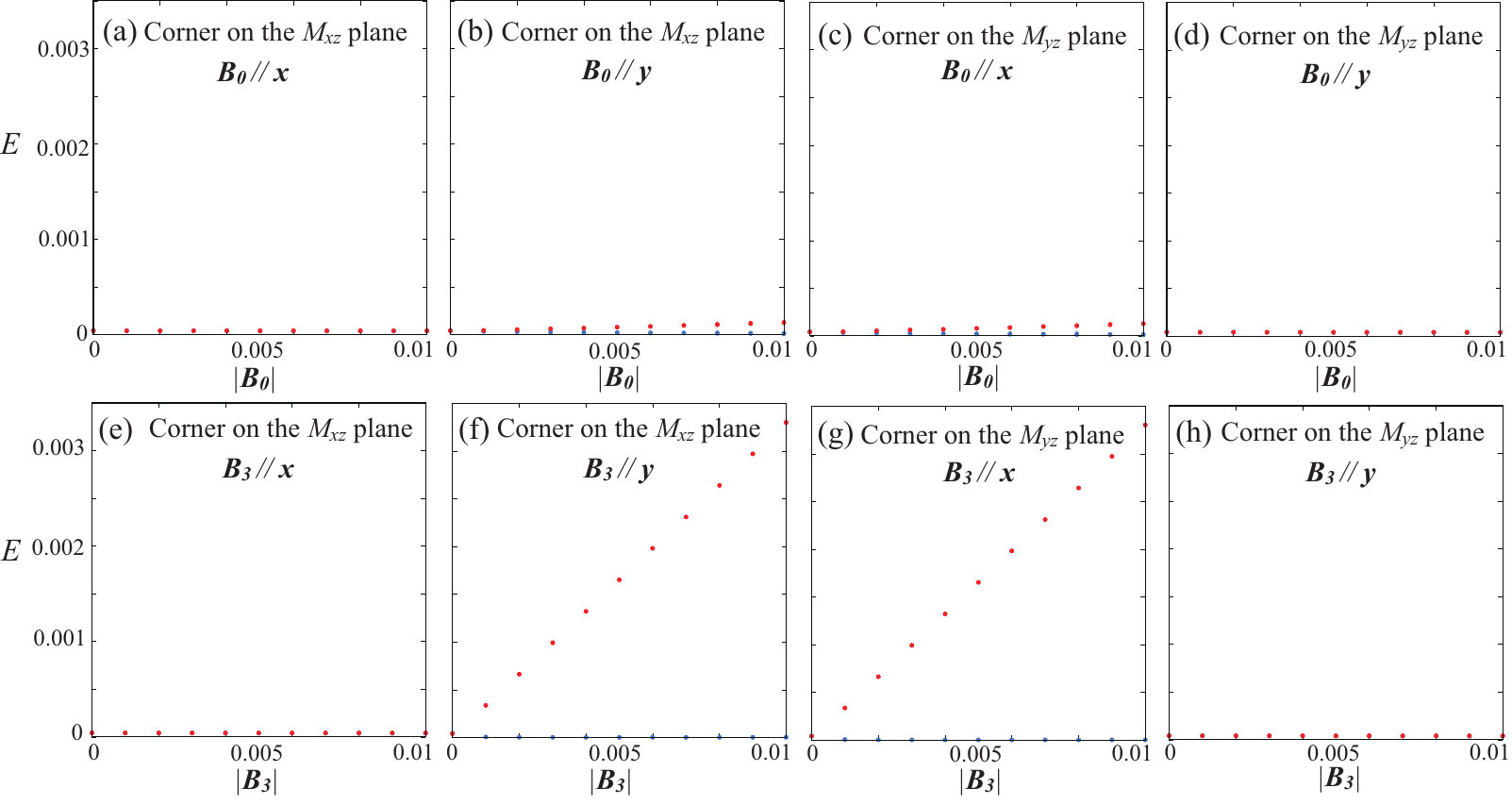}
\caption{ The energy spectra of the corner MKPs are shown as a function of the magnetic field: the direction and the types of the magnetic field are chosen as (a) $\bm{B}_0 \parallel \hat{x}$, (b) $\bm{B}_0 \parallel \hat{y}$, (e) $\bm{B}_3 \parallel \hat{x}$, and (f) $\bm{B}_3 \parallel \hat{y}$, where the BdG Hamiltonian with full open boundary condition with the lattice size being $N_{(11)}=N_{(1\bar{1})}=80$ is numerically diagonalized and the Zeeman term is added only at around the corners on the $M_{xz}$ mirror plane up to the five sites. Similarly, when the Zeeman term is added at around the corners on the $M_{yz}$ mirror plane, the energy spectra of the corner MKPs are shown in (c) $\bm{B}_0 \parallel \hat{x}$, (d) $\bm{B}_0 \parallel \hat{y}$, (g) $\bm{B}_3 \parallel \hat{x}$, and (h) $\bm{B}_3 \parallel \hat{y}$.
}
\label{fig:effectivemodel}
\end{figure}

To check the magnetic response of the corner MKP, we add two types of the Zeeman term: a sublattice-independent Zeeman term $\eta_0 \bm{B}_0 \cdot \bm{s} $ and a sublattice-dependent Zeeman term $\eta_3 \bm{B}_3 \cdot \bm{s}$ to the normal-state Hamiltonian. As we discussed above, the corner MKP has a nonzero expectation value only for the sublattice-dependent Zeeman term. Thus, only the sublattice-dependent Zeeman term generates an energy gap at the corner MKP. To verify this, we examine the energy spectrum of the corner MKP in four different cases: (i) $B_0$ at the corners preserving $M_{yz}$, (ii) $B_0$ at the corners preserving $M_{xz}$, (iii) $B_3$ at the corners preserving $M_{yz}$, (iv) $B_3$ at the corners preserving $M_{xz}$. Figure~\ref{fig:effectivemodel} shows the energy spectra of the corner MKP as a function of the Zeeman magnetic field $B_0$ or $B_3$: [(a),(b)] correspond to case (i), [(c),(d)] to case (ii), [(e),(f)] to case (iii), and [(g),(h)] to case (iv). As expected, the magnetic response of the corner MKP appears in cases (iii) and (iv), i.e., a sublattice-dependent Zeeman term, when the magnetic field is applied in the direction perpendicular to the mirror plane. On the other hand, in cases (i) and (ii),  there is little change in the energy spectra. 
Note that a small change occurs in Fig.~\ref{fig:effectivemodel} (b),(c) due to the mixing of each corner mode attributed to the finite size effect. 
The results are consistent with the analytical calculation in Appendix~\ref{app:effectivemodel}.

\let\doi\relax
\bibliography{multipole}

\begin{thebibliography}{81}%
\makeatletter
\providecommand \@ifxundefined [1]{%
 \@ifx{#1\undefined}
}%
\providecommand \@ifnum [1]{%
 \ifnum #1\expandafter \@firstoftwo
 \else \expandafter \@secondoftwo
 \fi
}%
\providecommand \@ifx [1]{%
 \ifx #1\expandafter \@firstoftwo
 \else \expandafter \@secondoftwo
 \fi
}%
\providecommand \natexlab [1]{#1}%
\providecommand \enquote  [1]{``#1''}%
\providecommand \bibnamefont  [1]{#1}%
\providecommand \bibfnamefont [1]{#1}%
\providecommand \citenamefont [1]{#1}%
\providecommand \href@noop [0]{\@secondoftwo}%
\providecommand \href [0]{\begingroup \@sanitize@url \@href}%
\providecommand \@href[1]{\@@startlink{#1}\@@href}%
\providecommand \@@href[1]{\endgroup#1\@@endlink}%
\providecommand \@sanitize@url [0]{\catcode `\\12\catcode `\$12\catcode
  `\&12\catcode `\#12\catcode `\^12\catcode `\_12\catcode `\%12\relax}%
\providecommand \@@startlink[1]{}%
\providecommand \@@endlink[0]{}%
\providecommand \url  [0]{\begingroup\@sanitize@url \@url }%
\providecommand \@url [1]{\endgroup\@href {#1}{\urlprefix }}%
\providecommand \urlprefix  [0]{URL }%
\providecommand \Eprint [0]{\href }%
\providecommand \doibase [0]{https://doi.org/}%
\providecommand \selectlanguage [0]{\@gobble}%
\providecommand \bibinfo  [0]{\@secondoftwo}%
\providecommand \bibfield  [0]{\@secondoftwo}%
\providecommand \translation [1]{[#1]}%
\providecommand \BibitemOpen [0]{}%
\providecommand \bibitemStop [0]{}%
\providecommand \bibitemNoStop [0]{.\EOS\space}%
\providecommand \EOS [0]{\spacefactor3000\relax}%
\providecommand \BibitemShut  [1]{\csname bibitem#1\endcsname}%
\let\auto@bib@innerbib\@empty
\bibitem [{\citenamefont {Qi}\ and\ \citenamefont {Zhang}(2011)}]{Qi11}%
  \BibitemOpen
  \bibfield  {author} {\bibinfo {author} {\bibfnamefont {X.-L.}\ \bibnamefont
  {Qi}}\ and\ \bibinfo {author} {\bibfnamefont {S.-C.}\ \bibnamefont {Zhang}},\
  }\bibfield  {title} {\bibinfo {title} {Topological insulators and
  superconductors},\ }\href {https://doi.org/10.1103/RevModPhys.83.1057}
  {\bibfield  {journal} {\bibinfo  {journal} {Rev. Mod. Phys.}\ }\textbf
  {\bibinfo {volume} {83}},\ \bibinfo {pages} {1057} (\bibinfo {year}
  {2011})}\BibitemShut {NoStop}%
\bibitem [{\citenamefont {Tanaka}\ \emph {et~al.}(2012)\citenamefont {Tanaka},
  \citenamefont {Sato},\ and\ \citenamefont {Nagaosa}}]{Tanaka12}%
  \BibitemOpen
  \bibfield  {author} {\bibinfo {author} {\bibfnamefont {Y.}~\bibnamefont
  {Tanaka}}, \bibinfo {author} {\bibfnamefont {M.}~\bibnamefont {Sato}},\ and\
  \bibinfo {author} {\bibfnamefont {N.}~\bibnamefont {Nagaosa}},\ }\bibfield
  {title} {\bibinfo {title} {Symmetry and topology in superconductors
  -odd-frequency pairing and edge states-},\ }\href
  {https://doi.org/10.1143/JPSJ.81.011013} {\bibfield  {journal} {\bibinfo
  {journal} {J. Phys. Soc. Jpn.}\ }\textbf {\bibinfo {volume} {81}},\ \bibinfo
  {pages} {011013} (\bibinfo {year} {2012})}\BibitemShut {NoStop}%
\bibitem [{\citenamefont {Sato}\ and\ \citenamefont {Ando}(2017)}]{Sato17}%
  \BibitemOpen
  \bibfield  {author} {\bibinfo {author} {\bibfnamefont {M.}~\bibnamefont
  {Sato}}\ and\ \bibinfo {author} {\bibfnamefont {Y.}~\bibnamefont {Ando}},\
  }\bibfield  {title} {\bibinfo {title} {Topological superconductors: a
  review},\ }\href {http://stacks.iop.org/0034-4885/80/i=7/a=076501} {\bibfield
   {journal} {\bibinfo  {journal} {Rep. Prog. Phys.}\ }\textbf {\bibinfo
  {volume} {80}},\ \bibinfo {pages} {076501} (\bibinfo {year}
  {2017})}\BibitemShut {NoStop}%
\bibitem [{\citenamefont {Read}\ and\ \citenamefont {Green}(2000)}]{Read00}%
  \BibitemOpen
  \bibfield  {author} {\bibinfo {author} {\bibfnamefont {N.}~\bibnamefont
  {Read}}\ and\ \bibinfo {author} {\bibfnamefont {D.}~\bibnamefont {Green}},\
  }\bibfield  {title} {\bibinfo {title} {Paired states of fermions in two
  dimensions with breaking of parity and time-reversal symmetries and the
  fractional quantum hall effect},\ }\href
  {https://doi.org/10.1103/PhysRevB.61.10267} {\bibfield  {journal} {\bibinfo
  {journal} {Phys. Rev. B}\ }\textbf {\bibinfo {volume} {61}},\ \bibinfo
  {pages} {10267} (\bibinfo {year} {2000})}\BibitemShut {NoStop}%
\bibitem [{\citenamefont {Kitaev}(2001)}]{Kitaev01}%
  \BibitemOpen
  \bibfield  {author} {\bibinfo {author} {\bibfnamefont {A.~Y.}\ \bibnamefont
  {Kitaev}},\ }\bibfield  {title} {\bibinfo {title} {Unpaired majorana fermions
  in quantum wires},\ }\href {https://doi.org/10.1070/1063-7869/44/10S/S29}
  {\bibfield  {journal} {\bibinfo  {journal} {Physics-Uspekhi}\ }\textbf
  {\bibinfo {volume} {44}},\ \bibinfo {pages} {131} (\bibinfo {year}
  {2001})}\BibitemShut {NoStop}%
\bibitem [{\citenamefont {Sato}(2003)}]{Sato03}%
  \BibitemOpen
  \bibfield  {author} {\bibinfo {author} {\bibfnamefont {M.}~\bibnamefont
  {Sato}},\ }\bibfield  {title} {\bibinfo {title} {Non-abelian statistics of
  axion strings},\ }\href {https://doi.org/10.1016/j.physletb.2003.09.047}
  {\bibfield  {journal} {\bibinfo  {journal} {Physics Letters B}\ }\textbf
  {\bibinfo {volume} {575}},\ \bibinfo {pages} {126–130} (\bibinfo {year}
  {2003})}\BibitemShut {NoStop}%
\bibitem [{\citenamefont {Fu}\ and\ \citenamefont {Kane}(2008)}]{Fu08}%
  \BibitemOpen
  \bibfield  {author} {\bibinfo {author} {\bibfnamefont {L.}~\bibnamefont
  {Fu}}\ and\ \bibinfo {author} {\bibfnamefont {C.~L.}\ \bibnamefont {Kane}},\
  }\bibfield  {title} {\bibinfo {title} {Superconducting proximity effect and
  majorana fermions at the surface of a topological insulator},\ }\href
  {https://doi.org/10.1103/PhysRevLett.100.096407} {\bibfield  {journal}
  {\bibinfo  {journal} {Phys. Rev. Lett.}\ }\textbf {\bibinfo {volume} {100}},\
  \bibinfo {pages} {096407} (\bibinfo {year} {2008})}\BibitemShut {NoStop}%
\bibitem [{\citenamefont {Sato}\ \emph {et~al.}(2009)\citenamefont {Sato},
  \citenamefont {Takahashi},\ and\ \citenamefont {Fujimoto}}]{SFT09}%
  \BibitemOpen
  \bibfield  {author} {\bibinfo {author} {\bibfnamefont {M.}~\bibnamefont
  {Sato}}, \bibinfo {author} {\bibfnamefont {Y.}~\bibnamefont {Takahashi}},\
  and\ \bibinfo {author} {\bibfnamefont {S.}~\bibnamefont {Fujimoto}},\
  }\bibfield  {title} {\bibinfo {title} {Non-abelian topological order in
  $s$-wave superfluids of ultracold fermionic atoms},\ }\href
  {https://doi.org/10.1103/PhysRevLett.103.020401} {\bibfield  {journal}
  {\bibinfo  {journal} {Phys. Rev. Lett.}\ }\textbf {\bibinfo {volume} {103}},\
  \bibinfo {pages} {020401} (\bibinfo {year} {2009})}\BibitemShut {NoStop}%
\bibitem [{\citenamefont {Alicea}(2010)}]{Jason10}%
  \BibitemOpen
  \bibfield  {author} {\bibinfo {author} {\bibfnamefont {J.}~\bibnamefont
  {Alicea}},\ }\bibfield  {title} {\bibinfo {title} {Majorana fermions in a
  tunable semiconductor device},\ }\href
  {https://doi.org/10.1103/PhysRevB.81.125318} {\bibfield  {journal} {\bibinfo
  {journal} {Phys. Rev. B}\ }\textbf {\bibinfo {volume} {81}},\ \bibinfo
  {pages} {125318} (\bibinfo {year} {2010})}\BibitemShut {NoStop}%
\bibitem [{\citenamefont {Sau}\ \emph {et~al.}(2010)\citenamefont {Sau},
  \citenamefont {Lutchyn}, \citenamefont {Tewari},\ and\ \citenamefont
  {Das~Sarma}}]{Sau10}%
  \BibitemOpen
  \bibfield  {author} {\bibinfo {author} {\bibfnamefont {J.~D.}\ \bibnamefont
  {Sau}}, \bibinfo {author} {\bibfnamefont {R.~M.}\ \bibnamefont {Lutchyn}},
  \bibinfo {author} {\bibfnamefont {S.}~\bibnamefont {Tewari}},\ and\ \bibinfo
  {author} {\bibfnamefont {S.}~\bibnamefont {Das~Sarma}},\ }\bibfield  {title}
  {\bibinfo {title} {Generic new platform for topological quantum computation
  using semiconductor heterostructures},\ }\href
  {https://doi.org/10.1103/PhysRevLett.104.040502} {\bibfield  {journal}
  {\bibinfo  {journal} {Phys. Rev. Lett.}\ }\textbf {\bibinfo {volume} {104}},\
  \bibinfo {pages} {040502} (\bibinfo {year} {2010})}\BibitemShut {NoStop}%
\bibitem [{\citenamefont {Lutchyn}\ \emph {et~al.}(2010)\citenamefont
  {Lutchyn}, \citenamefont {Sau},\ and\ \citenamefont {Das~Sarma}}]{Lutchyn10}%
  \BibitemOpen
  \bibfield  {author} {\bibinfo {author} {\bibfnamefont {R.~M.}\ \bibnamefont
  {Lutchyn}}, \bibinfo {author} {\bibfnamefont {J.~D.}\ \bibnamefont {Sau}},\
  and\ \bibinfo {author} {\bibfnamefont {S.}~\bibnamefont {Das~Sarma}},\
  }\bibfield  {title} {\bibinfo {title} {Majorana fermions and a topological
  phase transition in semiconductor-superconductor heterostructures},\ }\href
  {https://doi.org/10.1103/PhysRevLett.105.077001} {\bibfield  {journal}
  {\bibinfo  {journal} {Phys. Rev. Lett.}\ }\textbf {\bibinfo {volume} {105}},\
  \bibinfo {pages} {077001} (\bibinfo {year} {2010})}\BibitemShut {NoStop}%
\bibitem [{\citenamefont {Oreg}\ \emph {et~al.}(2010)\citenamefont {Oreg},
  \citenamefont {Refael},\ and\ \citenamefont {von Oppen}}]{Oreg10}%
  \BibitemOpen
  \bibfield  {author} {\bibinfo {author} {\bibfnamefont {Y.}~\bibnamefont
  {Oreg}}, \bibinfo {author} {\bibfnamefont {G.}~\bibnamefont {Refael}},\ and\
  \bibinfo {author} {\bibfnamefont {F.}~\bibnamefont {von Oppen}},\ }\bibfield
  {title} {\bibinfo {title} {Helical liquids and majorana bound states in
  quantum wires},\ }\href {https://doi.org/10.1103/PhysRevLett.105.177002}
  {\bibfield  {journal} {\bibinfo  {journal} {Phys. Rev. Lett.}\ }\textbf
  {\bibinfo {volume} {105}},\ \bibinfo {pages} {177002} (\bibinfo {year}
  {2010})}\BibitemShut {NoStop}%
\bibitem [{\citenamefont {Wang}\ \emph {et~al.}(2015)\citenamefont {Wang},
  \citenamefont {Zhang}, \citenamefont {Xu}, \citenamefont {Zeng},
  \citenamefont {Miao}, \citenamefont {Xu}, \citenamefont {Qian}, \citenamefont
  {Weng}, \citenamefont {Richard}, \citenamefont {Fedorov}, \citenamefont
  {Ding}, \citenamefont {Dai},\ and\ \citenamefont {Fang}}]{Wang15}%
  \BibitemOpen
  \bibfield  {author} {\bibinfo {author} {\bibfnamefont {Z.}~\bibnamefont
  {Wang}}, \bibinfo {author} {\bibfnamefont {P.}~\bibnamefont {Zhang}},
  \bibinfo {author} {\bibfnamefont {G.}~\bibnamefont {Xu}}, \bibinfo {author}
  {\bibfnamefont {L.~K.}\ \bibnamefont {Zeng}}, \bibinfo {author}
  {\bibfnamefont {H.}~\bibnamefont {Miao}}, \bibinfo {author} {\bibfnamefont
  {X.}~\bibnamefont {Xu}}, \bibinfo {author} {\bibfnamefont {T.}~\bibnamefont
  {Qian}}, \bibinfo {author} {\bibfnamefont {H.}~\bibnamefont {Weng}}, \bibinfo
  {author} {\bibfnamefont {P.}~\bibnamefont {Richard}}, \bibinfo {author}
  {\bibfnamefont {A.~V.}\ \bibnamefont {Fedorov}}, \bibinfo {author}
  {\bibfnamefont {H.}~\bibnamefont {Ding}}, \bibinfo {author} {\bibfnamefont
  {X.}~\bibnamefont {Dai}},\ and\ \bibinfo {author} {\bibfnamefont
  {Z.}~\bibnamefont {Fang}},\ }\bibfield  {title} {\bibinfo {title}
  {Topological nature of the ${\mathrm{fese}}_{0.5}{\mathrm{te}}_{0.5}$
  superconductor},\ }\href {https://doi.org/10.1103/PhysRevB.92.115119}
  {\bibfield  {journal} {\bibinfo  {journal} {Phys. Rev. B}\ }\textbf {\bibinfo
  {volume} {92}},\ \bibinfo {pages} {115119} (\bibinfo {year}
  {2015})}\BibitemShut {NoStop}%
\bibitem [{\citenamefont {Wu}\ \emph {et~al.}(2016)\citenamefont {Wu},
  \citenamefont {Qin}, \citenamefont {Liang}, \citenamefont {Fan},\ and\
  \citenamefont {Hu}}]{Wu16}%
  \BibitemOpen
  \bibfield  {author} {\bibinfo {author} {\bibfnamefont {X.}~\bibnamefont
  {Wu}}, \bibinfo {author} {\bibfnamefont {S.}~\bibnamefont {Qin}}, \bibinfo
  {author} {\bibfnamefont {Y.}~\bibnamefont {Liang}}, \bibinfo {author}
  {\bibfnamefont {H.}~\bibnamefont {Fan}},\ and\ \bibinfo {author}
  {\bibfnamefont {J.}~\bibnamefont {Hu}},\ }\bibfield  {title} {\bibinfo
  {title} {Topological characters in
  $\mathrm{Fe}({\mathrm{te}}_{1\ensuremath{-}x}{\mathrm{se}}_{x})$ thin
  films},\ }\href {https://doi.org/10.1103/PhysRevB.93.115129} {\bibfield
  {journal} {\bibinfo  {journal} {Phys. Rev. B}\ }\textbf {\bibinfo {volume}
  {93}},\ \bibinfo {pages} {115129} (\bibinfo {year} {2016})}\BibitemShut
  {NoStop}%
\bibitem [{\citenamefont {Xu}\ \emph {et~al.}(2016)\citenamefont {Xu},
  \citenamefont {Lian}, \citenamefont {Tang}, \citenamefont {Qi},\ and\
  \citenamefont {Zhang}}]{Xu16}%
  \BibitemOpen
  \bibfield  {author} {\bibinfo {author} {\bibfnamefont {G.}~\bibnamefont
  {Xu}}, \bibinfo {author} {\bibfnamefont {B.}~\bibnamefont {Lian}}, \bibinfo
  {author} {\bibfnamefont {P.}~\bibnamefont {Tang}}, \bibinfo {author}
  {\bibfnamefont {X.-L.}\ \bibnamefont {Qi}},\ and\ \bibinfo {author}
  {\bibfnamefont {S.-C.}\ \bibnamefont {Zhang}},\ }\bibfield  {title} {\bibinfo
  {title} {Topological superconductivity on the surface of fe-based
  superconductors},\ }\href {https://doi.org/10.1103/PhysRevLett.117.047001}
  {\bibfield  {journal} {\bibinfo  {journal} {Phys. Rev. Lett.}\ }\textbf
  {\bibinfo {volume} {117}},\ \bibinfo {pages} {047001} (\bibinfo {year}
  {2016})}\BibitemShut {NoStop}%
\bibitem [{\citenamefont {Zhang}\ \emph {et~al.}(2018)\citenamefont {Zhang},
  \citenamefont {Yaji}, \citenamefont {Hashimoto}, \citenamefont {Ota},
  \citenamefont {Kondo}, \citenamefont {Okazaki}, \citenamefont {Wang},
  \citenamefont {Wen}, \citenamefont {Gu}, \citenamefont {Ding} \emph
  {et~al.}}]{Zhang18}%
  \BibitemOpen
  \bibfield  {author} {\bibinfo {author} {\bibfnamefont {P.}~\bibnamefont
  {Zhang}}, \bibinfo {author} {\bibfnamefont {K.}~\bibnamefont {Yaji}},
  \bibinfo {author} {\bibfnamefont {T.}~\bibnamefont {Hashimoto}}, \bibinfo
  {author} {\bibfnamefont {Y.}~\bibnamefont {Ota}}, \bibinfo {author}
  {\bibfnamefont {T.}~\bibnamefont {Kondo}}, \bibinfo {author} {\bibfnamefont
  {K.}~\bibnamefont {Okazaki}}, \bibinfo {author} {\bibfnamefont
  {Z.}~\bibnamefont {Wang}}, \bibinfo {author} {\bibfnamefont {J.}~\bibnamefont
  {Wen}}, \bibinfo {author} {\bibfnamefont {G.}~\bibnamefont {Gu}}, \bibinfo
  {author} {\bibfnamefont {H.}~\bibnamefont {Ding}}, \emph {et~al.},\
  }\bibfield  {title} {\bibinfo {title} {Observation of topological
  superconductivity on the surface of an iron-based superconductor},\
  }\href@noop {} {\bibfield  {journal} {\bibinfo  {journal} {Science}\ }\textbf
  {\bibinfo {volume} {360}},\ \bibinfo {pages} {182} (\bibinfo {year}
  {2018})}\BibitemShut {NoStop}%
\bibitem [{\citenamefont {Wang}\ \emph {et~al.}(2018)\citenamefont {Wang},
  \citenamefont {Kong}, \citenamefont {Fan}, \citenamefont {Chen},
  \citenamefont {Zhu}, \citenamefont {Liu}, \citenamefont {Cao}, \citenamefont
  {Sun}, \citenamefont {Du}, \citenamefont {Schneeloch} \emph
  {et~al.}}]{Wang18}%
  \BibitemOpen
  \bibfield  {author} {\bibinfo {author} {\bibfnamefont {D.}~\bibnamefont
  {Wang}}, \bibinfo {author} {\bibfnamefont {L.}~\bibnamefont {Kong}}, \bibinfo
  {author} {\bibfnamefont {P.}~\bibnamefont {Fan}}, \bibinfo {author}
  {\bibfnamefont {H.}~\bibnamefont {Chen}}, \bibinfo {author} {\bibfnamefont
  {S.}~\bibnamefont {Zhu}}, \bibinfo {author} {\bibfnamefont {W.}~\bibnamefont
  {Liu}}, \bibinfo {author} {\bibfnamefont {L.}~\bibnamefont {Cao}}, \bibinfo
  {author} {\bibfnamefont {Y.}~\bibnamefont {Sun}}, \bibinfo {author}
  {\bibfnamefont {S.}~\bibnamefont {Du}}, \bibinfo {author} {\bibfnamefont
  {J.}~\bibnamefont {Schneeloch}}, \emph {et~al.},\ }\bibfield  {title}
  {\bibinfo {title} {Evidence for majorana bound states in an iron-based
  superconductor},\ }\href@noop {} {\bibfield  {journal} {\bibinfo  {journal}
  {Science}\ }\textbf {\bibinfo {volume} {362}},\ \bibinfo {pages} {333}
  (\bibinfo {year} {2018})}\BibitemShut {NoStop}%
\bibitem [{\citenamefont {Machida}\ \emph {et~al.}(2019)\citenamefont
  {Machida}, \citenamefont {Sun}, \citenamefont {Pyon}, \citenamefont {Takeda},
  \citenamefont {Kohsaka}, \citenamefont {Hanaguri}, \citenamefont {Sasagawa},\
  and\ \citenamefont {Tamegai}}]{Machida19}%
  \BibitemOpen
  \bibfield  {author} {\bibinfo {author} {\bibfnamefont {T.}~\bibnamefont
  {Machida}}, \bibinfo {author} {\bibfnamefont {Y.}~\bibnamefont {Sun}},
  \bibinfo {author} {\bibfnamefont {S.}~\bibnamefont {Pyon}}, \bibinfo {author}
  {\bibfnamefont {S.}~\bibnamefont {Takeda}}, \bibinfo {author} {\bibfnamefont
  {Y.}~\bibnamefont {Kohsaka}}, \bibinfo {author} {\bibfnamefont
  {T.}~\bibnamefont {Hanaguri}}, \bibinfo {author} {\bibfnamefont
  {T.}~\bibnamefont {Sasagawa}},\ and\ \bibinfo {author} {\bibfnamefont
  {T.}~\bibnamefont {Tamegai}},\ }\bibfield  {title} {\bibinfo {title}
  {Zero-energy vortex bound state in the superconducting topological surface
  state of fe(se,te)},\ }\href {https://doi.org/10.1038/s41563-019-0397-1}
  {\bibfield  {journal} {\bibinfo  {journal} {Nature Materials}\ }\textbf
  {\bibinfo {volume} {18}},\ \bibinfo {pages} {811–815} (\bibinfo {year}
  {2019})}\BibitemShut {NoStop}%
\bibitem [{\citenamefont {Ryu}\ and\ \citenamefont {Hatsugai}(2002)}]{Ryu02}%
  \BibitemOpen
  \bibfield  {author} {\bibinfo {author} {\bibfnamefont {S.}~\bibnamefont
  {Ryu}}\ and\ \bibinfo {author} {\bibfnamefont {Y.}~\bibnamefont {Hatsugai}},\
  }\bibfield  {title} {\bibinfo {title} {Topological origin of zero-energy edge
  states in particle-hole symmetric systems},\ }\href
  {https://doi.org/10.1103/PhysRevLett.89.077002} {\bibfield  {journal}
  {\bibinfo  {journal} {Phys. Rev. Lett.}\ }\textbf {\bibinfo {volume} {89}},\
  \bibinfo {pages} {077002} (\bibinfo {year} {2002})}\BibitemShut {NoStop}%
\bibitem [{\citenamefont {Sato}\ \emph {et~al.}(2011)\citenamefont {Sato},
  \citenamefont {Tanaka}, \citenamefont {Yada},\ and\ \citenamefont
  {Yokoyama}}]{Sato11}%
  \BibitemOpen
  \bibfield  {author} {\bibinfo {author} {\bibfnamefont {M.}~\bibnamefont
  {Sato}}, \bibinfo {author} {\bibfnamefont {Y.}~\bibnamefont {Tanaka}},
  \bibinfo {author} {\bibfnamefont {K.}~\bibnamefont {Yada}},\ and\ \bibinfo
  {author} {\bibfnamefont {T.}~\bibnamefont {Yokoyama}},\ }\bibfield  {title}
  {\bibinfo {title} {Topology of {Andreev} bound states with flat dispersion},\
  }\href {https://doi.org/10.1103/PhysRevB.83.224511} {\bibfield  {journal}
  {\bibinfo  {journal} {Phys. Rev. B}\ }\textbf {\bibinfo {volume} {83}},\
  \bibinfo {pages} {224511} (\bibinfo {year} {2011})}\BibitemShut {NoStop}%
\bibitem [{\citenamefont {Kashiwaya}\ and\ \citenamefont
  {Tanaka}(2000)}]{Kashiwaya2000}%
  \BibitemOpen
  \bibfield  {author} {\bibinfo {author} {\bibfnamefont {S.}~\bibnamefont
  {Kashiwaya}}\ and\ \bibinfo {author} {\bibfnamefont {Y.}~\bibnamefont
  {Tanaka}},\ }\bibfield  {title} {\bibinfo {title} {Tunnelling effects on
  surface bound states in unconventional superconductors},\ }\href
  {https://doi.org/10.1088/0034-4885/63/10/202} {\bibfield  {journal} {\bibinfo
   {journal} {Rep. Prog. Phys.}\ }\textbf {\bibinfo {volume} {63}},\ \bibinfo
  {pages} {1641} (\bibinfo {year} {2000})}\BibitemShut {NoStop}%
\bibitem [{\citenamefont {Streater}\ and\ \citenamefont
  {Wightman}(2000)}]{Streater00}%
  \BibitemOpen
  \bibfield  {author} {\bibinfo {author} {\bibfnamefont {R.~F.}\ \bibnamefont
  {Streater}}\ and\ \bibinfo {author} {\bibfnamefont {A.~S.}\ \bibnamefont
  {Wightman}},\ }\href@noop {} {\emph {\bibinfo {title} {PCT, spin and
  statistics, and all that}}},\ Vol.~\bibinfo {volume} {52}\ (\bibinfo
  {publisher} {Princeton University Press},\ \bibinfo {year}
  {2000})\BibitemShut {NoStop}%
\bibitem [{Note1()}]{Note1}%
  \BibitemOpen
  \bibinfo {note} {A physical Majorana neutrino can not be invariant under $C$
  once $C$-violating weak interactions are turn on~\cite {Kayser82}. Instead,
  it should be an eigenstate of $CPT$, which is not broken by the
  $CPT$-theorem.}\BibitemShut {Stop}%
\bibitem [{\citenamefont {Kayser}\ and\ \citenamefont
  {Goldhaber}(1983)}]{Kayser83}%
  \BibitemOpen
  \bibfield  {author} {\bibinfo {author} {\bibfnamefont {B.}~\bibnamefont
  {Kayser}}\ and\ \bibinfo {author} {\bibfnamefont {A.~S.}\ \bibnamefont
  {Goldhaber}},\ }\bibfield  {title} {\bibinfo {title} {{$\mathrm{CPT}$} and
  {$\mathrm{CP}$} properties of {Majorana} particles, and the consequences},\
  }\href {https://doi.org/10.1103/PhysRevD.28.2341} {\bibfield  {journal}
  {\bibinfo  {journal} {Phys. Rev. D}\ }\textbf {\bibinfo {volume} {28}},\
  \bibinfo {pages} {2341} (\bibinfo {year} {1983})}\BibitemShut {NoStop}%
\bibitem [{Note2()}]{Note2}%
  \BibitemOpen
  \bibinfo {note} {Note that the charge conjugation $C$ in the quantum field
  theory is unitary.}\BibitemShut {Stop}%
\bibitem [{\citenamefont {Schnyder}\ \emph {et~al.}(2008)\citenamefont
  {Schnyder}, \citenamefont {Ryu}, \citenamefont {Furusaki},\ and\
  \citenamefont {Ludwig}}]{Schnyder08}%
  \BibitemOpen
  \bibfield  {author} {\bibinfo {author} {\bibfnamefont {A.~P.}\ \bibnamefont
  {Schnyder}}, \bibinfo {author} {\bibfnamefont {S.}~\bibnamefont {Ryu}},
  \bibinfo {author} {\bibfnamefont {A.}~\bibnamefont {Furusaki}},\ and\
  \bibinfo {author} {\bibfnamefont {A.~W.~W.}\ \bibnamefont {Ludwig}},\
  }\bibfield  {title} {\bibinfo {title} {Classification of topological
  insulators and superconductors in three spatial dimensions},\ }\href
  {https://doi.org/10.1103/PhysRevB.78.195125} {\bibfield  {journal} {\bibinfo
  {journal} {Phys. Rev. B}\ }\textbf {\bibinfo {volume} {78}},\ \bibinfo
  {pages} {195125} (\bibinfo {year} {2008})}\BibitemShut {NoStop}%
\bibitem [{\citenamefont {Kitaev}(2009)}]{Kitaev09}%
  \BibitemOpen
  \bibfield  {author} {\bibinfo {author} {\bibfnamefont {A.}~\bibnamefont
  {Kitaev}},\ }\bibfield  {title} {\bibinfo {title} {Periodic table for
  topological insulators and superconductors},\ }\href
  {https://doi.org/10.1063/1.3149495} {\bibfield  {journal} {\bibinfo
  {journal} {AIP Conference Proceedings}\ }\textbf {\bibinfo {volume} {1134}},\
  \bibinfo {pages} {22} (\bibinfo {year} {2009})}\BibitemShut {NoStop}%
\bibitem [{\citenamefont {Schnyder}\ \emph {et~al.}(2009)\citenamefont
  {Schnyder}, \citenamefont {Ryu}, \citenamefont {Furusaki},\ and\
  \citenamefont {Ludwig}}]{Schnyder09}%
  \BibitemOpen
  \bibfield  {author} {\bibinfo {author} {\bibfnamefont {A.~P.}\ \bibnamefont
  {Schnyder}}, \bibinfo {author} {\bibfnamefont {S.}~\bibnamefont {Ryu}},
  \bibinfo {author} {\bibfnamefont {A.}~\bibnamefont {Furusaki}},\ and\
  \bibinfo {author} {\bibfnamefont {A.~W.~W.}\ \bibnamefont {Ludwig}},\
  }\bibfield  {title} {\bibinfo {title} {Classification of topological
  insulators and superconductors},\ }\href {https://doi.org/10.1063/1.3149481}
  {\bibfield  {journal} {\bibinfo  {journal} {AIP Conference Proceedings}\
  }\textbf {\bibinfo {volume} {1134}},\ \bibinfo {pages} {10} (\bibinfo {year}
  {2009})}\BibitemShut {NoStop}%
\bibitem [{\citenamefont {Ryu}\ \emph {et~al.}(2010)\citenamefont {Ryu},
  \citenamefont {Schnyder}, \citenamefont {Furusaki},\ and\ \citenamefont
  {Ludwig}}]{Ryu10}%
  \BibitemOpen
  \bibfield  {author} {\bibinfo {author} {\bibfnamefont {S.}~\bibnamefont
  {Ryu}}, \bibinfo {author} {\bibfnamefont {A.~P.}\ \bibnamefont {Schnyder}},
  \bibinfo {author} {\bibfnamefont {A.}~\bibnamefont {Furusaki}},\ and\
  \bibinfo {author} {\bibfnamefont {A.~W.~W.}\ \bibnamefont {Ludwig}},\
  }\bibfield  {title} {\bibinfo {title} {Topological insulators and
  superconductors: tenfold way and dimensional hierarchy},\ }\href
  {http://stacks.iop.org/1367-2630/12/i=6/a=065010} {\bibfield  {journal}
  {\bibinfo  {journal} {New Journal of Physics}\ }\textbf {\bibinfo {volume}
  {12}},\ \bibinfo {pages} {065010} (\bibinfo {year} {2010})}\BibitemShut
  {NoStop}%
\bibitem [{\citenamefont {Shiozaki}\ \emph {et~al.}(2022)\citenamefont
  {Shiozaki}, \citenamefont {Sato},\ and\ \citenamefont {Gomi}}]{Shiozaki22}%
  \BibitemOpen
  \bibfield  {author} {\bibinfo {author} {\bibfnamefont {K.}~\bibnamefont
  {Shiozaki}}, \bibinfo {author} {\bibfnamefont {M.}~\bibnamefont {Sato}},\
  and\ \bibinfo {author} {\bibfnamefont {K.}~\bibnamefont {Gomi}},\ }\bibfield
  {title} {\bibinfo {title} {Atiyah-hirzebruch spectral sequence in band
  topology: General formalism and topological invariants for 230 space
  groups},\ }\href {https://doi.org/10.1103/PhysRevB.106.165103} {\bibfield
  {journal} {\bibinfo  {journal} {Phys. Rev. B}\ }\textbf {\bibinfo {volume}
  {106}},\ \bibinfo {pages} {165103} (\bibinfo {year} {2022})}\BibitemShut
  {NoStop}%
\bibitem [{\citenamefont {Sato}(2009)}]{Sato09}%
  \BibitemOpen
  \bibfield  {author} {\bibinfo {author} {\bibfnamefont {M.}~\bibnamefont
  {Sato}},\ }\bibfield  {title} {\bibinfo {title} {Topological properties of
  spin-triplet superconductors and {Fermi} surface topology in the normal
  state},\ }\href {https://doi.org/10.1103/PhysRevB.79.214526} {\bibfield
  {journal} {\bibinfo  {journal} {Phys. Rev. B}\ }\textbf {\bibinfo {volume}
  {79}},\ \bibinfo {pages} {214526} (\bibinfo {year} {2009})}\BibitemShut
  {NoStop}%
\bibitem [{\citenamefont {Fu}\ and\ \citenamefont {Berg}(2010)}]{Fu10}%
  \BibitemOpen
  \bibfield  {author} {\bibinfo {author} {\bibfnamefont {L.}~\bibnamefont
  {Fu}}\ and\ \bibinfo {author} {\bibfnamefont {E.}~\bibnamefont {Berg}},\
  }\bibfield  {title} {\bibinfo {title} {Odd-parity topological
  superconductors: Theory and application to
  {${\mathrm{Cu}}_{x}{\mathrm{Bi}}_{2}{\mathrm{Se}}_{3}$}},\ }\href
  {https://doi.org/10.1103/PhysRevLett.105.097001} {\bibfield  {journal}
  {\bibinfo  {journal} {Phys. Rev. Lett.}\ }\textbf {\bibinfo {volume} {105}},\
  \bibinfo {pages} {097001} (\bibinfo {year} {2010})}\BibitemShut {NoStop}%
\bibitem [{\citenamefont {Sato}(2010)}]{Sato10}%
  \BibitemOpen
  \bibfield  {author} {\bibinfo {author} {\bibfnamefont {M.}~\bibnamefont
  {Sato}},\ }\bibfield  {title} {\bibinfo {title} {Topological odd-parity
  superconductors},\ }\href {https://doi.org/10.1103/PhysRevB.81.220504}
  {\bibfield  {journal} {\bibinfo  {journal} {Phys. Rev. B}\ }\textbf {\bibinfo
  {volume} {81}},\ \bibinfo {pages} {220504(R)} (\bibinfo {year}
  {2010})}\BibitemShut {NoStop}%
\bibitem [{\citenamefont {Kobayashi}\ \emph {et~al.}(2019)\citenamefont
  {Kobayashi}, \citenamefont {Yamakage}, \citenamefont {Tanaka},\ and\
  \citenamefont {Sato}}]{Kobayashi2019}%
  \BibitemOpen
  \bibfield  {author} {\bibinfo {author} {\bibfnamefont {S.}~\bibnamefont
  {Kobayashi}}, \bibinfo {author} {\bibfnamefont {A.}~\bibnamefont {Yamakage}},
  \bibinfo {author} {\bibfnamefont {Y.}~\bibnamefont {Tanaka}},\ and\ \bibinfo
  {author} {\bibfnamefont {M.}~\bibnamefont {Sato}},\ }\bibfield  {title}
  {\bibinfo {title} {Majorana multipole response of topological
  superconductors},\ }\href {https://doi.org/10.1103/PhysRevLett.123.097002}
  {\bibfield  {journal} {\bibinfo  {journal} {Phys. Rev. Lett.}\ }\textbf
  {\bibinfo {volume} {123}},\ \bibinfo {pages} {097002} (\bibinfo {year}
  {2019})}\BibitemShut {NoStop}%
\bibitem [{\citenamefont {Kobayashi}\ \emph {et~al.}(2021)\citenamefont
  {Kobayashi}, \citenamefont {Yamazaki}, \citenamefont {Yamakage},\ and\
  \citenamefont {Sato}}]{Kobayashi2021}%
  \BibitemOpen
  \bibfield  {author} {\bibinfo {author} {\bibfnamefont {S.}~\bibnamefont
  {Kobayashi}}, \bibinfo {author} {\bibfnamefont {Y.}~\bibnamefont {Yamazaki}},
  \bibinfo {author} {\bibfnamefont {A.}~\bibnamefont {Yamakage}},\ and\
  \bibinfo {author} {\bibfnamefont {M.}~\bibnamefont {Sato}},\ }\bibfield
  {title} {\bibinfo {title} {Majorana multipole response: General theory and
  application to wallpaper groups},\ }\href
  {https://doi.org/10.1103/PhysRevB.103.224504} {\bibfield  {journal} {\bibinfo
   {journal} {Phys. Rev. B}\ }\textbf {\bibinfo {volume} {103}},\ \bibinfo
  {pages} {224504} (\bibinfo {year} {2021})}\BibitemShut {NoStop}%
\bibitem [{\citenamefont {Bradley}\ and\ \citenamefont
  {Cracknell}(2003)}]{Bradley72}%
  \BibitemOpen
  \bibfield  {author} {\bibinfo {author} {\bibfnamefont {C.~J.}\ \bibnamefont
  {Bradley}}\ and\ \bibinfo {author} {\bibfnamefont {A.~P.}\ \bibnamefont
  {Cracknell}},\ }\href@noop {} {\emph {\bibinfo {title} {The Mathematical
  Theory of Symmetry in Solids}}}\ (\bibinfo  {publisher} {Oxford University
  Press},\ \bibinfo {address} {New York},\ \bibinfo {year} {2003})\BibitemShut
  {NoStop}%
\bibitem [{\citenamefont {Chung}\ and\ \citenamefont {Zhang}(2009)}]{Chung09}%
  \BibitemOpen
  \bibfield  {author} {\bibinfo {author} {\bibfnamefont {S.~B.}\ \bibnamefont
  {Chung}}\ and\ \bibinfo {author} {\bibfnamefont {S.-C.}\ \bibnamefont
  {Zhang}},\ }\bibfield  {title} {\bibinfo {title} {Detecting the {Majorana}
  fermion surface state of {$^{3}\mathrm{He}\mathrm{\text{\ensuremath{-}}}B$}
  through spin relaxation},\ }\href
  {https://doi.org/10.1103/PhysRevLett.103.235301} {\bibfield  {journal}
  {\bibinfo  {journal} {Phys. Rev. Lett.}\ }\textbf {\bibinfo {volume} {103}},\
  \bibinfo {pages} {235301} (\bibinfo {year} {2009})}\BibitemShut {NoStop}%
\bibitem [{\citenamefont {Mizushima}\ \emph {et~al.}(2012)\citenamefont
  {Mizushima}, \citenamefont {Sato},\ and\ \citenamefont
  {Machida}}]{Mizushima12}%
  \BibitemOpen
  \bibfield  {author} {\bibinfo {author} {\bibfnamefont {T.}~\bibnamefont
  {Mizushima}}, \bibinfo {author} {\bibfnamefont {M.}~\bibnamefont {Sato}},\
  and\ \bibinfo {author} {\bibfnamefont {K.}~\bibnamefont {Machida}},\
  }\bibfield  {title} {\bibinfo {title} {Symmetry protected topological order
  and spin susceptibility in superfluid
  {$^{3}\mathrm{He}\mathrm{\text{\ensuremath{-}}}B$}},\ }\href
  {https://doi.org/10.1103/PhysRevLett.109.165301} {\bibfield  {journal}
  {\bibinfo  {journal} {Phys. Rev. Lett.}\ }\textbf {\bibinfo {volume} {109}},\
  \bibinfo {pages} {165301} (\bibinfo {year} {2012})}\BibitemShut {NoStop}%
\bibitem [{\citenamefont {Qin}\ \emph {et~al.}(2022)\citenamefont {Qin},
  \citenamefont {Fang}, \citenamefont {Zhang},\ and\ \citenamefont
  {Hu}}]{Qin2022}%
  \BibitemOpen
  \bibfield  {author} {\bibinfo {author} {\bibfnamefont {S.}~\bibnamefont
  {Qin}}, \bibinfo {author} {\bibfnamefont {C.}~\bibnamefont {Fang}}, \bibinfo
  {author} {\bibfnamefont {F.-C.}\ \bibnamefont {Zhang}},\ and\ \bibinfo
  {author} {\bibfnamefont {J.}~\bibnamefont {Hu}},\ }\bibfield  {title}
  {\bibinfo {title} {Topological superconductivity in an extended $s$-wave
  superconductor and its implication to iron-based superconductors},\ }\href
  {https://doi.org/10.1103/PhysRevX.12.011030} {\bibfield  {journal} {\bibinfo
  {journal} {Phys. Rev. X}\ }\textbf {\bibinfo {volume} {12}},\ \bibinfo
  {pages} {011030} (\bibinfo {year} {2022})}\BibitemShut {NoStop}%
\bibitem [{\citenamefont {Tei}\ \emph {et~al.}(2023)\citenamefont {Tei},
  \citenamefont {Mizushima},\ and\ \citenamefont {Fujimoto}}]{Tei2023}%
  \BibitemOpen
  \bibfield  {author} {\bibinfo {author} {\bibfnamefont {J.}~\bibnamefont
  {Tei}}, \bibinfo {author} {\bibfnamefont {T.}~\bibnamefont {Mizushima}},\
  and\ \bibinfo {author} {\bibfnamefont {S.}~\bibnamefont {Fujimoto}},\
  }\bibfield  {title} {\bibinfo {title} {Possible realization of topological
  crystalline superconductivity with time-reversal symmetry in
  ${\mathrm{ute}}_{2}$},\ }\href {https://doi.org/10.1103/PhysRevB.107.144517}
  {\bibfield  {journal} {\bibinfo  {journal} {Phys. Rev. B}\ }\textbf {\bibinfo
  {volume} {107}},\ \bibinfo {pages} {144517} (\bibinfo {year}
  {2023})}\BibitemShut {NoStop}%
\bibitem [{\citenamefont {Fu}(2014)}]{Fu14}%
  \BibitemOpen
  \bibfield  {author} {\bibinfo {author} {\bibfnamefont {L.}~\bibnamefont
  {Fu}},\ }\bibfield  {title} {\bibinfo {title} {Odd-parity topological
  superconductor with nematic order: Application to
  {${\mathrm{Cu}}_{x}{\mathrm{Bi}}_{2}{\mathrm{Se}}_{3}$}},\ }\href
  {https://doi.org/10.1103/PhysRevB.90.100509} {\bibfield  {journal} {\bibinfo
  {journal} {Phys. Rev. B}\ }\textbf {\bibinfo {volume} {90}},\ \bibinfo
  {pages} {100509(R)} (\bibinfo {year} {2014})}\BibitemShut {NoStop}%
\bibitem [{\citenamefont {Brydon}\ \emph {et~al.}(2016)\citenamefont {Brydon},
  \citenamefont {Wang}, \citenamefont {Weinert},\ and\ \citenamefont
  {Agterberg}}]{Brydon16}%
  \BibitemOpen
  \bibfield  {author} {\bibinfo {author} {\bibfnamefont {P.~M.~R.}\
  \bibnamefont {Brydon}}, \bibinfo {author} {\bibfnamefont {L.}~\bibnamefont
  {Wang}}, \bibinfo {author} {\bibfnamefont {M.}~\bibnamefont {Weinert}},\ and\
  \bibinfo {author} {\bibfnamefont {D.~F.}\ \bibnamefont {Agterberg}},\
  }\bibfield  {title} {\bibinfo {title} {Pairing of $j=3/2$ fermions in
  {Half-Heusler} superconductors},\ }\href
  {https://doi.org/10.1103/PhysRevLett.116.177001} {\bibfield  {journal}
  {\bibinfo  {journal} {Phys. Rev. Lett.}\ }\textbf {\bibinfo {volume} {116}},\
  \bibinfo {pages} {177001} (\bibinfo {year} {2016})}\BibitemShut {NoStop}%
\bibitem [{\citenamefont {Shimizu}\ \emph {et~al.}(2017)\citenamefont
  {Shimizu}, \citenamefont {Kittaka}, \citenamefont {Nakamura}, \citenamefont
  {Sakakibara}, \citenamefont {Aoki}, \citenamefont {Homma}, \citenamefont
  {Nakamura},\ and\ \citenamefont {Machida}}]{Shimizu2017}%
  \BibitemOpen
  \bibfield  {author} {\bibinfo {author} {\bibfnamefont {Y.}~\bibnamefont
  {Shimizu}}, \bibinfo {author} {\bibfnamefont {S.}~\bibnamefont {Kittaka}},
  \bibinfo {author} {\bibfnamefont {S.}~\bibnamefont {Nakamura}}, \bibinfo
  {author} {\bibfnamefont {T.}~\bibnamefont {Sakakibara}}, \bibinfo {author}
  {\bibfnamefont {D.}~\bibnamefont {Aoki}}, \bibinfo {author} {\bibfnamefont
  {Y.}~\bibnamefont {Homma}}, \bibinfo {author} {\bibfnamefont
  {A.}~\bibnamefont {Nakamura}},\ and\ \bibinfo {author} {\bibfnamefont
  {K.}~\bibnamefont {Machida}},\ }\bibfield  {title} {\bibinfo {title}
  {Quasiparticle excitations and evidence for superconducting double
  transitions in monocrystalline
  {${\mathrm{U}}_{0.97}{\mathrm{Th}}_{0.03}{\mathrm{Be}}_{13}$}},\ }\href
  {https://doi.org/10.1103/PhysRevB.96.100505} {\bibfield  {journal} {\bibinfo
  {journal} {Phys. Rev. B}\ }\textbf {\bibinfo {volume} {96}},\ \bibinfo
  {pages} {100505} (\bibinfo {year} {2017})}\BibitemShut {NoStop}%
\bibitem [{\citenamefont {Machida}(2018)}]{Machida2018}%
  \BibitemOpen
  \bibfield  {author} {\bibinfo {author} {\bibfnamefont {K.}~\bibnamefont
  {Machida}},\ }\bibfield  {title} {\bibinfo {title} {Spin triplet nematic
  pairing symmetry and superconducting double transition in
  {${\mathrm{U}}_{1\mathrm{\ensuremath{-}}\mathrm{x}}$${\mathrm{Th}}_{\mathrm{x}}$${\mathrm{Be}}_{13}$}},\
  }\href {https://doi.org/10.7566/JPSJ.87.033703} {\bibfield  {journal}
  {\bibinfo  {journal} {Journal of the Physical Society of Japan}\ }\textbf
  {\bibinfo {volume} {87}},\ \bibinfo {pages} {033703} (\bibinfo {year}
  {2018})}\BibitemShut {NoStop}%
\bibitem [{\citenamefont {Mizushima}\ and\ \citenamefont
  {Nitta}(2018)}]{Mizushima2018}%
  \BibitemOpen
  \bibfield  {author} {\bibinfo {author} {\bibfnamefont {T.}~\bibnamefont
  {Mizushima}}\ and\ \bibinfo {author} {\bibfnamefont {M.}~\bibnamefont
  {Nitta}},\ }\bibfield  {title} {\bibinfo {title} {Topology and symmetry of
  surface majorana arcs in cyclic superconductors},\ }\href
  {https://doi.org/10.1103/PhysRevB.97.024506} {\bibfield  {journal} {\bibinfo
  {journal} {Phys. Rev. B}\ }\textbf {\bibinfo {volume} {97}},\ \bibinfo
  {pages} {024506} (\bibinfo {year} {2018})}\BibitemShut {NoStop}%
\bibitem [{\citenamefont {Sigrist}\ and\ \citenamefont
  {Rice}(1989)}]{Sigrist1989}%
  \BibitemOpen
  \bibfield  {author} {\bibinfo {author} {\bibfnamefont {M.}~\bibnamefont
  {Sigrist}}\ and\ \bibinfo {author} {\bibfnamefont {T.~M.}\ \bibnamefont
  {Rice}},\ }\bibfield  {title} {\bibinfo {title} {Phenomenological theory of
  the superconductivity phase diagram of
  {${\mathrm{U}}_{1\mathrm{\ensuremath{-}}\mathrm{x}}$${\mathrm{Th}}_{\mathrm{x}}$${\mathrm{Be}}_{13}$}},\
  }\href {https://doi.org/10.1103/PhysRevB.39.2200} {\bibfield  {journal}
  {\bibinfo  {journal} {Phys. Rev. B}\ }\textbf {\bibinfo {volume} {39}},\
  \bibinfo {pages} {2200} (\bibinfo {year} {1989})}\BibitemShut {NoStop}%
\bibitem [{\citenamefont {Yamazaki}\ \emph
  {et~al.}(2021{\natexlab{a}})\citenamefont {Yamazaki}, \citenamefont
  {Kobayashi},\ and\ \citenamefont {Yamakage}}]{Yamazaki2021prb}%
  \BibitemOpen
  \bibfield  {author} {\bibinfo {author} {\bibfnamefont {Y.}~\bibnamefont
  {Yamazaki}}, \bibinfo {author} {\bibfnamefont {S.}~\bibnamefont
  {Kobayashi}},\ and\ \bibinfo {author} {\bibfnamefont {A.}~\bibnamefont
  {Yamakage}},\ }\bibfield  {title} {\bibinfo {title} {Magnetic response of
  {Majorana} kramers pairs with an order-two symmetry},\ }\href
  {https://doi.org/10.1103/PhysRevB.103.094508} {\bibfield  {journal} {\bibinfo
   {journal} {Phys. Rev. B}\ }\textbf {\bibinfo {volume} {103}},\ \bibinfo
  {pages} {094508} (\bibinfo {year} {2021}{\natexlab{a}})}\BibitemShut
  {NoStop}%
\bibitem [{\citenamefont {Shiozaki}\ and\ \citenamefont
  {Sato}(2014)}]{Shiozaki14}%
  \BibitemOpen
  \bibfield  {author} {\bibinfo {author} {\bibfnamefont {K.}~\bibnamefont
  {Shiozaki}}\ and\ \bibinfo {author} {\bibfnamefont {M.}~\bibnamefont
  {Sato}},\ }\bibfield  {title} {\bibinfo {title} {Topology of crystalline
  insulators and superconductors},\ }\href
  {https://doi.org/10.1103/PhysRevB.90.165114} {\bibfield  {journal} {\bibinfo
  {journal} {Phys. Rev. B}\ }\textbf {\bibinfo {volume} {90}},\ \bibinfo
  {pages} {165114} (\bibinfo {year} {2014})}\BibitemShut {NoStop}%
\bibitem [{\citenamefont {Xiong}\ \emph {et~al.}(2017)\citenamefont {Xiong},
  \citenamefont {Yamakage}, \citenamefont {Kobayashi}, \citenamefont {Sato},\
  and\ \citenamefont {Tanaka}}]{Xiong2017}%
  \BibitemOpen
  \bibfield  {author} {\bibinfo {author} {\bibfnamefont {Y.}~\bibnamefont
  {Xiong}}, \bibinfo {author} {\bibfnamefont {A.}~\bibnamefont {Yamakage}},
  \bibinfo {author} {\bibfnamefont {S.}~\bibnamefont {Kobayashi}}, \bibinfo
  {author} {\bibfnamefont {M.}~\bibnamefont {Sato}},\ and\ \bibinfo {author}
  {\bibfnamefont {Y.}~\bibnamefont {Tanaka}},\ }\bibfield  {title} {\bibinfo
  {title} {Anisotropic magnetic responses of topological crystalline
  superconductors},\ }\bibfield  {journal} {\bibinfo  {journal} {Crystals}\
  }\textbf {\bibinfo {volume} {7}},\ \href
  {https://doi.org/10.3390/cryst7020058} {10.3390/cryst7020058} (\bibinfo
  {year} {2017})\BibitemShut {NoStop}%
\bibitem [{\citenamefont {Hor}\ \emph {et~al.}(2010)\citenamefont {Hor},
  \citenamefont {Williams}, \citenamefont {Checkelsky}, \citenamefont
  {Roushan}, \citenamefont {Seo}, \citenamefont {Xu}, \citenamefont
  {Zandbergen}, \citenamefont {Yazdani}, \citenamefont {Ong},\ and\
  \citenamefont {Cava}}]{Hor10}%
  \BibitemOpen
  \bibfield  {author} {\bibinfo {author} {\bibfnamefont {Y.~S.}\ \bibnamefont
  {Hor}}, \bibinfo {author} {\bibfnamefont {A.~J.}\ \bibnamefont {Williams}},
  \bibinfo {author} {\bibfnamefont {J.~G.}\ \bibnamefont {Checkelsky}},
  \bibinfo {author} {\bibfnamefont {P.}~\bibnamefont {Roushan}}, \bibinfo
  {author} {\bibfnamefont {J.}~\bibnamefont {Seo}}, \bibinfo {author}
  {\bibfnamefont {Q.}~\bibnamefont {Xu}}, \bibinfo {author} {\bibfnamefont
  {H.~W.}\ \bibnamefont {Zandbergen}}, \bibinfo {author} {\bibfnamefont
  {A.}~\bibnamefont {Yazdani}}, \bibinfo {author} {\bibfnamefont {N.~P.}\
  \bibnamefont {Ong}},\ and\ \bibinfo {author} {\bibfnamefont {R.~J.}\
  \bibnamefont {Cava}},\ }\bibfield  {title} {\bibinfo {title}
  {Superconductivity in {${\mathrm{Cu}}_{x}{\mathrm{Bi}}_{2}{\mathrm{Se}}_{3}$}
  and its implications for pairing in the undoped topological insulator},\
  }\href {https://doi.org/10.1103/PhysRevLett.104.057001} {\bibfield  {journal}
  {\bibinfo  {journal} {Phys. Rev. Lett.}\ }\textbf {\bibinfo {volume} {104}},\
  \bibinfo {pages} {057001} (\bibinfo {year} {2010})}\BibitemShut {NoStop}%
\bibitem [{\citenamefont {Sasaki}\ \emph {et~al.}(2011)\citenamefont {Sasaki},
  \citenamefont {Kriener}, \citenamefont {Segawa}, \citenamefont {Yada},
  \citenamefont {Tanaka}, \citenamefont {Sato},\ and\ \citenamefont
  {Ando}}]{Sasaki11}%
  \BibitemOpen
  \bibfield  {author} {\bibinfo {author} {\bibfnamefont {S.}~\bibnamefont
  {Sasaki}}, \bibinfo {author} {\bibfnamefont {M.}~\bibnamefont {Kriener}},
  \bibinfo {author} {\bibfnamefont {K.}~\bibnamefont {Segawa}}, \bibinfo
  {author} {\bibfnamefont {K.}~\bibnamefont {Yada}}, \bibinfo {author}
  {\bibfnamefont {Y.}~\bibnamefont {Tanaka}}, \bibinfo {author} {\bibfnamefont
  {M.}~\bibnamefont {Sato}},\ and\ \bibinfo {author} {\bibfnamefont
  {Y.}~\bibnamefont {Ando}},\ }\bibfield  {title} {\bibinfo {title}
  {Topological superconductivity in
  {${\mathrm{Cu}}_{x}{\mathrm{Bi}}_{2}{\mathrm{Se}}_{3}$}},\ }\href
  {https://doi.org/10.1103/PhysRevLett.107.217001} {\bibfield  {journal}
  {\bibinfo  {journal} {Phys. Rev. Lett.}\ }\textbf {\bibinfo {volume} {107}},\
  \bibinfo {pages} {217001} (\bibinfo {year} {2011})}\BibitemShut {NoStop}%
\bibitem [{\citenamefont {Matano}\ \emph {et~al.}(2016)\citenamefont {Matano},
  \citenamefont {Kriener}, \citenamefont {Segawa}, \citenamefont {Ando},\ and\
  \citenamefont {Zheng}}]{Matano16}%
  \BibitemOpen
  \bibfield  {author} {\bibinfo {author} {\bibfnamefont {K.}~\bibnamefont
  {Matano}}, \bibinfo {author} {\bibfnamefont {M.}~\bibnamefont {Kriener}},
  \bibinfo {author} {\bibfnamefont {K.}~\bibnamefont {Segawa}}, \bibinfo
  {author} {\bibfnamefont {Y.}~\bibnamefont {Ando}},\ and\ \bibinfo {author}
  {\bibfnamefont {G.-q.}\ \bibnamefont {Zheng}},\ }\bibfield  {title} {\bibinfo
  {title} {Spin-rotation symmetry breaking in the superconducting state of
  {${\mathrm{Cu}}_{x}{\mathrm{Bi}}_{2}{\mathrm{Se}}_{3}$}},\ }\href
  {https://doi.org/http://dx.doi.org/10.1038/nphys3781} {\bibfield  {journal}
  {\bibinfo  {journal} {Nature Physics}\ }\textbf {\bibinfo {volume} {12}},\
  \bibinfo {pages} {852} (\bibinfo {year} {2016})}\BibitemShut {NoStop}%
\bibitem [{\citenamefont {Yonezawa}\ \emph {et~al.}(2017)\citenamefont
  {Yonezawa}, \citenamefont {Tajiri}, \citenamefont {Nakata}, \citenamefont
  {Nagai}, \citenamefont {Wang}, \citenamefont {Segawa}, \citenamefont {Ando},\
  and\ \citenamefont {Maeno}}]{Yonezawa17}%
  \BibitemOpen
  \bibfield  {author} {\bibinfo {author} {\bibfnamefont {S.}~\bibnamefont
  {Yonezawa}}, \bibinfo {author} {\bibfnamefont {K.}~\bibnamefont {Tajiri}},
  \bibinfo {author} {\bibfnamefont {S.}~\bibnamefont {Nakata}}, \bibinfo
  {author} {\bibfnamefont {Y.}~\bibnamefont {Nagai}}, \bibinfo {author}
  {\bibfnamefont {Z.}~\bibnamefont {Wang}}, \bibinfo {author} {\bibfnamefont
  {K.}~\bibnamefont {Segawa}}, \bibinfo {author} {\bibfnamefont
  {Y.}~\bibnamefont {Ando}},\ and\ \bibinfo {author} {\bibfnamefont
  {Y.}~\bibnamefont {Maeno}},\ }\bibfield  {title} {\bibinfo {title}
  {Thermodynamic evidence for nematic superconductivity in
  {${\mathrm{Cu}}_{x}{\mathrm{Bi}}_{2}{\mathrm{Se}}_{3}$}},\ }\href
  {https://doi.org/http://dx.doi.org/10.1038/nphys3907} {\bibfield  {journal}
  {\bibinfo  {journal} {Nature Physics}\ }\textbf {\bibinfo {volume} {13}},\
  \bibinfo {pages} {123} (\bibinfo {year} {2017})}\BibitemShut {NoStop}%
\bibitem [{\citenamefont {Goll}\ \emph {et~al.}(2008)\citenamefont {Goll},
  \citenamefont {Marz}, \citenamefont {Hamann}, \citenamefont {Tomanic},
  \citenamefont {Grube}, \citenamefont {Yoshino},\ and\ \citenamefont
  {Takabatake}}]{Goll08}%
  \BibitemOpen
  \bibfield  {author} {\bibinfo {author} {\bibfnamefont {G.}~\bibnamefont
  {Goll}}, \bibinfo {author} {\bibfnamefont {M.}~\bibnamefont {Marz}}, \bibinfo
  {author} {\bibfnamefont {A.}~\bibnamefont {Hamann}}, \bibinfo {author}
  {\bibfnamefont {T.}~\bibnamefont {Tomanic}}, \bibinfo {author} {\bibfnamefont
  {K.}~\bibnamefont {Grube}}, \bibinfo {author} {\bibfnamefont
  {T.}~\bibnamefont {Yoshino}},\ and\ \bibinfo {author} {\bibfnamefont
  {T.}~\bibnamefont {Takabatake}},\ }\bibfield  {title} {\bibinfo {title}
  {Thermodynamic and transport properties of the non-centrosymmetric
  superconductor {LaBiPt}},\ }\href
  {https://doi.org/https://doi.org/10.1016/j.physb.2007.10.089} {\bibfield
  {journal} {\bibinfo  {journal} {Physica B: Condensed Matter}\ }\textbf
  {\bibinfo {volume} {403}},\ \bibinfo {pages} {1065 } (\bibinfo {year}
  {2008})}\BibitemShut {NoStop}%
\bibitem [{\citenamefont {Butch}\ \emph {et~al.}(2011)\citenamefont {Butch},
  \citenamefont {Syers}, \citenamefont {Kirshenbaum}, \citenamefont {Hope},\
  and\ \citenamefont {Paglione}}]{Butch11}%
  \BibitemOpen
  \bibfield  {author} {\bibinfo {author} {\bibfnamefont {N.~P.}\ \bibnamefont
  {Butch}}, \bibinfo {author} {\bibfnamefont {P.}~\bibnamefont {Syers}},
  \bibinfo {author} {\bibfnamefont {K.}~\bibnamefont {Kirshenbaum}}, \bibinfo
  {author} {\bibfnamefont {A.~P.}\ \bibnamefont {Hope}},\ and\ \bibinfo
  {author} {\bibfnamefont {J.}~\bibnamefont {Paglione}},\ }\bibfield  {title}
  {\bibinfo {title} {Superconductivity in the topological semimetal {YPtBi}},\
  }\href {https://doi.org/10.1103/PhysRevB.84.220504} {\bibfield  {journal}
  {\bibinfo  {journal} {Phys. Rev. B}\ }\textbf {\bibinfo {volume} {84}},\
  \bibinfo {pages} {220504(R)} (\bibinfo {year} {2011})}\BibitemShut {NoStop}%
\bibitem [{\citenamefont {Tafti}\ \emph {et~al.}(2013)\citenamefont {Tafti},
  \citenamefont {Fujii}, \citenamefont {Juneau-Fecteau}, \citenamefont
  {Ren\'e~de Cotret}, \citenamefont {Doiron-Leyraud}, \citenamefont
  {Asamitsu},\ and\ \citenamefont {Taillefer}}]{Tafti13}%
  \BibitemOpen
  \bibfield  {author} {\bibinfo {author} {\bibfnamefont {F.~F.}\ \bibnamefont
  {Tafti}}, \bibinfo {author} {\bibfnamefont {T.}~\bibnamefont {Fujii}},
  \bibinfo {author} {\bibfnamefont {A.}~\bibnamefont {Juneau-Fecteau}},
  \bibinfo {author} {\bibfnamefont {S.}~\bibnamefont {Ren\'e~de Cotret}},
  \bibinfo {author} {\bibfnamefont {N.}~\bibnamefont {Doiron-Leyraud}},
  \bibinfo {author} {\bibfnamefont {A.}~\bibnamefont {Asamitsu}},\ and\
  \bibinfo {author} {\bibfnamefont {L.}~\bibnamefont {Taillefer}},\ }\bibfield
  {title} {\bibinfo {title} {Superconductivity in the noncentrosymmetric
  {half-Heusler} compound {LuPtBi}: A candidate for topological
  superconductivity},\ }\href {https://doi.org/10.1103/PhysRevB.87.184504}
  {\bibfield  {journal} {\bibinfo  {journal} {Phys. Rev. B}\ }\textbf {\bibinfo
  {volume} {87}},\ \bibinfo {pages} {184504} (\bibinfo {year}
  {2013})}\BibitemShut {NoStop}%
\bibitem [{\citenamefont {Xu}\ \emph {et~al.}(2014)\citenamefont {Xu},
  \citenamefont {Wang}, \citenamefont {Zhang}, \citenamefont {Du},
  \citenamefont {Liu}, \citenamefont {Wang}, \citenamefont {Wu}, \citenamefont
  {Liu},\ and\ \citenamefont {Zhang}}]{GXu16}%
  \BibitemOpen
  \bibfield  {author} {\bibinfo {author} {\bibfnamefont {G.}~\bibnamefont
  {Xu}}, \bibinfo {author} {\bibfnamefont {W.}~\bibnamefont {Wang}}, \bibinfo
  {author} {\bibfnamefont {X.}~\bibnamefont {Zhang}}, \bibinfo {author}
  {\bibfnamefont {Y.}~\bibnamefont {Du}}, \bibinfo {author} {\bibfnamefont
  {E.}~\bibnamefont {Liu}}, \bibinfo {author} {\bibfnamefont {S.}~\bibnamefont
  {Wang}}, \bibinfo {author} {\bibfnamefont {G.}~\bibnamefont {Wu}}, \bibinfo
  {author} {\bibfnamefont {Z.}~\bibnamefont {Liu}},\ and\ \bibinfo {author}
  {\bibfnamefont {X.~X.}\ \bibnamefont {Zhang}},\ }\bibfield  {title} {\bibinfo
  {title} {Weak antilocalization effect and noncentrosymmetric
  superconductivity in a topologically nontrivial semimetal {LuPdBi}},\ }\href
  {https://doi.org/http://dx.doi.org/10.1038/srep05709} {\bibfield  {journal}
  {\bibinfo  {journal} {Scientific Reports}\ }\textbf {\bibinfo {volume} {4}},\
  \bibinfo {pages} {5709} (\bibinfo {year} {2014})}\BibitemShut {NoStop}%
\bibitem [{\citenamefont {Bay}\ \emph {et~al.}(2012)\citenamefont {Bay},
  \citenamefont {Naka}, \citenamefont {Huang},\ and\ \citenamefont
  {de~Visser}}]{Bay12}%
  \BibitemOpen
  \bibfield  {author} {\bibinfo {author} {\bibfnamefont {T.~V.}\ \bibnamefont
  {Bay}}, \bibinfo {author} {\bibfnamefont {T.}~\bibnamefont {Naka}}, \bibinfo
  {author} {\bibfnamefont {Y.~K.}\ \bibnamefont {Huang}},\ and\ \bibinfo
  {author} {\bibfnamefont {A.}~\bibnamefont {de~Visser}},\ }\bibfield  {title}
  {\bibinfo {title} {Superconductivity in noncentrosymmetric {YPtBi} under
  pressure},\ }\href {https://doi.org/10.1103/PhysRevB.86.064515} {\bibfield
  {journal} {\bibinfo  {journal} {Phys. Rev. B}\ }\textbf {\bibinfo {volume}
  {86}},\ \bibinfo {pages} {064515} (\bibinfo {year} {2012})}\BibitemShut
  {NoStop}%
\bibitem [{\citenamefont {Kim}\ \emph {et~al.}(2018)\citenamefont {Kim},
  \citenamefont {Wang}, \citenamefont {Nakajima}, \citenamefont {Hu},
  \citenamefont {Ziemak}, \citenamefont {Syers}, \citenamefont {Wang},
  \citenamefont {Hodovanets}, \citenamefont {Denlinger}, \citenamefont
  {Brydon}, \citenamefont {Agterberg}, \citenamefont {Tanatar}, \citenamefont
  {Prozorov},\ and\ \citenamefont {Paglione}}]{Kim18}%
  \BibitemOpen
  \bibfield  {author} {\bibinfo {author} {\bibfnamefont {H.}~\bibnamefont
  {Kim}}, \bibinfo {author} {\bibfnamefont {K.}~\bibnamefont {Wang}}, \bibinfo
  {author} {\bibfnamefont {Y.}~\bibnamefont {Nakajima}}, \bibinfo {author}
  {\bibfnamefont {R.}~\bibnamefont {Hu}}, \bibinfo {author} {\bibfnamefont
  {S.}~\bibnamefont {Ziemak}}, \bibinfo {author} {\bibfnamefont
  {P.}~\bibnamefont {Syers}}, \bibinfo {author} {\bibfnamefont
  {L.}~\bibnamefont {Wang}}, \bibinfo {author} {\bibfnamefont {H.}~\bibnamefont
  {Hodovanets}}, \bibinfo {author} {\bibfnamefont {J.~D.}\ \bibnamefont
  {Denlinger}}, \bibinfo {author} {\bibfnamefont {P.~M.~R.}\ \bibnamefont
  {Brydon}}, \bibinfo {author} {\bibfnamefont {D.~F.}\ \bibnamefont
  {Agterberg}}, \bibinfo {author} {\bibfnamefont {M.~A.}\ \bibnamefont
  {Tanatar}}, \bibinfo {author} {\bibfnamefont {R.}~\bibnamefont {Prozorov}},\
  and\ \bibinfo {author} {\bibfnamefont {J.}~\bibnamefont {Paglione}},\
  }\bibfield  {title} {\bibinfo {title} {Beyond triplet: Unconventional
  superconductivity in a spin-3/2 topological semimetal},\ }\href
  {https://doi.org/10.1126/sciadv.aao4513} {\bibfield  {journal} {\bibinfo
  {journal} {Science Advances}\ }\textbf {\bibinfo {volume} {4}},\ \bibinfo
  {pages} {eaao4513} (\bibinfo {year} {2018})}\BibitemShut {NoStop}%
\bibitem [{\citenamefont {Yamazaki}\ \emph {et~al.}(2020)\citenamefont
  {Yamazaki}, \citenamefont {Kobayashi},\ and\ \citenamefont
  {Yamakage}}]{Yamazaki2019}%
  \BibitemOpen
  \bibfield  {author} {\bibinfo {author} {\bibfnamefont {Y.}~\bibnamefont
  {Yamazaki}}, \bibinfo {author} {\bibfnamefont {S.}~\bibnamefont
  {Kobayashi}},\ and\ \bibinfo {author} {\bibfnamefont {A.}~\bibnamefont
  {Yamakage}},\ }\bibfield  {title} {\bibinfo {title} {Magnetic response of
  {Majorana} kramers pairs protected by $\mathbb{Z}_{2}$ invariants},\ }\href
  {https://doi.org/10.7566/JPSJ.89.043703} {\bibfield  {journal} {\bibinfo
  {journal} {J. Phys. Soc. Jpn.}\ }\textbf {\bibinfo {volume} {89}},\ \bibinfo
  {pages} {043703} (\bibinfo {year} {2020})}\BibitemShut {NoStop}%
\bibitem [{\citenamefont {Yamazaki}\ \emph
  {et~al.}(2021{\natexlab{b}})\citenamefont {Yamazaki}, \citenamefont
  {Kobayashi},\ and\ \citenamefont {Yamakage}}]{Yamazaki2021jpsj}%
  \BibitemOpen
  \bibfield  {author} {\bibinfo {author} {\bibfnamefont {Y.}~\bibnamefont
  {Yamazaki}}, \bibinfo {author} {\bibfnamefont {S.}~\bibnamefont
  {Kobayashi}},\ and\ \bibinfo {author} {\bibfnamefont {A.}~\bibnamefont
  {Yamakage}},\ }\bibfield  {title} {\bibinfo {title} {Electric multipoles of
  double majorana kramers pairs},\ }\href
  {https://doi.org/10.7566/JPSJ.90.073701} {\bibfield  {journal} {\bibinfo
  {journal} {J. Phys. Soc. Jpn.}\ }\textbf {\bibinfo {volume} {90}},\ \bibinfo
  {pages} {073701} (\bibinfo {year} {2021}{\natexlab{b}})}\BibitemShut
  {NoStop}%
\bibitem [{\citenamefont {Daido}\ \emph {et~al.}(2019)\citenamefont {Daido},
  \citenamefont {Yoshida},\ and\ \citenamefont {Yanase}}]{Daido2019}%
  \BibitemOpen
  \bibfield  {author} {\bibinfo {author} {\bibfnamefont {A.}~\bibnamefont
  {Daido}}, \bibinfo {author} {\bibfnamefont {T.}~\bibnamefont {Yoshida}},\
  and\ \bibinfo {author} {\bibfnamefont {Y.}~\bibnamefont {Yanase}},\
  }\bibfield  {title} {\bibinfo {title} {$\mathbb{Z}_{4}$ topological
  superconductivity in {UCoGe}},\ }\href
  {https://doi.org/10.1103/PhysRevLett.122.227001} {\bibfield  {journal}
  {\bibinfo  {journal} {Phys. Rev. Lett.}\ }\textbf {\bibinfo {volume} {122}},\
  \bibinfo {pages} {227001} (\bibinfo {year} {2019})}\BibitemShut {NoStop}%
\bibitem [{\citenamefont {Kawakami}\ and\ \citenamefont
  {Sato}(2019)}]{Kawakam2019}%
  \BibitemOpen
  \bibfield  {author} {\bibinfo {author} {\bibfnamefont {T.}~\bibnamefont
  {Kawakami}}\ and\ \bibinfo {author} {\bibfnamefont {M.}~\bibnamefont
  {Sato}},\ }\bibfield  {title} {\bibinfo {title} {Topological crystalline
  superconductivity in {Dirac} semimetal phase of iron-based superconductors},\
  }\href {https://doi.org/10.1103/PhysRevB.100.094520} {\bibfield  {journal}
  {\bibinfo  {journal} {Phys. Rev. B}\ }\textbf {\bibinfo {volume} {100}},\
  \bibinfo {pages} {094520} (\bibinfo {year} {2019})}\BibitemShut {NoStop}%
\bibitem [{\citenamefont {Langbehn}\ \emph {et~al.}(2017)\citenamefont
  {Langbehn}, \citenamefont {Peng}, \citenamefont {Trifunovic}, \citenamefont
  {von Oppen},\ and\ \citenamefont {Brouwer}}]{Langbehn2017}%
  \BibitemOpen
  \bibfield  {author} {\bibinfo {author} {\bibfnamefont {J.}~\bibnamefont
  {Langbehn}}, \bibinfo {author} {\bibfnamefont {Y.}~\bibnamefont {Peng}},
  \bibinfo {author} {\bibfnamefont {L.}~\bibnamefont {Trifunovic}}, \bibinfo
  {author} {\bibfnamefont {F.}~\bibnamefont {von Oppen}},\ and\ \bibinfo
  {author} {\bibfnamefont {P.~W.}\ \bibnamefont {Brouwer}},\ }\bibfield
  {title} {\bibinfo {title} {Reflection-symmetric second-order topological
  insulators and superconductors},\ }\href
  {https://doi.org/10.1103/PhysRevLett.119.246401} {\bibfield  {journal}
  {\bibinfo  {journal} {Phys. Rev. Lett.}\ }\textbf {\bibinfo {volume} {119}},\
  \bibinfo {pages} {246401} (\bibinfo {year} {2017})}\BibitemShut {NoStop}%
\bibitem [{\citenamefont {Geier}\ \emph {et~al.}(2018)\citenamefont {Geier},
  \citenamefont {Trifunovic}, \citenamefont {Hoskam},\ and\ \citenamefont
  {Brouwer}}]{Geier2018}%
  \BibitemOpen
  \bibfield  {author} {\bibinfo {author} {\bibfnamefont {M.}~\bibnamefont
  {Geier}}, \bibinfo {author} {\bibfnamefont {L.}~\bibnamefont {Trifunovic}},
  \bibinfo {author} {\bibfnamefont {M.}~\bibnamefont {Hoskam}},\ and\ \bibinfo
  {author} {\bibfnamefont {P.~W.}\ \bibnamefont {Brouwer}},\ }\bibfield
  {title} {\bibinfo {title} {Second-order topological insulators and
  superconductors with an order-two crystalline symmetry},\ }\href
  {https://doi.org/10.1103/PhysRevB.97.205135} {\bibfield  {journal} {\bibinfo
  {journal} {Phys. Rev. B}\ }\textbf {\bibinfo {volume} {97}},\ \bibinfo
  {pages} {205135} (\bibinfo {year} {2018})}\BibitemShut {NoStop}%
\bibitem [{\citenamefont {Volpez}\ \emph {et~al.}(2019)\citenamefont {Volpez},
  \citenamefont {Loss},\ and\ \citenamefont {Klinovaja}}]{Volpez19}%
  \BibitemOpen
  \bibfield  {author} {\bibinfo {author} {\bibfnamefont {Y.}~\bibnamefont
  {Volpez}}, \bibinfo {author} {\bibfnamefont {D.}~\bibnamefont {Loss}},\ and\
  \bibinfo {author} {\bibfnamefont {J.}~\bibnamefont {Klinovaja}},\ }\bibfield
  {title} {\bibinfo {title} {Second-order topological superconductivity in
  $\ensuremath{\pi}$-junction {Rashba} layers},\ }\href
  {https://doi.org/10.1103/PhysRevLett.122.126402} {\bibfield  {journal}
  {\bibinfo  {journal} {Phys. Rev. Lett.}\ }\textbf {\bibinfo {volume} {122}},\
  \bibinfo {pages} {126402} (\bibinfo {year} {2019})}\BibitemShut {NoStop}%
\bibitem [{\citenamefont {Wang}\ \emph {et~al.}(2012)\citenamefont {Wang},
  \citenamefont {Li}, \citenamefont {Zhang}, \citenamefont {Zhang},
  \citenamefont {Zhang}, \citenamefont {Li}, \citenamefont {Ding},
  \citenamefont {Ou}, \citenamefont {Deng}, \citenamefont {Chang} \emph
  {et~al.}}]{Wang2012interface}%
  \BibitemOpen
  \bibfield  {author} {\bibinfo {author} {\bibfnamefont {Q.-Y.}\ \bibnamefont
  {Wang}}, \bibinfo {author} {\bibfnamefont {Z.}~\bibnamefont {Li}}, \bibinfo
  {author} {\bibfnamefont {W.-H.}\ \bibnamefont {Zhang}}, \bibinfo {author}
  {\bibfnamefont {Z.-C.}\ \bibnamefont {Zhang}}, \bibinfo {author}
  {\bibfnamefont {J.-S.}\ \bibnamefont {Zhang}}, \bibinfo {author}
  {\bibfnamefont {W.}~\bibnamefont {Li}}, \bibinfo {author} {\bibfnamefont
  {H.}~\bibnamefont {Ding}}, \bibinfo {author} {\bibfnamefont {Y.-B.}\
  \bibnamefont {Ou}}, \bibinfo {author} {\bibfnamefont {P.}~\bibnamefont
  {Deng}}, \bibinfo {author} {\bibfnamefont {K.}~\bibnamefont {Chang}}, \emph
  {et~al.},\ }\bibfield  {title} {\bibinfo {title} {Interface-induced
  high-temperature superconductivity in single unit-cell fese films on
  {SrTiO$_3$}},\ }\href {https://doi.org/10.1088/0256-307X/29/3/037402}
  {\bibfield  {journal} {\bibinfo  {journal} {Chinese Physics Letters}\
  }\textbf {\bibinfo {volume} {29}},\ \bibinfo {pages} {037402} (\bibinfo
  {year} {2012})}\BibitemShut {NoStop}%
\bibitem [{\citenamefont {Wang}\ \emph {et~al.}(2017)\citenamefont {Wang},
  \citenamefont {Liu}, \citenamefont {Liu},\ and\ \citenamefont
  {Wang}}]{Wang2017high}%
  \BibitemOpen
  \bibfield  {author} {\bibinfo {author} {\bibfnamefont {Z.}~\bibnamefont
  {Wang}}, \bibinfo {author} {\bibfnamefont {C.}~\bibnamefont {Liu}}, \bibinfo
  {author} {\bibfnamefont {Y.}~\bibnamefont {Liu}},\ and\ \bibinfo {author}
  {\bibfnamefont {J.}~\bibnamefont {Wang}},\ }\bibfield  {title} {\bibinfo
  {title} {High-temperature superconductivity in one-unit-cell fese films},\
  }\href {https://doi.org/10.1088/1361-648X/aa5f26} {\bibfield  {journal}
  {\bibinfo  {journal} {Journal of Physics: Condensed Matter}\ }\textbf
  {\bibinfo {volume} {29}},\ \bibinfo {pages} {153001} (\bibinfo {year}
  {2017})}\BibitemShut {NoStop}%
\bibitem [{Note3()}]{Note3}%
  \BibitemOpen
  \bibinfo {note} {Here the Seitz symbol $\{p|\protect \bm {a}_p\}$ represents
  a space group operation consisting of a point-group operation $p$ and a
  translation $\protect \bm {a}_p$.}\BibitemShut {Stop}%
\bibitem [{\citenamefont {Sato}\ and\ \citenamefont
  {Fujimoto}(2016)}]{SatoFujimoto16}%
  \BibitemOpen
  \bibfield  {author} {\bibinfo {author} {\bibfnamefont {M.}~\bibnamefont
  {Sato}}\ and\ \bibinfo {author} {\bibfnamefont {S.}~\bibnamefont
  {Fujimoto}},\ }\bibfield  {title} {\bibinfo {title} {Majorana fermions and
  topology in superconductors},\ }\href
  {https://doi.org/10.7566/JPSJ.85.072001} {\bibfield  {journal} {\bibinfo
  {journal} {J. Phys. Soc. Jpn.}\ }\textbf {\bibinfo {volume} {85}},\ \bibinfo
  {pages} {072001} (\bibinfo {year} {2016})}\BibitemShut {NoStop}%
\bibitem [{\citenamefont {Jeon}\ \emph {et~al.}(2017)\citenamefont {Jeon},
  \citenamefont {Xie}, \citenamefont {Li}, \citenamefont {Wang}, \citenamefont
  {Bernevig},\ and\ \citenamefont {Yazdani}}]{Jeon17}%
  \BibitemOpen
  \bibfield  {author} {\bibinfo {author} {\bibfnamefont {S.}~\bibnamefont
  {Jeon}}, \bibinfo {author} {\bibfnamefont {Y.}~\bibnamefont {Xie}}, \bibinfo
  {author} {\bibfnamefont {J.}~\bibnamefont {Li}}, \bibinfo {author}
  {\bibfnamefont {Z.}~\bibnamefont {Wang}}, \bibinfo {author} {\bibfnamefont
  {B.~A.}\ \bibnamefont {Bernevig}},\ and\ \bibinfo {author} {\bibfnamefont
  {A.}~\bibnamefont {Yazdani}},\ }\bibfield  {title} {\bibinfo {title}
  {Distinguishing a {Majorana} zero mode using spin-resolved measurements},\
  }\href {https://doi.org/10.1126/science.aan3670} {\bibfield  {journal}
  {\bibinfo  {journal} {Science}\ }\textbf {\bibinfo {volume} {358}},\ \bibinfo
  {pages} {772} (\bibinfo {year} {2017})}\BibitemShut {NoStop}%
\bibitem [{\citenamefont {Cornils}\ \emph {et~al.}(2017)\citenamefont
  {Cornils}, \citenamefont {Kamlapure}, \citenamefont {Zhou}, \citenamefont
  {Pradhan}, \citenamefont {Khajetoorians}, \citenamefont {Fransson},
  \citenamefont {Wiebe},\ and\ \citenamefont {Wiesendanger}}]{Cornils17}%
  \BibitemOpen
  \bibfield  {author} {\bibinfo {author} {\bibfnamefont {L.}~\bibnamefont
  {Cornils}}, \bibinfo {author} {\bibfnamefont {A.}~\bibnamefont {Kamlapure}},
  \bibinfo {author} {\bibfnamefont {L.}~\bibnamefont {Zhou}}, \bibinfo {author}
  {\bibfnamefont {S.}~\bibnamefont {Pradhan}}, \bibinfo {author} {\bibfnamefont
  {A.~A.}\ \bibnamefont {Khajetoorians}}, \bibinfo {author} {\bibfnamefont
  {J.}~\bibnamefont {Fransson}}, \bibinfo {author} {\bibfnamefont
  {J.}~\bibnamefont {Wiebe}},\ and\ \bibinfo {author} {\bibfnamefont
  {R.}~\bibnamefont {Wiesendanger}},\ }\bibfield  {title} {\bibinfo {title}
  {Spin-resolved spectroscopy of the {Yu-Shiba-Rusinov} states of individual
  atoms},\ }\href {https://doi.org/10.1103/PhysRevLett.119.197002} {\bibfield
  {journal} {\bibinfo  {journal} {Phys. Rev. Lett.}\ }\textbf {\bibinfo
  {volume} {119}},\ \bibinfo {pages} {197002} (\bibinfo {year}
  {2017})}\BibitemShut {NoStop}%
\bibitem [{\citenamefont {Nagato}\ \emph {et~al.}(2009)\citenamefont {Nagato},
  \citenamefont {Higashitani},\ and\ \citenamefont {Nagai}}]{Nagato09}%
  \BibitemOpen
  \bibfield  {author} {\bibinfo {author} {\bibfnamefont {Y.}~\bibnamefont
  {Nagato}}, \bibinfo {author} {\bibfnamefont {S.}~\bibnamefont
  {Higashitani}},\ and\ \bibinfo {author} {\bibfnamefont {K.}~\bibnamefont
  {Nagai}},\ }\bibfield  {title} {\bibinfo {title} {Strong anisotropy in spin
  susceptibility of superfluid
  {$^{3}\mathrm{He}\mathrm{\text{\ensuremath{-}}}B$} film caused by surface
  bound states},\ }\href {https://doi.org/10.1143/JPSJ.78.123603} {\bibfield
  {journal} {\bibinfo  {journal} {J. Phys. Soc. Jpn.}\ }\textbf {\bibinfo
  {volume} {78}},\ \bibinfo {pages} {123603} (\bibinfo {year}
  {2009})}\BibitemShut {NoStop}%
\bibitem [{\citenamefont {Ominato}\ \emph {et~al.}(2024)\citenamefont
  {Ominato}, \citenamefont {Yamakage},\ and\ \citenamefont
  {Matsuo}}]{Ominato2024}%
  \BibitemOpen
  \bibfield  {author} {\bibinfo {author} {\bibfnamefont {Y.}~\bibnamefont
  {Ominato}}, \bibinfo {author} {\bibfnamefont {A.}~\bibnamefont {Yamakage}},\
  and\ \bibinfo {author} {\bibfnamefont {M.}~\bibnamefont {Matsuo}},\
  }\bibfield  {title} {\bibinfo {title} {Dynamical majorana ising spin response
  in a topological superconductor--magnet hybrid by microwave irradiation},\
  }\href {https://doi.org/10.1103/PhysRevB.109.L121405} {\bibfield  {journal}
  {\bibinfo  {journal} {Phys. Rev. B}\ }\textbf {\bibinfo {volume} {109}},\
  \bibinfo {pages} {L121405} (\bibinfo {year} {2024})}\BibitemShut {NoStop}%
\bibitem [{\citenamefont {Fang}\ \emph {et~al.}(2014)\citenamefont {Fang},
  \citenamefont {Gilbert},\ and\ \citenamefont {Bernevig}}]{CFang2014}%
  \BibitemOpen
  \bibfield  {author} {\bibinfo {author} {\bibfnamefont {C.}~\bibnamefont
  {Fang}}, \bibinfo {author} {\bibfnamefont {M.~J.}\ \bibnamefont {Gilbert}},\
  and\ \bibinfo {author} {\bibfnamefont {B.~A.}\ \bibnamefont {Bernevig}},\
  }\bibfield  {title} {\bibinfo {title} {New class of topological
  superconductors protected by magnetic group symmetries},\ }\href
  {https://doi.org/10.1103/PhysRevLett.112.106401} {\bibfield  {journal}
  {\bibinfo  {journal} {Phys. Rev. Lett.}\ }\textbf {\bibinfo {volume} {112}},\
  \bibinfo {pages} {106401} (\bibinfo {year} {2014})}\BibitemShut {NoStop}%
\bibitem [{\citenamefont {Liu}\ \emph {et~al.}(2014)\citenamefont {Liu},
  \citenamefont {He},\ and\ \citenamefont {Law}}]{XJLiu2014}%
  \BibitemOpen
  \bibfield  {author} {\bibinfo {author} {\bibfnamefont {X.-J.}\ \bibnamefont
  {Liu}}, \bibinfo {author} {\bibfnamefont {J.~J.}\ \bibnamefont {He}},\ and\
  \bibinfo {author} {\bibfnamefont {K.~T.}\ \bibnamefont {Law}},\ }\bibfield
  {title} {\bibinfo {title} {Demonstrating lattice symmetry protection in
  topological crystalline superconductors},\ }\href
  {https://doi.org/10.1103/PhysRevB.90.235141} {\bibfield  {journal} {\bibinfo
  {journal} {Phys. Rev. B}\ }\textbf {\bibinfo {volume} {90}},\ \bibinfo
  {pages} {235141} (\bibinfo {year} {2014})}\BibitemShut {NoStop}%
\bibitem [{\citenamefont {Kobayashi}\ and\ \citenamefont
  {Furusaki}(2020)}]{Kobayashi2020}%
  \BibitemOpen
  \bibfield  {author} {\bibinfo {author} {\bibfnamefont {S.}~\bibnamefont
  {Kobayashi}}\ and\ \bibinfo {author} {\bibfnamefont {A.}~\bibnamefont
  {Furusaki}},\ }\bibfield  {title} {\bibinfo {title} {Double majorana vortex
  zero modes in superconducting topological crystalline insulators with surface
  rotation anomaly},\ }\href {https://doi.org/10.1103/PhysRevB.102.180505}
  {\bibfield  {journal} {\bibinfo  {journal} {Phys. Rev. B}\ }\textbf {\bibinfo
  {volume} {102}},\ \bibinfo {pages} {180505} (\bibinfo {year}
  {2020})}\BibitemShut {NoStop}%
\bibitem [{\citenamefont {Hu}\ \emph {et~al.}(2022)\citenamefont {Hu},
  \citenamefont {Wu}, \citenamefont {Liu},\ and\ \citenamefont
  {Zhang}}]{Hu2022Competing}%
  \BibitemOpen
  \bibfield  {author} {\bibinfo {author} {\bibfnamefont {L.-H.}\ \bibnamefont
  {Hu}}, \bibinfo {author} {\bibfnamefont {X.}~\bibnamefont {Wu}}, \bibinfo
  {author} {\bibfnamefont {C.-X.}\ \bibnamefont {Liu}},\ and\ \bibinfo {author}
  {\bibfnamefont {R.-X.}\ \bibnamefont {Zhang}},\ }\bibfield  {title} {\bibinfo
  {title} {Competing vortex topologies in iron-based superconductors},\ }\href
  {https://doi.org/10.1103/PhysRevLett.129.277001} {\bibfield  {journal}
  {\bibinfo  {journal} {Phys. Rev. Lett.}\ }\textbf {\bibinfo {volume} {129}},\
  \bibinfo {pages} {277001} (\bibinfo {year} {2022})}\BibitemShut {NoStop}%
\bibitem [{\citenamefont {Hu}\ and\ \citenamefont
  {Zhang}(2023)}]{Hu2023topological}%
  \BibitemOpen
  \bibfield  {author} {\bibinfo {author} {\bibfnamefont {L.-H.}\ \bibnamefont
  {Hu}}\ and\ \bibinfo {author} {\bibfnamefont {R.-X.}\ \bibnamefont {Zhang}},\
  }\bibfield  {title} {\bibinfo {title} {Topological superconducting vortex
  from trivial electronic bands},\ }\href@noop {} {\bibfield  {journal}
  {\bibinfo  {journal} {Nature Communications}\ }\textbf {\bibinfo {volume}
  {14}},\ \bibinfo {pages} {640} (\bibinfo {year} {2023})}\BibitemShut
  {NoStop}%
\bibitem [{\citenamefont {Altland}\ and\ \citenamefont
  {Zirnbauer}(1997)}]{Altland97}%
  \BibitemOpen
  \bibfield  {author} {\bibinfo {author} {\bibfnamefont {A.}~\bibnamefont
  {Altland}}\ and\ \bibinfo {author} {\bibfnamefont {M.~R.}\ \bibnamefont
  {Zirnbauer}},\ }\bibfield  {title} {\bibinfo {title} {Nonstandard symmetry
  classes in mesoscopic normal-superconducting hybrid structures},\ }\href
  {https://doi.org/10.1103/PhysRevB.55.1142} {\bibfield  {journal} {\bibinfo
  {journal} {Phys. Rev. B}\ }\textbf {\bibinfo {volume} {55}},\ \bibinfo
  {pages} {1142} (\bibinfo {year} {1997})}\BibitemShut {NoStop}%
\bibitem [{\citenamefont {Kayser}(1982)}]{Kayser82}%
  \BibitemOpen
  \bibfield  {author} {\bibinfo {author} {\bibfnamefont {B.}~\bibnamefont
  {Kayser}},\ }\bibfield  {title} {\bibinfo {title} {Majorana neutrinos and
  their electromagnetic properties},\ }\href
  {https://doi.org/10.1103/PhysRevD.26.1662} {\bibfield  {journal} {\bibinfo
  {journal} {Phys. Rev. D}\ }\textbf {\bibinfo {volume} {26}},\ \bibinfo
  {pages} {1662} (\bibinfo {year} {1982})}\BibitemShut {NoStop}%
\end{thebibliography}%
\end{document}